\documentclass[usenatbib,fleqn]{mnras}

\usepackage{newtxtext,newtxmath}
\usepackage[T1]{fontenc}
\usepackage{ae,aecompl}

\usepackage{graphicx}
\usepackage{times}
\usepackage{color}
\usepackage{amsmath}
\usepackage{amssymb}	

\usepackage{epstopdf}

\usepackage{geometry}
\title[Escape of LyC and \Lya\ in Clouds]
{Understanding the escape of LyC and Ly$\alpha$ photons from turbulent clouds}
\author[Taysun Kimm et al.]{  
\parbox[t]{\textwidth}{
Taysun Kimm$^{1}$\thanks{e-mail: tkimm@yonsei.ac.kr},
J\'er\'emy Blaizot$^{2}$,
Thibault Garel$^{2,4}$,
L\'eo Michel-Dansac$^{2}$,
Harley Katz$^{3}$,
Joakim Rosdahl$^{2}$, 
Anne Verhamme$^{4}$,
Martin Haehnelt$^{5}$
}
\vspace*{6pt} \\
$^1$ Department of Astronomy, Yonsei University, 50 Yonsei-ro, Seodaemun-gu, Seoul 03722, Republic of Korea\\
$^2$ Univ Lyon, Univ Lyon1, Ens de Lyon, CNRS, Centre de Recherche Astrophysique de Lyon UMR5574, F-69230, Saint-Genis- Laval, France \\
$^3$ Astrophysics, University of Oxford, Denys Wilkinson Building, Keble Road, Oxford OX1 3RH, UK\\
$^4$ Observatoire de Gen\'eve, Universit\'e de Gen\'eve, 51 Ch. des Maillettes, 1290 Versoix, Switzerland\\
$^5$ Kavli Institute for Cosmology and Institute of Astronomy, Madingley Road, Cambridge, CB3 0HA, UK\\
}

\begin{document}
\maketitle

\newcommand{\fbar}{\mbox{$f_{\rm bar}$}}
\newcommand{\mbar}{\mbox{$m_{\rm bar}$}}
\newcommand{\nH}{\mbox{$n_{\rm H}$}}
\newcommand{\kms}{\mbox{${\rm km\,s^{-1}}$}}
\newcommand{\msun}{\mbox{$\rm M_\odot$}}
\newcommand{\Zsun}{\mbox{$\rm Z_\odot$}}
\newcommand{\msunyr}{\mbox{$\rm M_\odot\,{\rm yr^{-1}}$}}
\newcommand{\mvir}{\mbox{$M_{\rm vir}$}}
\newcommand{\mgas}{\mbox{$M_{\rm gas}$}}
\newcommand{\mstar}{\mbox{$M_{\rm star}$}}
\newcommand{\sfr}{\mbox{$\dot{M}_{\rm star}$}}
\newcommand{\mhalo}{\mbox{$M_{\rm halo}$}}
\newcommand{\rvir}{\mbox{$R_{\rm vir}$}}
\newcommand{\mn}{\mbox{{\sc \small Horizon}-MareNostrum}}
\newcommand{\nut}{\mbox{{\sc \small Nut}}}
\newcommand{\ramses}{\mbox{{\sc \small Ramses}}}
\newcommand{\nth}{\mbox{$n_{\rm SF}$}}
\newcommand{\cmq}{\mbox{${\rm cm^{-3}}$}}
\newcommand{\fesc}{\mbox{${\rm f_{\rm esc}}$}}
\newcommand{\lya}{Ly$\alpha$}
\newcommand{\Lya}{Ly$\alpha$}
\newcommand{\LyC}{\mbox{${\rm LyC}$}}
\newcommand{\fescLya}{\mbox{$f_{\rm esc}^{\rm Lya}$}}
\newcommand{\fescLyC}{\mbox{$f_{\rm esc}^{\rm LyC}$}}



\begin{abstract}
Understanding the escape of Lyman continuum (LyC) and Lyman alpha (\Lya) photons from molecular clouds is one of the keys to constraining the reionization history of the Universe and the evolution of galaxies at high redshift. Using a set of radiation-hydrodynamic simulations with adaptive mesh refinement, we investigate how photons propagate and escape from turbulent clouds with different masses, star formation efficiencies (SFEs), and metallicities, as well as with different models of stellar spectra and supernova feedback. We find that the escape fractions in both LyC and \Lya\ are generally increasing with time if the cloud is efficiently dispersed by radiation and supernova feedback. When the total SFE is low (1\% of the cloud mass), $0.1-5\%$ of LyC photons leave the metal-poor cloud, whereas the fractions increase to  $20-70\%$ in clouds with a 10\% SFE. LyC photons escape more efficiently if gas metallicity is lower, if the upper mass limit in the stellar initial mass function is higher, if binary interactions are allowed in the evolution of stars, or if additional strong radiation pressure, such as \Lya\ pressure, is present. As a result, the number of escaping LyC photons can easily vary by a factor of $\sim4$ on cloud scales. The escape fractions of \Lya\ photons are systemically higher ($60-80\%$) than those of LyC photons despite large optical depths at line centre ($\tau_0\sim10^6-10^{9}$). Scattering of \Lya\ photons is already significant on cloud scales, leading to double-peaked profiles with peak separations of $v_{\rm sep}\sim400\,\kms$ during the initial stage of the cloud evolution, while it becomes narrower than $v_{\rm sep} \la 150 \, \kms$ in the LyC bright phase. Comparisons with observations of low-redshift galaxies suggest that \Lya\ photons require further interactions with neutral hydrogen to reproduce their velocity offset for a given LyC escape fraction.
\end{abstract}

\begin{keywords}
Cosmology: reionization -- galaxies: high-redshift 
\end{keywords}

\voffset=-0.6in
\hoffset=0.2in

\section{Introduction}
In a $\Lambda$CDM paradigm, the initial density perturbations develop into the large-scale cosmic web through gravitational interactions. Dark matter haloes (DMHs) form at the intersection of the filaments inside of which gas collapses and forms stars via radiative cooling. During this process, massive OB stars in dwarf-sized galaxies produce a large number of Lyman Continuum (LyC) photons that ionize neutral hydrogen in the Universe \citep{madau99}. As more structures collapse, the ionized bubbles expand and percolate, and the Universe is believed to be fully re-ionized by $z\approx6$. The detection of strong Lyman absorption troughs in the spectra of quasi-stellar objects around $z=6$ \citep{fan01,fan06} confirmed that the Universe was indeed opaque to ionizing radiation at $z\ga 6$, supporting this picture.

The physics behind the propagation of ionizing radiation in an expanding Universe is straightforward, and can be modelled using contemporary cosmological radiation-hydrodynamics (RHD) techniques \citep{wise09,kimm14,wise14,gnedin14,pawlik15,ocvirk16,xu16,kimm17,rosdahl18,finlator18}. Studies show that the first ionized bubbles appear with the emergence of massive Pop III stars in haloes of mass $\sim10^{6-7}\,\msun$, and then expand as subsequent metal-enriched populations provide additional LyC photons. Because the star formation histories in low-mass galaxies are quite intermittent,  the ionized hydrogen in the dense circumgalactic (CGM) and intergalactic medium (IGM) recombines quickly, sometimes reducing the volume filling fraction of the ionized regions. Once galaxies become massive enough to host a number of star-forming clouds, they provide LyC photons continuously, and the bubbles grow from the over-dense regions into the void, while relatively dense filamentary gas remains self-shielded from the background radiation fields \citep[e.g.][]{faucher-giguere10,rosdahl12,chardin18}. These excess ionizing photons photo-heat gas and prevent it from collapsing on to small DMHs, which have a virial velocity less than $\sim10\,\kms$, possibly delaying the growth of stellar mass in dwarf galaxies that we observe in the local Universe \citep[e.g.][]{efstathiou92,gnedin00,somerville02,okamoto08,geen13}.

An important conclusion from numerical experiments is that the majority of the ionizing photons arise from dwarf-sized haloes. Based on the stellar mass-to-halo mass relations and the escape fractions obtained from RHDs, \citet{kimm17} solved the simple equation for reionization \citep{madau99} and found that LyC photons from the halos of mass $10^8\,\msun \la \mvir \la 10^{11}\msun$ must be included to match the end of the reionization epoch as well as the Thompson optical depth measured from the polarization signals of the cosmic microwave background. Improving upon this, \citet{katz18a,katz19} developed a photon tracer algorithm that can follow the sources of ionization directly inside RHD simulations, and found that metal-poor galaxies ($\sim0.001-0.1 \,Z_\odot$) embedded in halos of mass $10^8\,\msun \la M_{\rm halo} \la 10^{10}\,\msun$ are likely to be mainly responsible for the reionization of the Universe. 

However, the detailed process of how LyC photons interact with the gas within the star-forming clouds remains elusive. Several studies point out that the optical depth on $\sim$ 100 pc scales is already quite significant \citep{dove94,kimjh13,kimm14,paardekooper15,trebitsch17}, meaning that a large fraction of LyC photons are absorbed inside the star-forming clouds. Currently, resolving the turbulent structure of molecular clouds is still challenging in galactic-scale simulations, and only a few simulations that adopt very high resolution ($\la$ 1 pc) begin to reproduce the basic observed properties of the star forming clouds, such as the linewidth-size relation \citep{larson81,heyer04}, in idealised disk simulations \citep{hopkins12b,grisdale18}. Considering that the typical resolution of a reionization simulations is even lower ($\sim 10-1000\,{\rm pc}$), one can expect that the complex turbulent structure and corresponding Stromgren sphere inside the star-forming clouds is under-resolved. Thus, the leakage of the ionizing radiation may have been crudely approximated, potentially affecting the conclusions on the escape of LyC photons from the DMHs.

Recently, \citet{dale12} performed smoothed particle hydrodynamics simulations with a photo-ionization code to model the evolution of metal-rich star-forming clouds with different masses. They concluded that the escape fractions of LyC photons can be as high as $\sim90\%$  if clouds of mass $10^4\,\msun < M_{\rm cloud} < 10^6\,\msun$ are efficiently dispersed. \citet{dale13} further showed that the escape fractions are usually very high ($\fescLyC\sim0.2-0.9$) if partially unbound clouds convert a significant fraction ($10-30\%$) of mass into stars. Similarly, \citet{howard18} ran RHD simulations with the {\sc flash} Eulerian code and argued that the leakage of LyC photons is very significant ($\sim65\%$ in clouds of mass $5\times10^4\,\msun < M_{\rm cloud} < 10^5\,\msun$ with $\sim20\%$ star formation efficiencies (SFEs)), although the escape fraction decreases to $\la 10\%$ in a $10^4$ or $10^6\,\msun$ cloud. Given that the majority of the ionizing radiation is produced before SNe explode \citep{leitherer99,bruzual03}, the overall high escape fractions indicate that photo-ionization heating plays a critical role in clearing channels for the LyC photons on cloud scales \citep[e.g.][]{krumholz07b,dale12,peters17,geen16,gavagnin17,kimjg18,kannan18}. Using high-resolution (0.7 pc) cosmological RHDs with and without photo-ionization heating, \citet{kimm17} also confirmed that star-forming gas in metal-poor ($Z\sim0.003\,Z_{\odot}$), dwarf-sized haloes ($M_{\rm halo}\sim10^8\,\msun$) is disrupted rapidly due to photo-heating even before SNe explode, leading to $\fescLyC\sim0.4$ \citep[see also][]{wise14,xu16}. Note that such a high escape fraction is indeed observed in compact starburst galaxies where a copious amount of ionizing radiation is being emitted  \citep[][c.f. \citealt{leitherer16,puschnig17}]{de-barros16,shapley16,bian17,izotov18,vanzella18}. These corroborate that the propagation of photons during the early stage of star formation needs to be better understood to make firm predictions on the escape of LyC photons from galaxies.

In principle, it would be best to directly observe the ionizing part of the stellar spectrum from galaxies at $z\ga6$ to determine the contribution of dwarf galaxies to the reionization of the Universe. However, few LyC photons would survive from absorption by neutral IGM at $z\ge6$. An alternative method has thus been put forward to select the LyC leaking candidates and study their properties based on the profile of Lyman $\alpha$ (\Lya) line \citep{verhamme15,dijkstra16}. The basic idea is that because \Lya\ photons resonantly scatter with neutral hydrogen, the emerging line width depends on how gas is distributed around the source. If there exists low-density channels in which few LyC photons would be absorbed, \Lya\ would escape from the medium with little shift in frequency, and thus LyC leaking galaxies are likely to show a velocity profile narrower than $300\,\kms$ in \Lya. Given that \Lya\ is one of the strongest lines observed in the spectra of galaxies at high redshifts \citep[e.g.][]{shapley03}, the approach seems promising and may be used in large observational programmes \citep[e.g.][]{marchi17,steidel18}. 

The beauty of \Lya\ is that it can be used not only to pre-select the LyC leakers, but also to infer the kinematics of the interstellar medium (ISM) in galaxies. As is well established in the literature \citep[e.g.][]{ahn03,verhamme06,dijkstra06,barnes11}, an expanding medium would preferentially absorb photons with frequencies shorter than the line centre, resulting in a spectrum with a pronounced red peak. This type of profile is often observed in star-forming, Lyman break galaxies \citep[e.g.][]{steidel10,kornei10}, and may be used to estimate the amount of galactic outflows. However, because \Lya\ photons primarily arise from young star-forming regions through recombination in gas ionized by LyC radiation, it is necessary to understand how \Lya\ photons are created and propagate on cloud scales before they interact with the ISM. Unfortunately, current models make use of either idealized environments, such as uniform or clumpy distributions, or gas distributions from galactic scale simulations where ISM structures are under-resolved. Little work has been done thus far based on cloud simulations where internal turbulent structures are well resolved. Despite the success at reproducing the overall features of \Lya\ profiles \citep{verhamme08,gronke17}, the simple models may be improved to better interpret the line shape and to make more accurate predictions to find the LyC leaking candidates by studying the propagation of \Lya\ photons inside star-forming clouds.

To this end, we perform high-resolution RHD simulations of turbulent gas clouds with feedback from supernovae and stellar radiation. The aim of these experiments is three-fold. First, we attempt to understand the absorption and propagation of LyC photons in clouds with complex turbulent structures by varying the mass of the cloud and the total SFE. Second, there are several uncertainties regarding stellar evolution, such as the maximum mass of stars \citep{crowther10} or the evolution of the spectral energy distributions (SEDs) due to binary interactions \citep[e.g.][]{stanway16}. These can affect the predictions of the reionization of the Universe as well as the SFE inside the cloud \citep{geen18}, and thus it is necessary to understand what level of uncertainty is implicitly inherited from our limited understanding of stellar evolution. Finally, we aim to examine the emergent \Lya\ profiles from the clouds and compare them with previous results so that the LyC candidates are more efficiently pre-selected in observations. The \Lya\ profiles obtained from this work may be used to model the fraction of bright Lyman alpha emitters (LAEs) in semi-numerical approaches \citep[e.g.][]{choudhury15,mesinger15,weinberger18}, and to infer the relevance of the demise in the fraction of strong LAEs during the epoch of reionization \citep[e.g.][]{stark10,treu13,schenker14}.

This paper is organized as follows. In Section 2, we present the initial conditions and input physics of the simulations. Section 3 presents how the escape fractions evolve in a cloud with different SFE and stellar spectra. Section 4 discusses the effects of turbulent structures, the connection between the escape fractions of LyC and \Lya\ photons,  the implications for reionization, and the impact of star formation and feedback schemes. The summary and conclusions are given in Section 5.

\begin{table}
   \caption{The eight photon groups used in our simulations. From left to right, each column describes the name of the photon group, minimum and maximum photon energy, dust opacity, and main purpose of each photon group. }
   \centering
   \begin{tabular}{lcccl}
   \hline
  Photon & $\epsilon_0$ & $\epsilon_1$ & $\kappa$  & Main function \\
  group           &      [eV]          &   [eV] &   [$\rm cm^2/g$]   & \\
     \hline
   EUV$_{\rm HeII}$ & 54.42 & $\infty$ & $10^3$ & HeII ionisation\\
   EUV$_{\rm HeI}$ & 24.59 & 54.42 & $10^3$ & HeI ionisation\\
   EUV$_{\rm HI,2}$ & 15.2 & 24.59 & $10^3$ & HI and $\rm H_2$ ionisation\\
   EUV$_{\rm HI,1}$ & 13.6 & 15.2 & $10^3$ & HI ionisation\\
   LW & 11.2 & 13.6 & $10^3$ & $\rm H_2$ dissociation\\
   FUV & 5.6 & 11.2 & $10^3$ & Photoelectric heating\\
   Optical & 1.0 & 5.6 & $10^3$ & Radiation pressure\\
   IR & 0.1 & 1.0 & 5 & Radiation pressure\\
        \hline
   \end{tabular}
   \label{tab:photon}
\end{table}

\begin{table*}
   \caption{List of simulations performed in this study. From left to right, each column shows the name of the simulation, the size of the simulated box, the maximum cell width, the mass and half-mass radius of the cloud, the total stellar mass, the initial gas metallicity, the type of SED, the upper mass limit to the IMF, the time of the last snapshot, the placement of stars in the initial conditions, and necessary remarks. }
   \centering
   \begin{tabular}{lcccccccccccc}
   \hline
  Name &  $L_{\rm box}$  & $\Delta x_{\rm min}$  & $M_{\rm cloud}$ & $r_{\rm 1/2}$ & $M_{\rm star}$ & $Z_{\rm gas}$ & SED & $M_{\rm max}$ &  $t_{\rm final}$ & SF & Remarks  \\
     & [pc] & [pc] & [$\msun$] &  [pc]  &[$\msun$]  &  & & [$\msun$] & [Myr]  &  &  \\
     \hline
   {\tt M6\_SFE10} &  512 & 0.25 &  $10^6$ & 35 & $10^5$   & 0.002 & BPASS  & 100 & 10 & random & \\
  {\tt M6\_SFE1}  &  512 & 0.25  & $10^6$ & 35 & $10^4$   & 0.002 & BPASS  & 100 & 20  & random & Fiducial\\
   {\tt M5\_SFE10} &  256 & 0.25 & $10^5$ & 13 &  $10^4$ & 0.002 & BPASS   & 100 & 7  & random &\\
   {\tt M5\_SFE1} &  256 & 0.25 & $10^5$ &13 & $10^3$ & 0.002 & BPASS & 100 & 20  & random &\\
     \hline
   {\tt M6\_SFE10\_sng} &  512 & 0.25 & $10^6$ & 35 & $10^5$  & 0.002 & BC03 & 100 &  10  & random & Single stellar SED\\
   {\tt M6\_SFE1\_sng} & 512 & 0.25 & $10^6$ & 35 & $10^4$  & 0.002 & BC03 & 100 & 20  & random & Single stellar SED \\
   {\tt M6\_SFE10\_300} &  512 &0.25 & $10^6$ & 35 & $10^5$  & 0.002 & BPASS & 300 & 10  & random &  \\
   {\tt M6\_SFE1\_300} &  512 & 0.25 & $10^6$ & 35 & $10^4$  & 0.002 & BPASS & 300 & 20  & random  & \\
   {\tt M6\_SFE10\_Zsun} &  512 & 0.25 & $10^6$ & 35 & $10^5$  & 0.02 & BPASS  & 100 & 10  & random & \\
   {\tt M6\_SFE1\_Zsun} & 512 & 0.25 & $10^6$ & 35 & $10^4$  & 0.02 & BPASS  & 100 & 20  & random & \\
       \hline
    {\tt M6\_SFE1\_noTurb} & 512 & 0.25 & $10^6$ & 35 & $10^4$  & 0.02 & BPASS  & 100 & 20  & random & No turbulence\\
     {\tt M6\_SFE10\_noSN}  &  512 & 0.25  & $10^6$ & 35 & $10^5$   & 0.002 & BPASS  & 100 & 10  & random & No SNe\\
    {\tt M6\_SFE1\_noSN}  &  512 & 0.25  & $10^6$ & 35 & $10^4$   & 0.002 & BPASS  & 100 & 20  & random & No SNe\\
         \hline
     {\tt M6\_SFE1\_dSF} & 512 & 0.25 & $10^6$ & 35 & $10^4$  & 0.02 & BPASS  & 100 & 20  & dense & \\
   {\tt M6\_SFE1\_dSF\_PLya} & 512 & 0.25 & $10^6$ & 35 & $10^4$  & 0.02 & BPASS  & 100 & 20  & dense & \Lya\ pressure\\
        \hline
   \end{tabular}
   \label{tab:sim}
\end{table*}

\section{Simulations}

We perform 15 RHD simulations with different cloud masses, SFEs, SEDs, metallicities, and feedback to examine the escape fraction of LyC and \Lya\ photons on cloud scales using {\sc ramses-rt} \citep{teyssier02,rosdahl13,rosdahl15}. The Euler equations are solved using the HLLC method with the positivity-conserving slope limiter \citep{toro94}. We adopt a Courant number of 0.7. The Poisson equation is evolved using the multi-grid method \citep{guillet11}. A uniform UV background is turned on with the self-shielding approximation, such that gas denser than $\nH \simeq 0.01\,\cmq$ is not affected by heating \citep{rosdahl12}. In order to model the photoelectric heating as well as the transport of ionizing radiation, we use the GLF solver with eight photon groups, as detailed in Table~\ref{tab:photon}. We adopt the frequency-dependent cross-sections \citep{katz17} based on \citet{rosdahl13} and \citet{baczynski15}. The evolution of seven chemical species (H{\sc I}, H{\sc II}, He{\sc I}, He{\sc II}, He{\sc III}, H$_{\rm 2}$, and {\sc $e^{-}$}) is followed by solving photo-chemistry equations \citep[see][for details]{katz17,kimm17}. The speed of light is reduced to $10^{-3}\,c$ to reduce the computation, where $c$ is the full speed of light. 

The initial conditions follow Gaussian density distributions with the maximum densities of $\nH=200\, \cmq$ and $72\, \cmq$ and a 1$\sigma$ radius of 10 pc and 30 pc for clouds of gas mass $10^5\,\msun$ and $10^6\, \msun$, respectively. We then add Kolmogorov turbulence with a power spectrum of the form $ \propto k^{-5/3}$, adopting the mixture of the solenoidal (60\%) and compressive (40\%) mode for one free-fall time (4 and 5 Myr, respectively). The resulting turbulent energy is about $\approx$80\% of the gravitational binding energy inside the half-mass radius, and thus the simulated clouds are marginally gravitationally bound initially. At later epochs, turbulence is generated either by radiation feedback or SN explosions. The clouds are assumed to be metal-poor ($Z_{\rm gas}=0.1\,Z_{\odot}$), motivated by the fact that the metallicity of galaxies typically observed at high redshift is low \citep[$\sim0.2\,Z_\odot$,][]{pettini00,song14,bouwens16,tamura18} and that reionization is likely to be driven by metal-poor dwarf galaxies \citep[e.g.][]{katz19}. Note that this is one of the main differences of our work compared to previous studies where simulated clouds are preferentially metal-rich \citep[e.g.][]{dale12,howard18,kimjg18,geen18}. However, we also test the solar metallicity case to determine the effects of metallicity on the escape of LyC and \Lya\ photons.

\begin{figure}
   \centering
   \includegraphics[width=8.6cm]{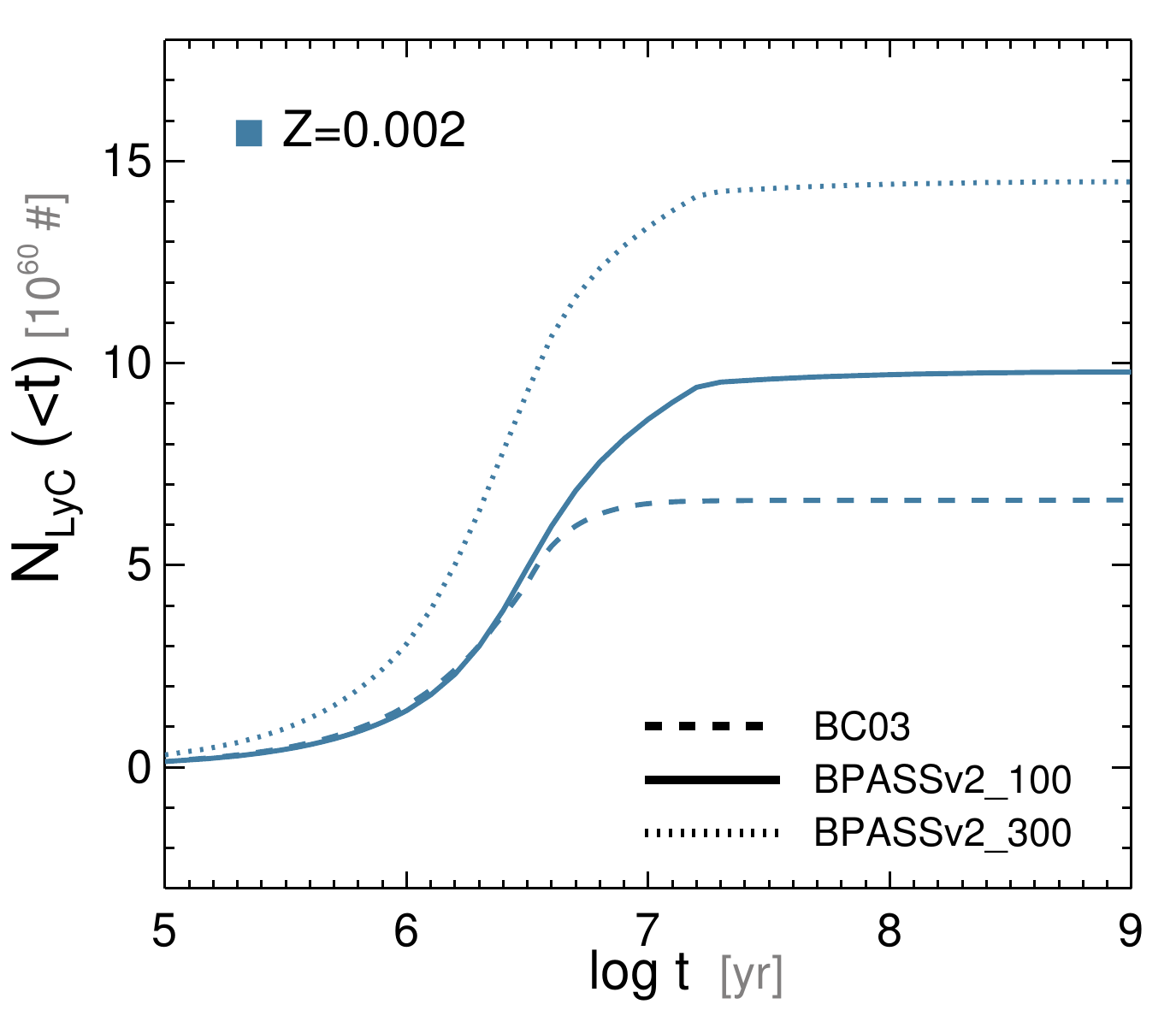} 
   \caption{The cumulative number of LyC photons generated from a simple stellar population of   1$\,\msun$. The number of LyC photons from single \citep[dashed,][]{bruzual03} and binary    \citep[solid and dotted,][]{stanway16} stellar evolution is shown with different line styles. We also test the binary SEDs with two different maximum cut-off masses, $100\,\msun$ (solid) and $300\,\msun$ (dotted), in order to examine the possible uncertainty in predicting the number of escaping photons.  }
   \label{fig:input}
\end{figure}

The simulations include five different forms of stellar feedback, i.e., photo-ionization heating by UV photons \citep{rosdahl13}, direct radiation pressure\footnote{We note that momentum transfer to the neighbouring cells may be under-estimated when the Stromgren sphere is not properly resolved. To alleviate this, we adopt a correction scheme that takes into account the isotropic flux (\texttt {rt\_isoPress}) \citep{rosdahl15}.} by UV and optical photons \citep{rosdahl15}, radiation pressure by multiple scatterings of IR photons \citep{rosdahl15}, photo-electric heating on dust \citep{katz17,kimm17}, and Type II supernova feedback. We model supernova feedback using the mechanical scheme \citep[][see also \citealt{hopkins14}, \citealt{smith18m}, \citealt{lupi19} for a similar method]{kimm14,kimm15}, which is designed to ensure the correct momentum input to the surroundings \citep{thornton98}, with a realistic time delay between 4 and 40 Myr based on the lifetime of massive ($8-100\,\msun$) main-sequence stars \citep{leitherer99}. We assume an SN frequency of $0.011\, \msun^{-1}$, appropriate for the Kroupa initial mass function \citep[IMF,][]{kroupa01}. Each SN event ejects metals of mass 1.4 $\msun$ (i.e., metallicity of the ejecta is set to 0.075), and we assume the initial ejecta energy to be $10^{51}\,{\rm erg}$. The effect of photo-electric heating on dust is included by explicitly following the propagation and absorption of the photons at the Habing band \citep[see Table~\ref{tab:photon},][]{kimm17}. We assume that the amount of dust is proportional to the amount of metals with a dust-to-metal ratio of 0.4 \citep[e.g.][]{dwek98,draine07b} at $T<10^6\,{\rm K}$. At higher temperatures, dust is assumed to be destroyed as a crude approximation for the thermal sputtering process.

The fiducial model uses the binary star SEDs from \citet{stanway16}, but we also examine the SEDs with single stellar evolution \citep{bruzual03}. These SEDs are generated with a lower (upper) limit of mass of $0.1\,\msun$ ($100\,\msun$) and are used to compute the instantaneous luminosity as a function of age, in 8 photon groups representing radiation energy (or frequency) intervals as shown in Table~\ref{tab:photon}. We also test SEDs generated with the upper mass limit of $300\,\msun$ with binary stellar evolution to examine the importance of the IMF on the predictions of the escape fractions. Note that the cumulative number of LyC photons from metal-poor populations can vary by a factor of $\sim2$ depending on this assumption of the stellar population (Figure~\ref{fig:input}) \citep[see also][for different estimates of the total number of LyC photons arising from variations in stellar IMF, metallicity, and rotation rates]{topping15}. We summarize the initial conditions of the simulations in Table~\ref{tab:sim}.

We consider two cases of SFE, 1\% and 10\% SFE, where we place star particles in the cloud comprising 1\% and 10\% of the cloud mass, respectively. Recent studies suggest that the total stellar mass formed in the simulations may depend on how radiation pressure is modelled \citep{hopkins18,kimjg18,krumholz18} and how turbulence is driven \citep{geen18}. It can also be affected by the inclusion of other feedback processes, such as radiation pressure by \Lya\ photons \citep{dijkstra08,smith17,kimm18} or re-processed infrared photons \citep{hopkins12a,krumholz12,rosdahl15b,skinner15,tsang15}. In order to circumvent these uncertainties, we simply start with stars as an initial condition instead of modelling formation and accretion onto sink particles \citep[e.g.][]{bate95}. Specifically, we randomly place star particles of mass $10^3\,\msun$ in the inner region of the cloud following the Gaussian distribution of the dispersion of 5 pc. The resulting mean radius of the stellar distributions is $\approx$ 3 pc, and the average gas density of their host cells is initially $\left<\nH\right>\sim10$--$20\,\cmq$. 

We stop the simulations when $\approx 98\%$ of the ionizing radiation is emitted from a simple stellar population with binaries ($t=20\,{\rm Myr}$) for the 1\% SFE cases. This is chosen to minimize the computational cost, as covering 99\% of the total LyC radiation would require the modelling of additional $\approx 20\,{\rm Myr}$. For the runs with 10\% SFEs, we choose the final snapshot in which the instantaneous escape fraction reaches near 100\%, which corresponds to 7 and 10 Myr for the $M_{\rm cloud}=10^5\,\msun$ and $10^6\,\msun$ cloud, respectively. Note that a large number of snapshots ($\approx 70-100$) is generated with a time interval of $\Delta t=0.1\,{\rm Myr}$, $0.2\,{\rm Myr}$, and $0.25\,{\rm Myr}$ at $0\le t \le 1\,{\rm Myr}$, $1\le t \le 5\,{\rm Myr}$, $5\le t \le 20\,{\rm Myr}$, respectively, for an accurate determination of the luminosity-weighted time average of escape fractions.

\begin{figure*}
   \centering
   \includegraphics[width=17.5cm]{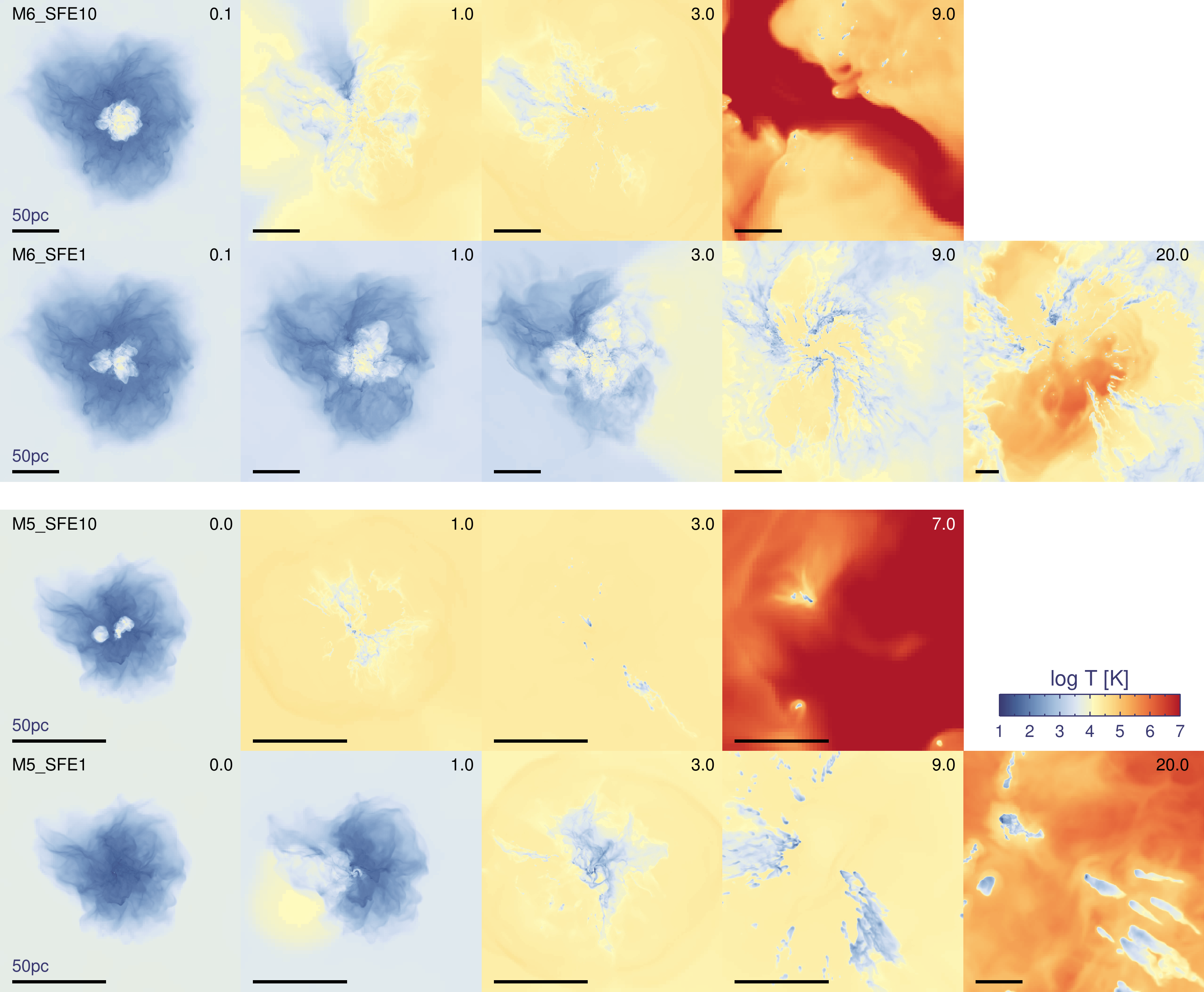} 
   \caption{Evolution of the simulated clouds with radiation and SN feedback. Each column shows projected mass-weighted temperature distributions at different times, as indicated in the top right corner in units of Myr. Different rows correspond to models with different cloud mass (\texttt{M6},   \texttt{M5}: $M_{\rm cloud}=10^6$, $10^5\,\msun$) and stellar mass (\texttt{SFE10}, \texttt{SFE1}: $M_{\rm star}=0.1 \,M_{\rm cloud}$, $0.01 \,M_{\rm cloud}$). The clouds are irradiated with SEDs generated with binary stellar evolution \citep[bpass\_v2,][]{stanway16}. The black bar in the bottom of each panel displays the scale of 50 pc. 
   }
   \label{fig:img}
\end{figure*}

\begin{figure}
   \centering
   \includegraphics[width=7.cm]{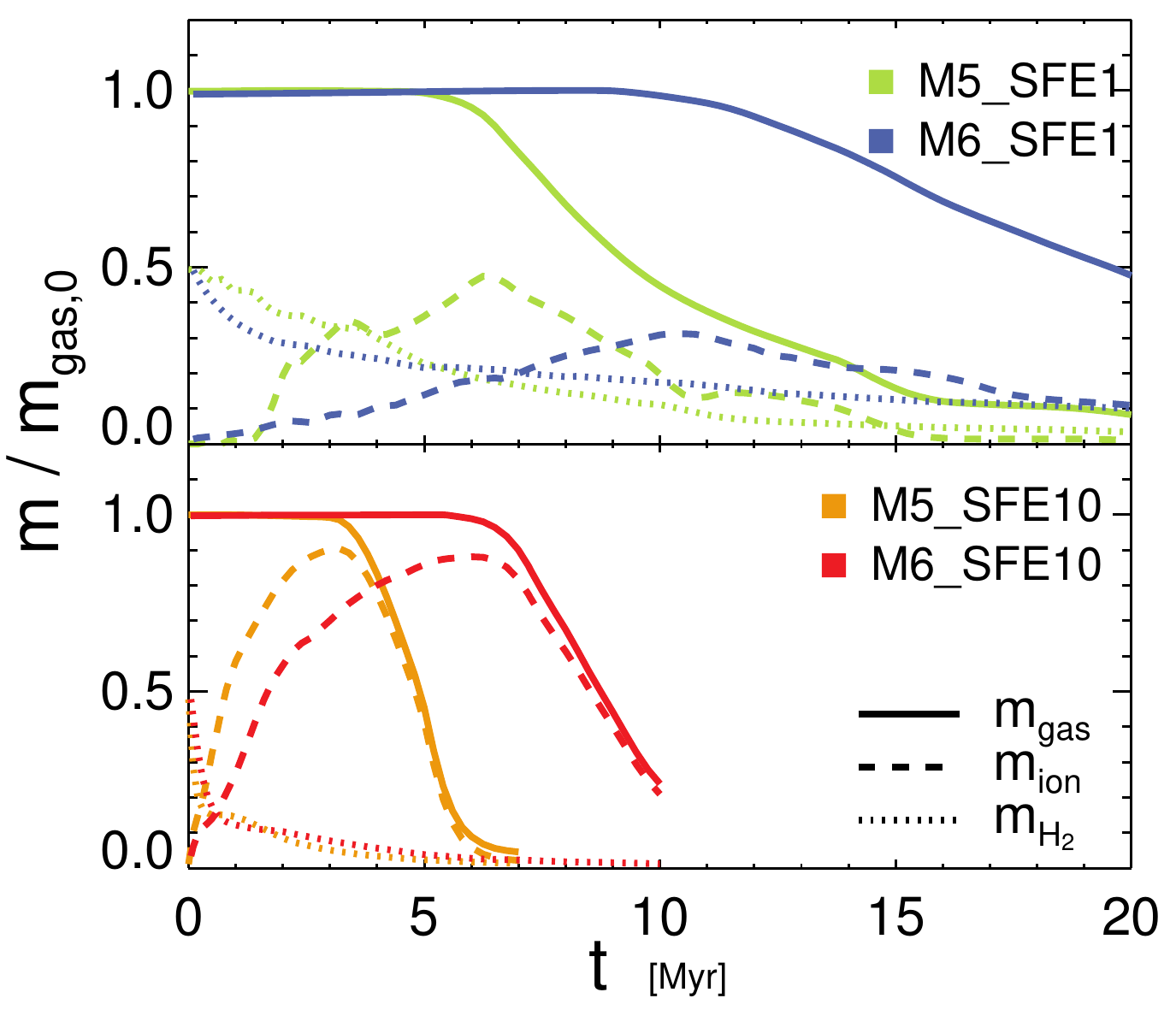} 
         \includegraphics[width=7.cm]{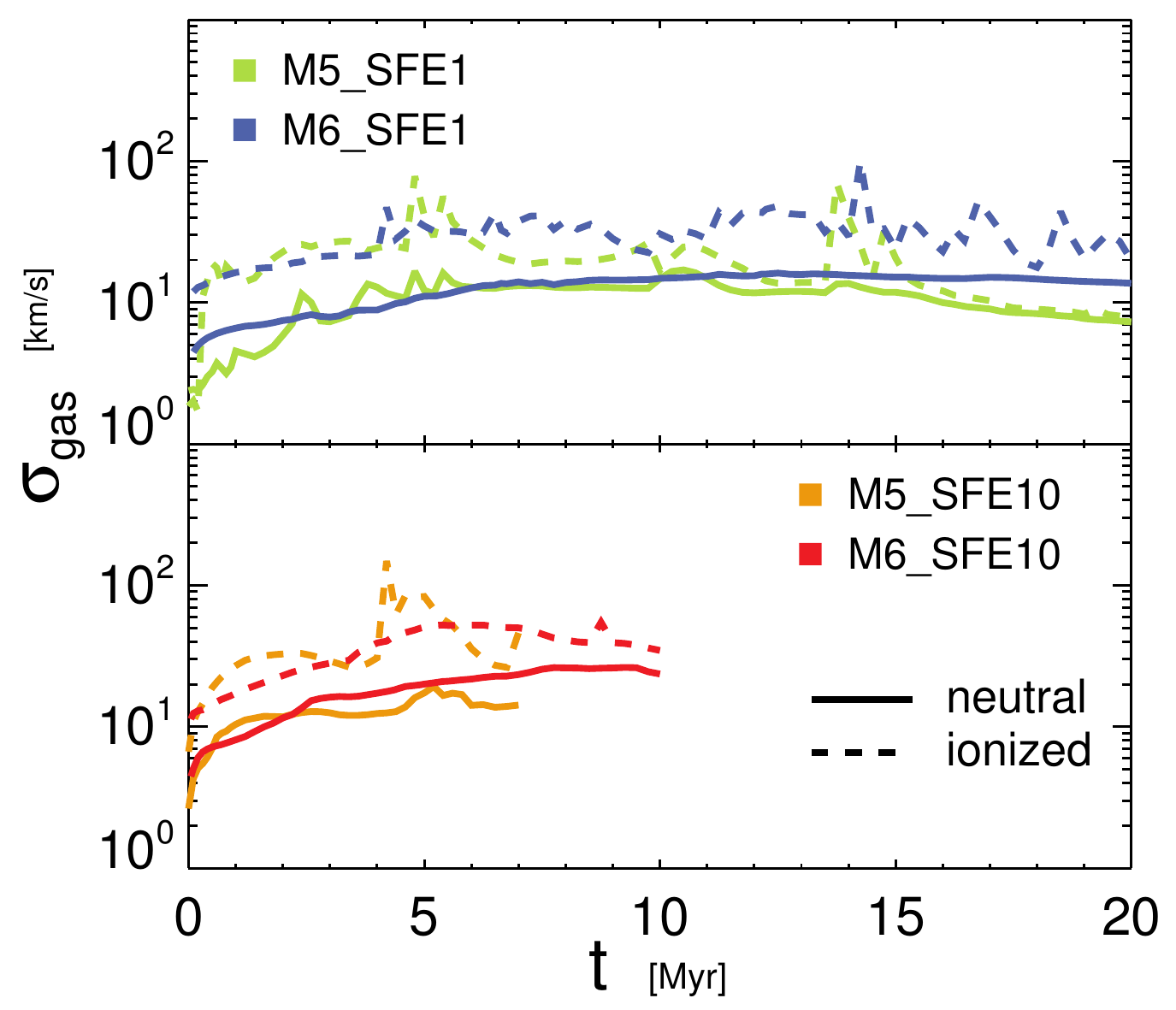} 
   \caption{Evolution of the mass (top; $m$) and the velocity dispersion (bottom; $\sigma_{\rm gas}$) in the simulated clouds. The top panel shows the mass fractions of the total (solid), ionized (dashed), and molecular gas (dotted), normalised to the initial total gas mass ($m_{\rm gas,0}$) in the simulated domain, in the clouds of a 1\% SFE, while the results from the clouds with a 10\% SFE are shown in the second panel. The velocity dispersion ($\sigma_{\rm gas}$) of the neutral and total gas are shown as solid and dashed lines, respectively, in the third (1\% SFE) and the fourth (10\% SFE) panels. Different colour codings denote the clouds of different masses, as indicated in the legend.   }
   \label{fig:cloud_mass}
\end{figure}

\begin{figure*}
  \centering
  \includegraphics[width=5.5cm]{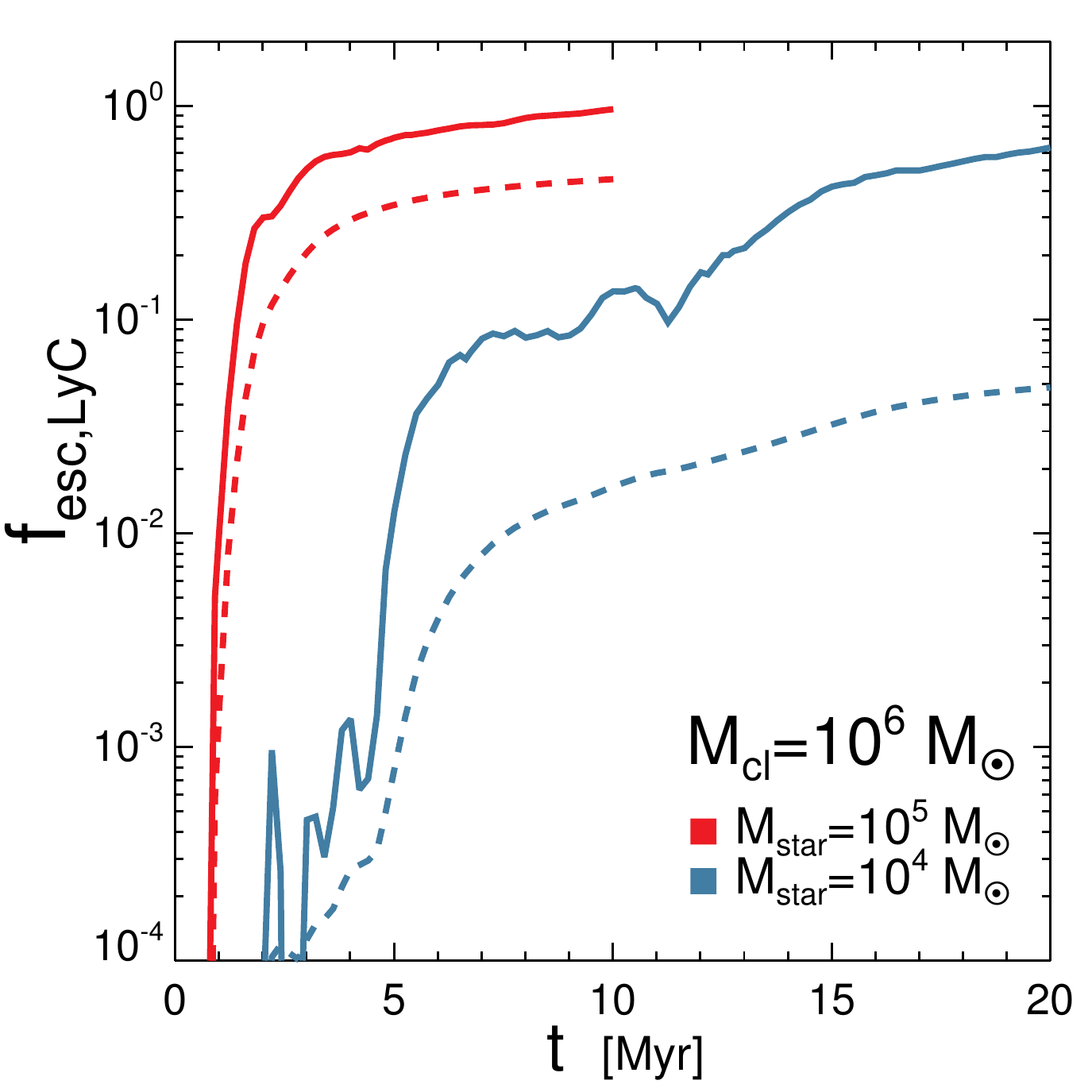} 
  \includegraphics[width=5.5cm]{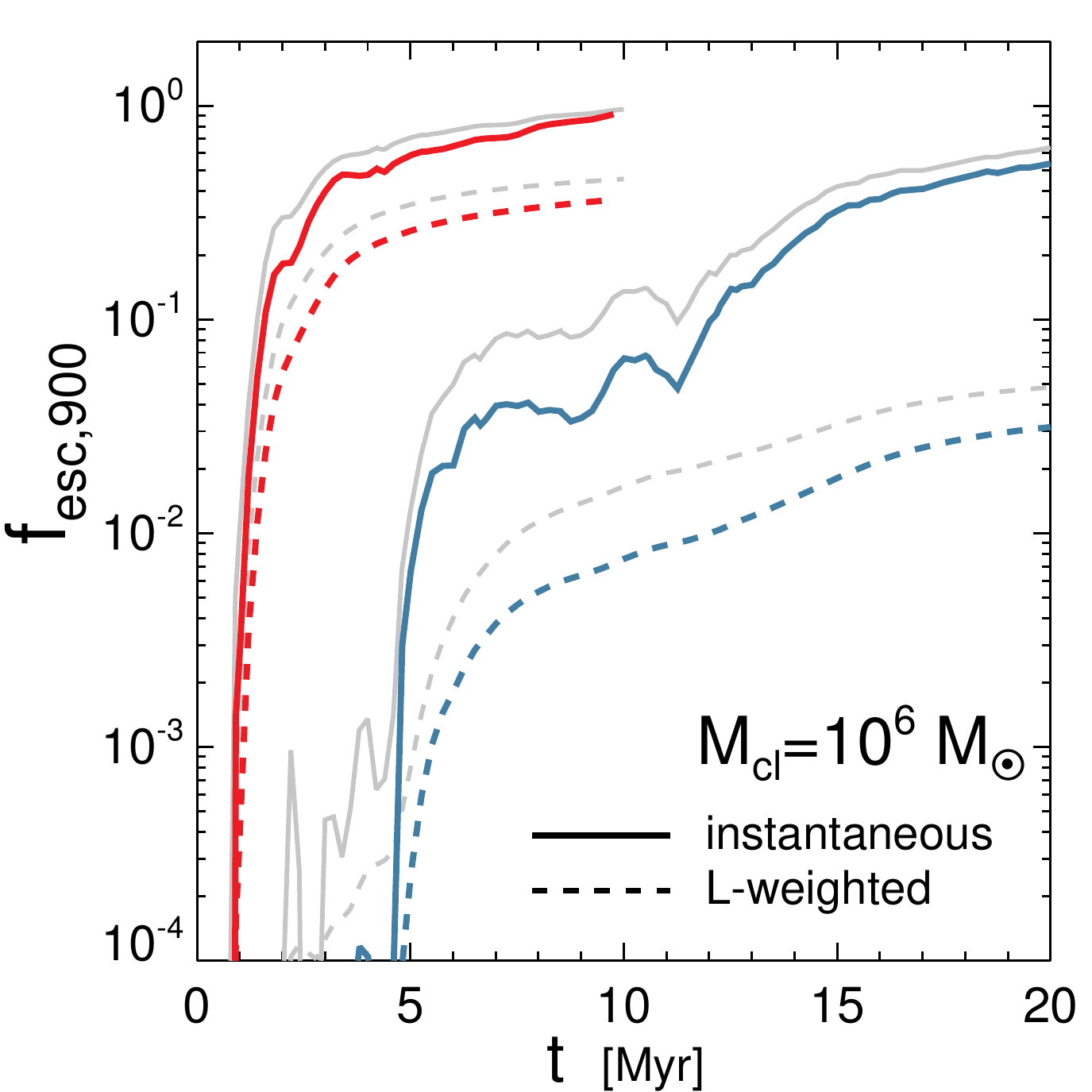}
    \includegraphics[width=5.5cm]{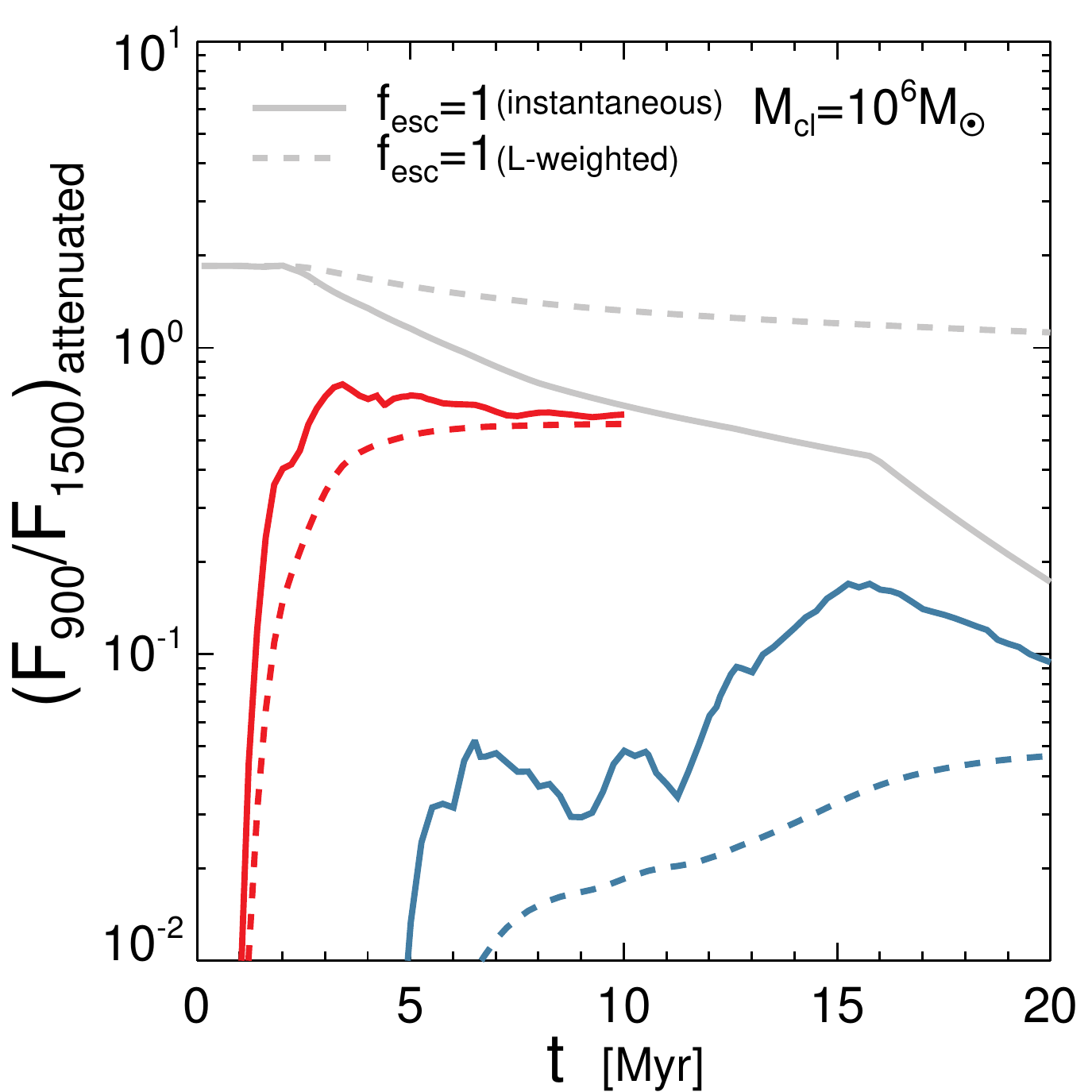}  
      \includegraphics[width=5.5cm]{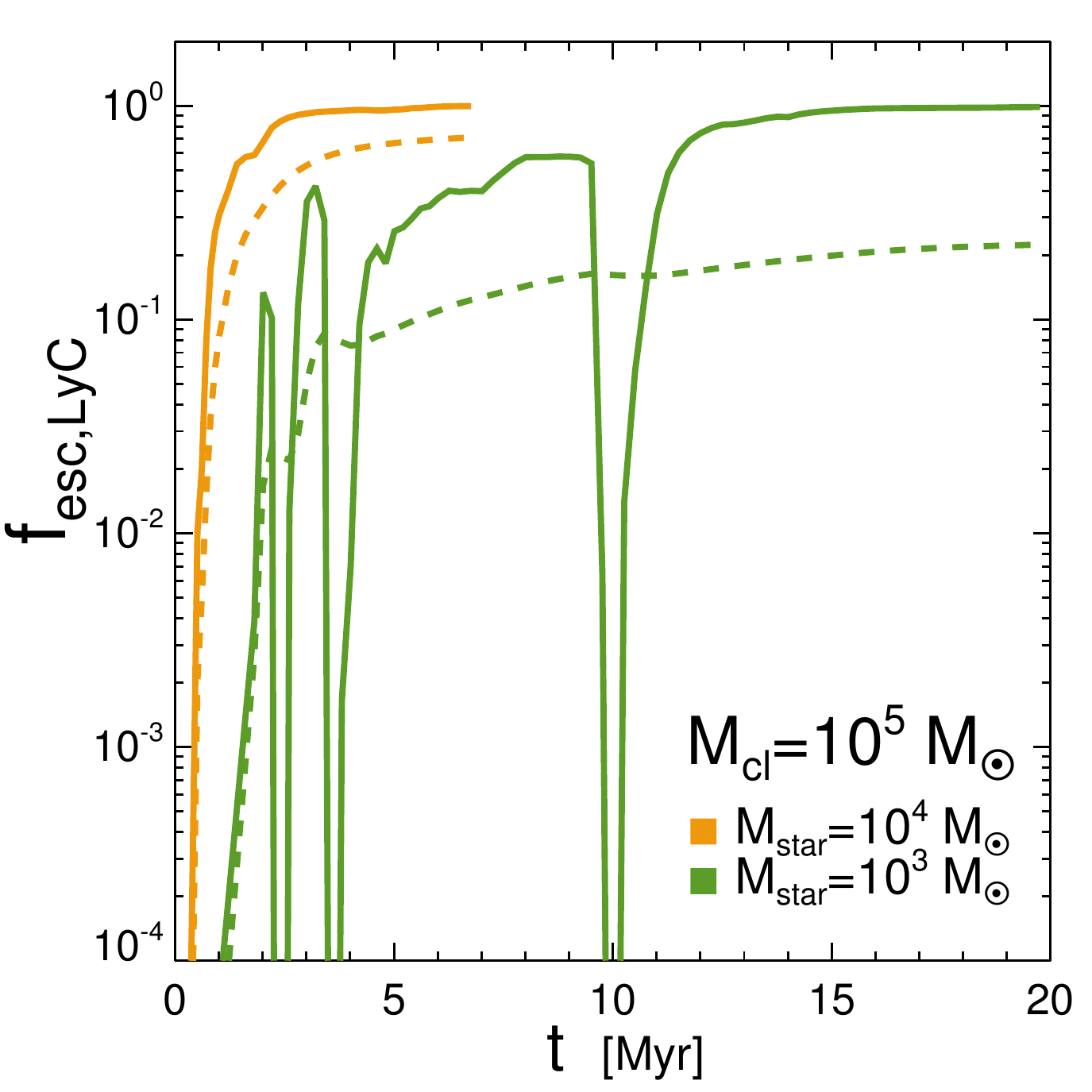} 
  \includegraphics[width=5.5cm]{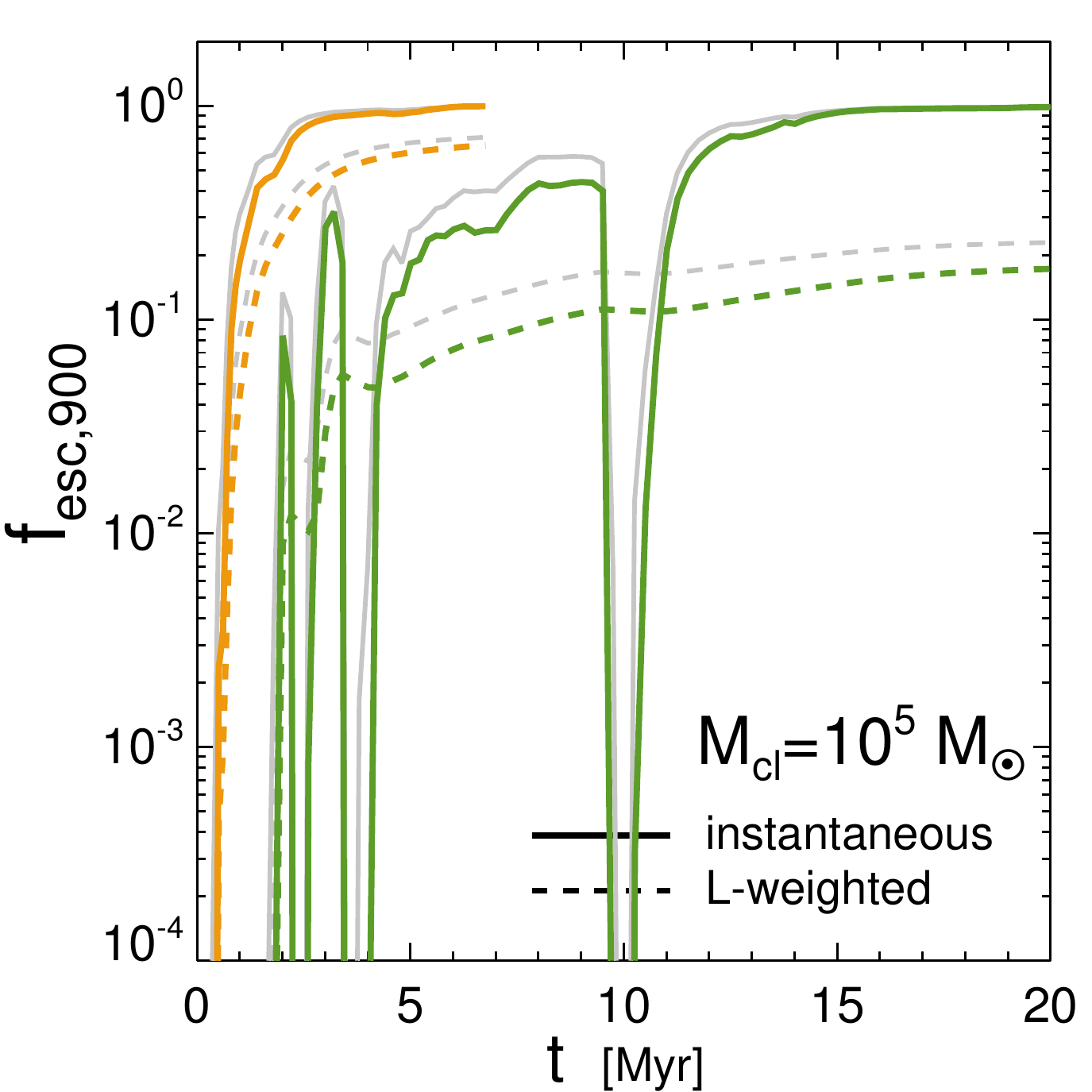}
      \includegraphics[width=5.5cm]{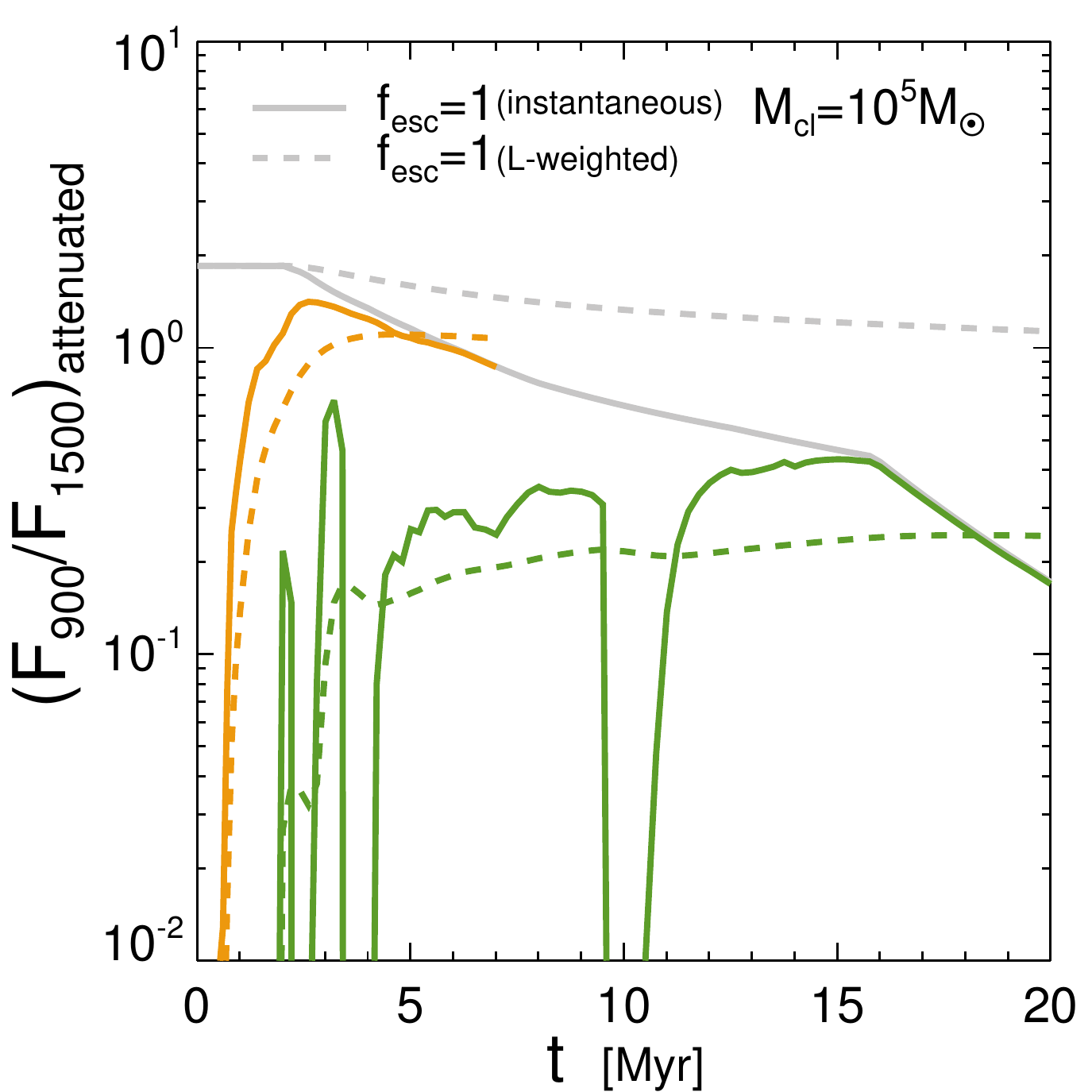}  
   \caption{Escape fractions of LyC photons and the attenuated flux density ratio between $900$ and $1500$ \AA\ ($F_{900}/F_{1500}$) in turbulent clouds. Left panels show the escape fractions measured based on the photons of the wavelength $0<\lambda<912$ \AA\ ($f_{\rm esc,LyC}$), whereas the middle panels indicate the escape fractions measured at $880<\lambda<910$ \AA\ ($f_{\rm esc}^{\rm 900}$). The top and bottom panels show the escape fraction in a $10^6\,\msun$ and $10^5\,\msun$ cloud, respectively. Different color-codings denote the runs with different stellar masses. Solid lines display the instantaneous quantities, while dashed lines show the time-averaged, luminosity-weighted escape fractions or flux ratio until the age of $t$. In the middle panels, we include $f_{\rm esc,LyC}$ as grey lines for comparison. Grey lines in the right panels indicate the intrinsic flux density ratio. All models are based on binary SEDs with an upper stellar mass limit of $100\,\msun$. The escape fraction increases with increasing SFE and decreasing cloud mass.}
   \label{fig:fescLyC}
\end{figure*}

\begin{figure}
   \centering
   \includegraphics[width=8cm]{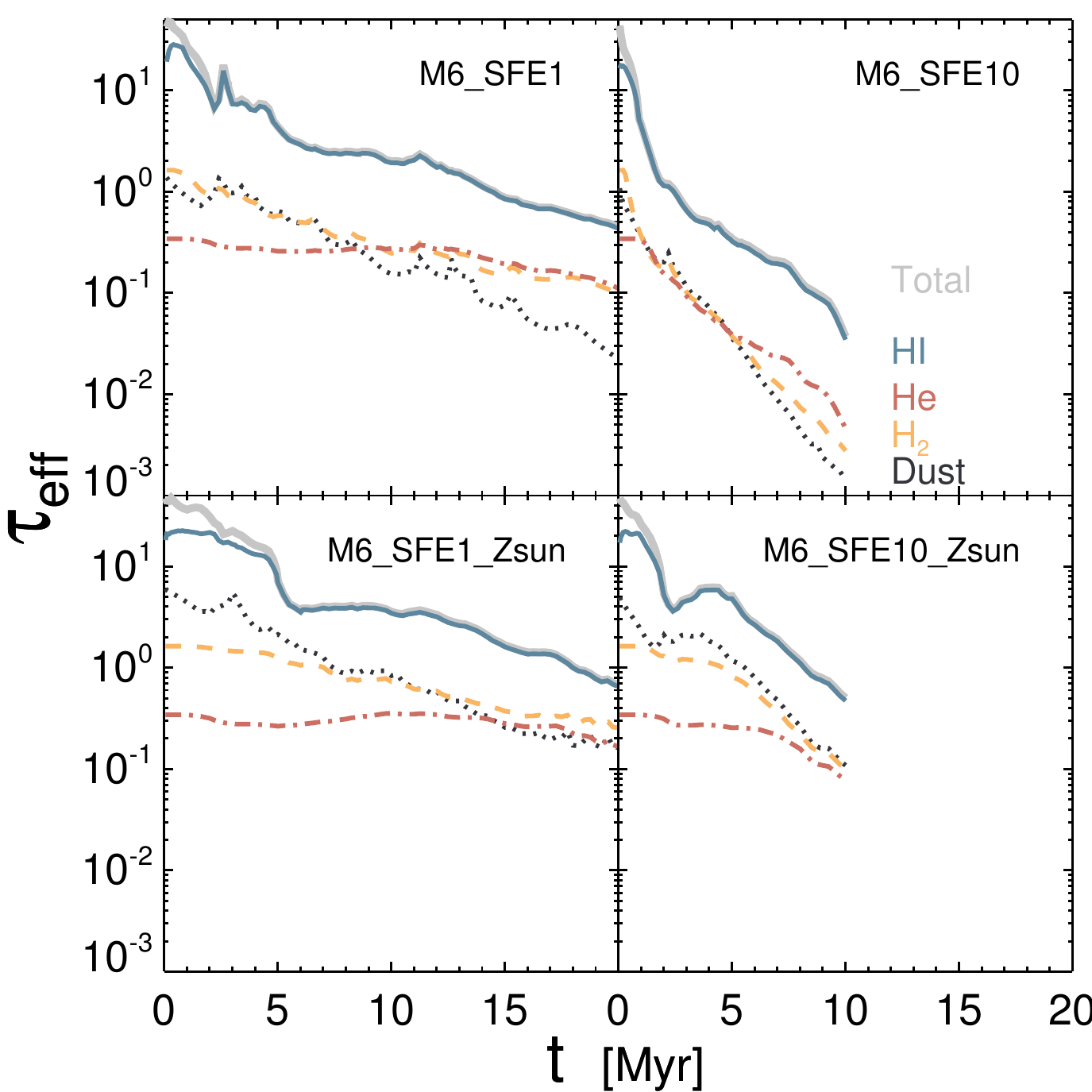}
   \caption{Effective optical depth to LyC photons ($\tau_{\rm eff}\equiv - \ln \fescLyC$) in the simulated clouds. Different colour codings and line styles indicate $\tau_{\rm eff}$ due to different elements (grey solid: total, blue solid: neutral hydrogen, red dot-dashed: neutral and single ionized helium, orange dashed: molecular hydrogen, black dotted: dust). The top panels show $\tau_{\rm eff}$ in the metal-poor massive clouds with different SFE, while the metal-rich counterparts are included in the bottom. Note that the absorption due to dust, helium, and molecular hydrogen is sub-dominant compared to that due to neutral hydrogen. }
   \label{fig:tau_break}
\end{figure}

\section{Results}
\subsection{Evolution of the clouds}
We begin our discussion by describing the general features of the evolution of the turbulent clouds. Figure~\ref{fig:img} shows the temperature distributions of the clouds with different SFEs and cloud masses. Ionizing photons are initially completely absorbed by the host cells of young star particles, imparting momentum of $L_{\rm ion}/c$, where $L_{\rm ion}$ is the luminosity of ionizing radiation.  The absorption of longer wavelength photons by dust also transfers momentum to the surroundings, but their contribution to the total direct radiation pressure is sub-dominant near young stellar populations \citep[e.g.][]{leitherer99}. The Stromgren sphere develops shortly after several recombination time scales ($\sim 10^{-3}-10^{-2}\,{\rm Myr}$), which over-pressurises the gas surrounding the radiation source. Once photo-ionization heating creates low-density channels, the ionization bubble expands faster, as gas at larger radii is more tenuous. Between 4 Myr and 40 Myr, SNe explode intermittently, enhancing outflows that were originally accelerated by photo-ionization heating and direct radiation pressure. 

We find that the cloud with a larger number of stars is disrupted more quickly than that with a low SFE. The turbulent cloud is destroyed as early as $\sim$ 1--3 Myr in the case of the model with a 10\% SFE (\texttt{M6\_SFE10} and \texttt{M5\_SFE10}), indicating that radiation feedback is strong enough to blow out the gas \citep{dale12,dale14,walch12,geen16,kim17}. Very dense, star-forming gas is all destroyed, and approximately 80\% and 95\% of the gas leaves the simulated domain by the end in the \texttt{M6\_SFE10} and \texttt{M5\_SFE10} runs, respectively (Figure~\ref{fig:cloud_mass}).  As discussed extensively in the literature, photo-ionization is mainly responsible for this process \citep[e.g.][]{matzner02}, while direct radiation pressure is more significant in the dense regime ($\nH \ga 10^{5}\,\cmq$)  \citep[e.g.][]{rosdahl15b,kimm17,kimjg18}. Figure~\ref{fig:cloud_mass} indeed shows that the majority of the gas in the cloud is quickly ionized and accelerated to the velocity that significantly exceeds the escape velocity of the clouds ($\sim 5-10\,\kms$). Non-thermal radiation pressure from multi-scattered IR photons is not expected to play an important role in regulating the overall dynamics of the cloud, as they are not effectively trapped due to the low metallicity and low optical depths in the cloud. \citet{skinner15} also show that even for the metal-rich environments, the opacity must be significantly higher ($\kappa_{\rm IR} > 15\,{\rm cm^2/g}$) to unbind the cloud and reduce the amount of star formation as observed in some super star clusters.

Figure~\ref{fig:img} also shows that some of the dense parts of the clouds survive the radiation feedback for a long period of time ($\sim10-20\,{\rm Myr}$), creating comet-like structures in the runs with a 1\% SFE (\texttt{M6\_SFE1} and \texttt{M5\_SFE1}). This happens because the pressure of the dense gas is significantly higher ($P/k_B \sim 10^{5-7}\,{\rm cm^{3}\,K^{-1}}$) than the ram pressure from the warm/hot gas in the HII bubble  ($P/k_B\sim 10^{3-4}\,{\rm cm^{3}\,K^{-1}}$). Once the comet-like structures move away from the centre where the bright sources are located, they become more difficult to photo-evaporate, as their solid angle on the sky becomes smaller and fewer LyC photons can interact with them. As a result, some of the clouds survive even though the majority of the gas is blown out from the system, and the velocity dispersion of the neutral gas ($\sigma\la 10\,\kms$) does not increase as significantly as that of the ionized gas ($\sigma\ga 10 - 50\,\kms$, Figure~\ref{fig:cloud_mass}). In contrast, if the stellar population produces enough LyC photons ($\approx10^{65}$) to keep the dense clumps ionized (i.e., 10\% SFE), the dense clumps are quickly evaporated (the fourth column of the first and third rows in Figure~\ref{fig:img}). 

\subsection{Escape of LyC photons}
Now that we understand the general features of the simulated turbulent clouds, we study where and when most LyC photons are absorbed in different environments. We also examine the effects of the shape of SEDs and the gas metallicity on the escape of LyC photons in this section.

\subsubsection{Escape fractions of LyC radiation}
In order to measure the escape of LyC photons, we post-process the simulation snapshots with a simple ray tracing method. This is done by casting 12288 ($=12\times32^2$) rays per star particle using the {\sc healpix} algorithm \citep{gorski05} and by measuring the remaining photons at the computational boundary after attenuation due to $\rm HI$, $\rm H_2$, $\rm HeI$, $\rm HeII$, and dust, as
\begin{gather}
\fescLyC(t) = \frac{ \int_{\nu_0}^{\infty}  d\nu  \int d\Omega \sum_i   \left\{ L_{i}(\nu,t)  \exp \left[- \tau_{i}(\nu,t; \Omega) \right] \right\} / 4\pi h \nu   }{   \int_{\nu_0}^{\infty}  d\nu \left[ \sum_i L_{i}(\nu,t) \right] / h \nu },  
\end{gather}
where $L_i (\nu,t)$ is the luminosity of the $i$-th star particle with age $t$, $\tau_{i}(\nu,t; \Omega)$ is the total optical depth due to dust and gas at a given frequency $\nu$  along the sight line $\Omega$, and $\nu_0=3.287\times10^{15}\,{\rm Hz}$ is the frequency at the Lyman limit. Each ray carries an SED depending on the age and metallicity, and also on the stellar evolutionary model assumed (i.e., single vs binary). The optical depth to neutral hydrogen and singly ionized helium is computed using the photo-ionization cross-section of a hydrogenic ion with the nuclear charge $Z$ \citep[][Equation~2.4]{osterbrock06}. For the photo-ionization cross-section of neutral helium and molecular hydrogen, we adopt the fitting formula from \citet{yan98,yan01}. Also included is the absorption due to either small Magellanic cloud-type dust based on the fitting formula to the effective cross-section \citep{gnedin08} or Milky Way-type dust \citep{weingartner01} with a dust-to-metal ratio of 0.4 \citep{draine07} for the metal-poor ($Z_{\rm gas}=0.002$) or metal-rich ($0.02$) cloud, respectively. Note that the amount of dust is assumed to depend on the fraction of ionized hydrogen, as $f_{\rm dust} = (1-x_{\rm HII}) + f_{\rm ion} x_{\rm HII} $, where $f_{\rm dust}$ is the relative mass fraction of dust with respect to the purely neutral case, $x_{\rm HII}$ is the mass fraction of ionized hydrogen, and $f_{\rm ion}=0.01$ is the free parameter that takes into account the observed abundance of dust in HII regions \citep[see][for detailed discussion]{laursen09}.\footnote{We note that this is not fully self-consistent with the modelling of dust in our RHD calculations. However, because our code does not follow the formation and destruction of dust explicitly, we post-process our simulation outputs using the simple dust model of \citet{laursen09}, which better reflects the observed dust abundance in a variety of environments and reproduces \Lya\ properties of high-z galaxies.}

In Figure~\ref{fig:fescLyC}, we show the escape of LyC photons in clouds of different SFEs. The escape fraction ($f_{\rm esc}^{\rm LyC}$) generally increases with time, reaching $f_{\rm esc}^{\rm LyC} \approx 100\%$ when the cloud is entirely disrupted. One can also see that the variation in $f_{\rm esc}^{\rm LyC}$ is more monotonic when a SFE is higher, as the disruption process is more efficient \citep[c.f.][]{howard18}. In contrast, the run with a 1\% SFE (producing only one star particle of mass $10^3\,\msun$, \texttt{M5\_SFE1}) predicts noisy $f_{\rm esc}^{\rm LyC}$. This happens because, although it is initially placed in a relatively low-density environment, the only star particle encounters and is swallowed by adjacent dense clumps inside the giant molecular cloud (GMC). Ionizing radiation is temporarily blocked, but because photo-ionization heating and radiation pressure from the star are strong enough to destroy these local clumps, the high escape fraction is quickly recovered. The sign of the noisy feature can also be found in the \texttt{M6\_SFE1} run where ten star particles are emitting ionizing radiation.

As expected, we find that the time-averaged, luminosity-weighted escape fraction, $\left<f_{\rm esc}^{\rm LyC}\right>$, is a strong function of SFE \citep{dove00}. In the massive cloud with a large number of stars (\texttt{M6\_SFE10}), $45.4\%$ of the LyC photons escape from the GMC, while only a small fraction ($4.8\%$) of LyC photons leave in the case of a 1\% SFE (\texttt{M6\_SFE1})\footnote{We have examined the uncertainty due to randomizing the initial positions of star particles by running two additional simulations with a 1\% SFE, and found that the luminosity-weighted escape fraction can change by $\sim 30\%$ ($\left<f_{\rm esc}^{\rm LyC}\right>=$3.2\% and 5.2\%).}. The same trend is seen for the cloud with less mass ($M_{\rm cloud}=10^{5}\,\msun$, $\left<f_{\rm esc}^{\rm LyC}\right>=71.6\%$ vs $22.9\%$). This can be attributed to the fact that the most important mechanism during the early phase, i.e., photo-ionization heating, can impart radial momentum in proportion to $N_{\rm ion}^{4/7}$ \citep{krumholz15}. This lends support to the claim that the escape fraction relies sensitively on the burstiness of star formation histories \citep{kimm17,trebitsch17}.  

It is also interesting to note that less massive clouds show higher escape fractions for a given SFE. Because the initial average density is chosen to be higher by a factor of $\sim2$ in the less massive cloud ($\left<n_{\rm H}\right>\approx 160 \,\cmq$), the radial momentum from photo-ionization heating \citep[$p_{\rm rad} \propto \nH^{-1/7}$,][]{krumholz15} as well as the momentum from SN explosion \citep[$p_{\rm rad} \propto \nH^{-2/17}$,][]{blondin98,thornton98} would be smaller in the less massive cloud; hence, one may expect the escape fraction to be lower. However, this is opposite to our findings. This may simply be due to the fact that the less massive cloud happens to have low-density channels around young star particles, given that they would encounter a fewer number of neutral hydrogen atoms than in the run with the massive cloud. The high escape fractions can also be attributed to the fact that ionization fronts reach the edge of the cloud earlier in the less massive cloud. \citet{geen15} show that, for a uniform medium, the propagation velocity of ionization fronts can be written as
\begin{equation}
\frac{1}{c_{\rm s,i} } \frac{d r_{\rm i}(t)}{dt} = \left( \frac{r_{\rm S}}{r_i (t)}\right)^{\frac{3}{4}} - \left(\frac{c_{\rm ext}}{c_{\rm s,i}}\right)^2 \left( \frac{r_{\rm S}}{r_i (t)}\right)^{-\frac{3}{4}} + \frac{v_{\rm ext}}{ c_{\rm s,i}},
\label{eq:hii}
\end{equation}
where $r$ is the radius, $r_S$ is the Stromgren sphere radius, $r_i$ is the radius of the ionization front, $t$ is the time, $c_{s,i}$ is the sound speed of the ionized medium, $v_{\rm ext}$ is the infall velocity of the ambient gas, and $c_{\rm ext}$ represents the velocity term due to the thermal and turbulent pressure. Then, the time for the front to reach the edge of the cloud ($r_{\rm cloud}$) may be written as  
\begin{equation}
\tau_{\rm ion}= \int_1^{r_{\rm cloud}/r_{\rm S}} \, \frac{r_{\rm S} / c_{\rm s,i}}{  y^{-3/4}- \left(c_{\rm ext}/c_{\rm s,i}\right)^2 y^{3/4}  } dy 
\end{equation} 
where $y\equiv r_{\rm i}  / r_{\rm S}$. As shown in Appendix (Figure~\ref{fig:analytic_ex}), although the number of ionizing photons per atom is the same for a given SFE, the time required for the ionization front to propagate to the edge of the cloud is shorter in the cloud with $M_{\rm cloud}=10^5 \,\msun$ compared to the one with $M_{\rm cloud}=10^6 \,\msun$. As a consequence, the bright phase of escaping ionizing radiation starts earlier in the less massive cloud, leading to a higher $\left<f_{\rm esc}^{\rm LyC}\right>$.

The escape fractions obtained from our simulations are useful to understand the reionization history of the Universe, but may not be practical to compare with observations. For example, because it is not possible to observe the whole wavelength range of LyC photons, \fescLyC\ is often measured by comparing the average flux densities at two different wavelengths, such as $F_{900}$ (flux density at wavelength $880 < \lambda < 910$ \AA\ ) and $F_{1500}$ (flux density at $1475 < \lambda < 1525$ \AA) \citep[e.g.][]{steidel18}. Motivated by this, we also present the escape fractions of the photons with wavelength $880 < \lambda < 910$ ($f_{\rm esc}^{900}$) in the middle panels of Figure~\ref{fig:fescLyC}. We find that $f_{\rm esc}^{900}$ is systemically smaller than $f_{\rm esc}^{\rm LyC}$ because the absorption cross-section to ionizing radiation due to neutral hydrogen is larger at longer wavelengths below the Lyman edge \citep{osterbrock06}. The runs with the massive metal-poor cloud ($M_{\rm cloud}=10^6\,\msun$) show $\left<f_{\rm esc}^{900}\right>$=3.1\% and 37\% for a 1\% and 10\% SFE, respectively, while 17\% and 65\% of the photons around 900\AA\ escape from the less massive cloud. Therefore, one should keep in mind that the luminosity-weighted time average of $f_{\rm esc}^{900}$ can be $\sim20-35\%$ lower than $\left<f_{\rm esc}^{\rm LyC}\right>$ when comparing the simulated escape fractions with observationally derived quantities. Corresponding luminosity-weighted flux density ratios ($F_{900}/F_{1500}$) are found to be 0.05, 0.56, 0.24, and 1.1 by the end of each simulation (\texttt{M6\_SFE1}, \texttt{M6\_SFE10}, \texttt{M5\_SFE1}, and \texttt{M5\_SFE10}, respectively). Note that the flux density ratios are somewhat higher than those observed in the compact starbursts \citep[e.g.][]{shapley16}, likely because our clouds are very young and have no underlying population that preferentially produces the flux density around 1500 \AA.

To compute the relative contributions to the absorption of LyC photons, we also measure the effective optical depth due to $\rm HI$, $\rm H_2$, $\rm HeI+HeII$, and dust, as $\tau_{\rm eff}\equiv - \ln f_{\rm esc,i}^{\rm LyC}$, where $f_{\rm esc,i}^{\rm LyC}$ is the instantaneous escape fraction after attenuation by each element $i$. Figure~\ref{fig:tau_break} (top panels) shows that the majority of the LyC photons are absorbed by neutral hydrogen in the metal-poor, massive clouds. Initially, the effective optical depth due to dust is also significant ($\tau_{\rm eff,dust}\sim 1$) but decreases steadily, as the fraction of ionized hydrogen increases and dust is assumed to be destroyed in the HII region. The absorption due to helium and molecular hydrogen is also minor compared to that due to neutral hydrogen, but we find that their total contribution is as equally important as dust even in the metal-rich environments (bottom panels).

\begin{figure}
   \centering
   \includegraphics[width=8.5cm]{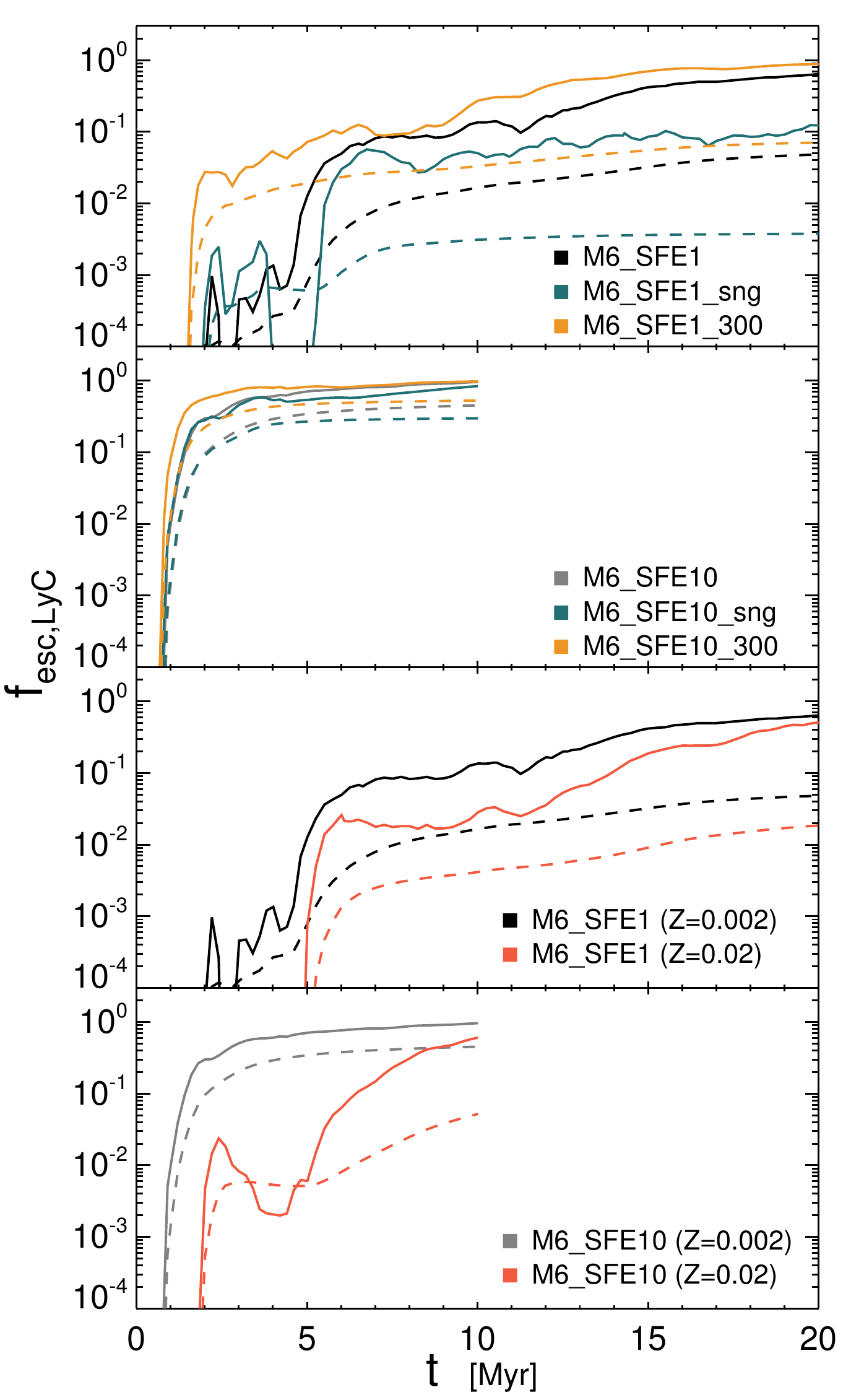} 
   \caption{Escape fraction of LyC photons ($\fescLyC$) for the massive cloud ($M_{\rm cloud}=10^6\,\msun$) with different assumptions about the SEDs and metallicity, as indicated in the legend. We compare three different SED models: \texttt{bin} and \texttt{sng} represents the run with the binary \citep{stanway16} and single stellar evolution \citep{bruzual03} with an upper mass cut-off of $100\,\msun$, while a \texttt{300} model is generated with the upper mass cut-off of $300\,\msun$ with binary stellar evolution. The solid lines represent the instantaneous escape fraction, while the luminosity-weighted time average is displayed with dashed lines. The top and middle panels show the escape fractions with a 1\% and 10\% SFE, respectively, and the bottom two panels show the effects of gas metallicities on the escape fraction. }
   \label{fig:fescLyCextra}
\end{figure}

\subsubsection{Impact of SED models on the escape of LyC radiation}

We find that the prediction of the escape fractions relies on the choice of the SEDs used in the simulations. Figure~\ref{fig:fescLyCextra} (top two panels) shows that $\left<f_{\rm esc}^{\rm LyC}\right>$ in the run adopting a single stellar evolution model \citep{bruzual03} is lower than the run with binaries. This can be attributed to the fact that the former predicts an ionizing photon production rate that is a factor of $\sim 2$ smaller than the model with binaries at $t\ga 5\,{\rm Myr}$ (Figure~\ref{fig:input}). In particular, mergers and transfer of gas between binaries results in enhanced ionizing emission at $t\ga3\,{\rm Myr}$ \citep[][see also Figure~\ref{fig:input}]{stanway16}. This allows LyC photons to escape more easily, as the clouds become significantly disrupted at this stage. In contrast, few ionizing photons are generated after $10\,{\rm Myr}$ in the single SED case (\texttt{M6\_SFE1\_sng}), leading to an order of magnitude smaller $\left<f_{\rm esc}^{\rm LyC}\right>$ of $\sim0.4\%$ than $\left<f_{\rm esc}^{\rm LyC}\right>=4.8\%$  of the \texttt{M6\_SFE1} model. The difference becomes less notable if the SFE is high enough to disrupt the cloud in the early phase during which a large number of LyC photons are still produced. For example, the runs with a 10\% SFE (\texttt{M6\_SFE10} and \texttt{M6\_SFE10\_sng}) yield $\left<f_{\rm esc}^{\rm LyC}\right>=45.4\%$ and $30.0\%$, respectively. 

Uncertainties persist in the SED models as the maximum stellar mass remains unknown \citep[e.g.][]{kroupa13}. Previous studies suggest that the initial mass functions do not extend to more than $120-200\,\msun$ \citep{massey98,oey05}, but re-analysis of O stars in 30 Doradus suggests the possibility of $300 \,\msun$ stars \citep{crowther10}. To assess the impact of the variation in the upper mass limit on escape fractions, we perform two additional simulations (the $10^6$M$_{\odot}$ cloud with 1\% and 10\% SFE) adopting the binary SEDs with a maximum mass of $300\,\msun$. Note that this SED provides an additional 50\% of LyC photons in the early phase ($t \la 5 \,{\rm Myr}$) compared to our fiducial case (Figure~\ref{fig:input}). We find that the choice of the upper mass limit seems to have a secondary effect compared to the inclusion of binaries (Figure~\ref{fig:fescLyCextra}). With the higher upper mass, the escape fractions are slightly enhanced from $\left<f_{\rm esc}^{\rm LyC}\right>=4.8\%$ to $5.2\%$ in the run with a 1\% SFE, and from $45.4\%$ to $53.1\%$ in the cloud with a 10\% SFE. This indicates that the additional photons from very massive stars ($100 \le M/\msun \le 300$) are preferentially used to build up the initial HII bubbles and help to create the low-density channels instead of leaving the clouds.

\subsubsection{Impact of gas metallicity on the escape of LyC radiation}

The bottom two panels of Figure~\ref{fig:fescLyCextra} show the effects of gas metallicity on the escape of LyC photons. Compared to the metal-poor case ($Z_{\rm gas}=0.002$), the escape from the massive cloud runs with solar metallicity (\texttt{M6\_SFEx\_Zsol}) is significantly reduced by a factor of 2.5 from  $\left<f_{\rm esc}^{\rm LyC}\right>=4.8\%$ to $1.9\% $ for a cloud with 1\% SFE or by an order of magnitude from $45.4\%$ to $5.2\%$ in the runs with a 10\% SFE. Note that we use the same metallicity ($Z_{\rm star}=0.002$) for star particles in order to keep the number of ionizing photons the same for all simulations with the same cloud mass and SFE; thus, the decrease in $\left<f_{\rm esc}^{\rm LyC}\right>$ is due to the enhanced attenuation by dust and/or more efficient metal cooling. To identify the cause of the significant reduction, we further examine $\left<f_{\rm esc}^{\rm LyC}\right>$ from the metal-rich runs, assuming that the dust-to-metal ratio is smaller by an order of magnitude (i.e. $D/M=0.04$) in the post-processing step, and find that the LyC escape is still considerably smaller ($\left<f_{\rm esc}^{\rm LyC}\right>=1.9\%$ (\texttt{SFE1}) or $5.0\%$ (\texttt{SFE10})) than in the low-metallicity clouds (see also Figure~\ref{fig:tau_break} for the relative contribution to the optical depth by dust). This demonstrates that the effects due to dust absorption are secondary on cloud scales and that the low escape fraction arises mainly because ionized hydrogen quickly recombines as a result of efficient metal cooling. Indeed, we find that the enhanced cooling limits the effects of photo-ionization heating and that the cloud is not dispersed as efficiently as in the metal-poor case (compare Figure~\ref{fig:img} with Figure~\ref{fig:img_all} in Appendix). These experiments imply that the observed escape fractions from the relatively metal-rich LBG galaxies at $z \sim 3$ \citep{steidel01,siana10,leitet13,mostardi15,grazian16,marchi17,steidel18} are likely to be a lower limit of the escape fractions of high-redshift metal-poor dwarf galaxies that likely reionized the Universe.

\begin{figure}
   \centering
   \includegraphics[width=8.5cm]{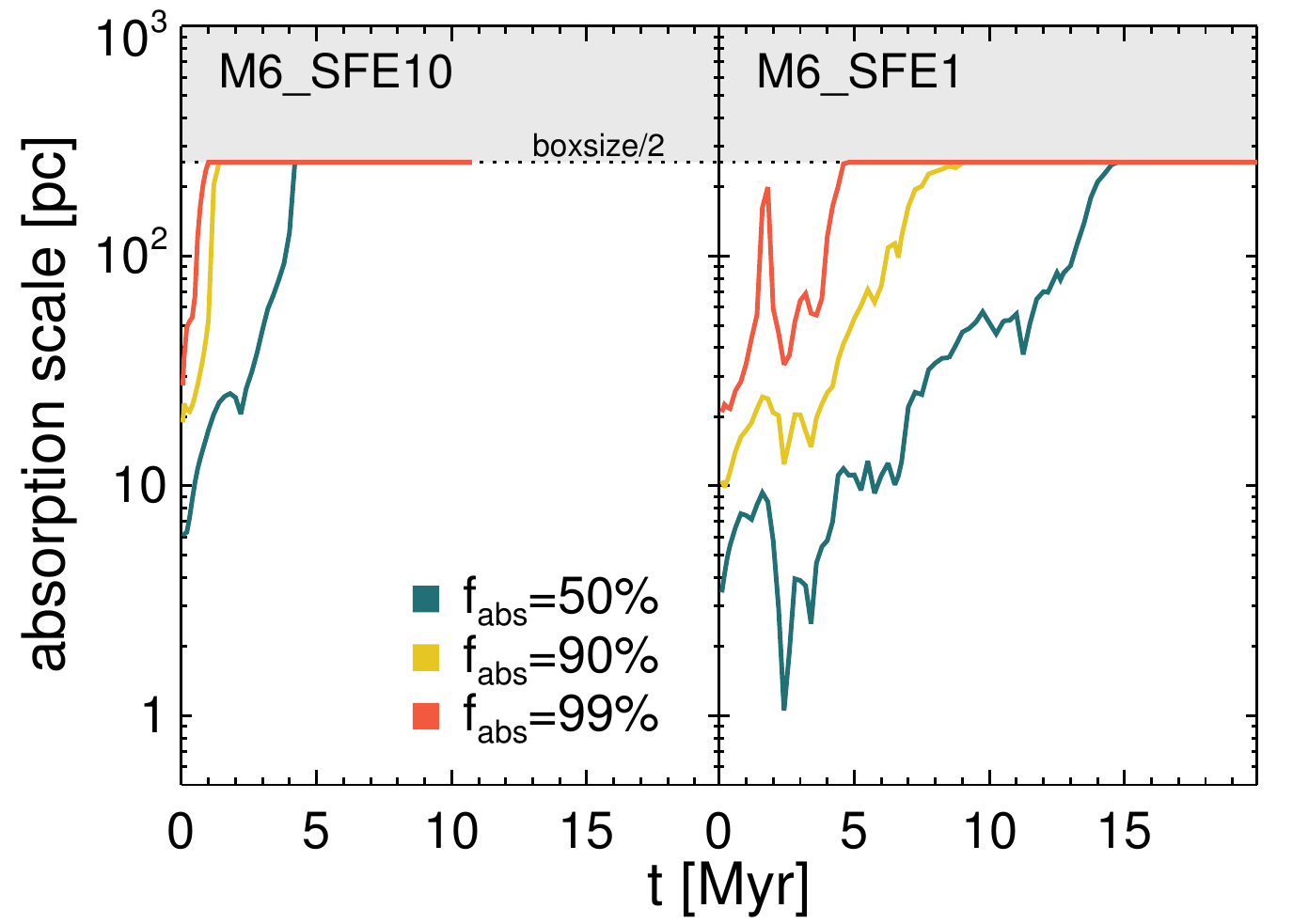} 
   \caption{Distance within which the majority of LyC photons are absorbed in the massive cloud with $M_{\rm cloud}=10^6\,\msun$. The absorption scales are estimated by computing how far photons can propagate when the instantaneous escape fractions are 50\%, 10\%, and 1\%. The left and right panels display the absorption scales for the runs with a high (10\%) and low (1\%) SFE. A large fraction of the photons are absorbed on $\sim10\,{\rm pc}$ scales during the early bright phases, especially if SFE is low ($t \la 5  \, {\rm Myr}$).}
   \label{fig:abs_scale}
\end{figure}

\subsubsection{Absorption scale of LyC radiation}

In galactic scale simulations with finite resolution, the turbulent structure of the star-forming clouds is under-resolved, and the estimation of LyC escape from these simulations is often uncertain \citep[e.g.][]{ma16}. In this regard, it is useful to compute where most of the LyC photons are absorbed in different environments. In Figure~\ref{fig:abs_scale}, we measure the absorption scale based on the ray tracing method described in the previous section and show that half of the LyC photons are absorbed on small scales, particularly when the SFE is low ($1\%$, \texttt{M6\_SFE1}). More specifically, the photon number-weighted absorption scales for this cloud are 8, 33, and 83 pc for $f_{\rm abs}=50$, $90$, and $99\%$, respectively. When the star formation becomes more intense (10\% SFE), 50\%, 90\%, and 99\% of the LyC photons are absorbed at much larger distances (49, 140, and 191 pc). This suggests that in order to determine the instantaneous escape fractions with reasonable accuracy (within a factor of two), one should adopt a computational resolution better than $\Delta x_{\rm min} \la {\rm  10 \, pc}$, provided that the formation and disruption of star-forming clouds are reasonably well captured in the simulations. Based on high-resolution (2 and 4 pc) cosmological RHD simulations adopting strong supernova feedback, \citet{kimm14} also show that the optical depth to LyC photons in their atomic-cooling haloes is large on 100 pc scales ($\tau \sim 2-4$) \citep[see also][]{paardekooper15,kim13} and that the luminosity-weighted escape fractions are converged at these resolutions.

\subsubsection{Variability of the LyC escape fractions}

Our simulations with SFEs greater than 1\%, which is what observations appear to support on average \citep[][c.f. \citealt{leroy17}]{lada10,evans14,vutisalchavakul16,lee_eve16}, suggest that $\fescLyC$ tends to increase monotonically once radiation blows the cloud gas away and develops low-density channels \citep[see also][]{kimjg18}. This may sound contradictory to previous findings that $\fescLyC$ fluctuates rapidly over time in galaxies \citep{wise14,kimm14,paardekooper15,ma16,trebitsch17}. However, the variation observed in the galactic scale simulations is on the timescale of $\sim10-30\,{\rm Myr}$, which is not inconsistent with the timescale of the variability in our simulations (c.f. the runs with a 1\% SFE). 

An exception is when stars form in deeply embedded environments where stellar feedback cannot disrupt the surrounding clumps early on \citep{howard18}. If this were the case, the escape fractions would be highly time-dependent even on Myr timescales, and the predictions from galactic scale simulations \citep[e.g.][]{wise09,kimm14,xu16,trebitsch17,rosdahl18}, where small-scale structures are unresolved, might be quite uncertain. However, as we discuss later in Section~\ref{sec:lya_pressure}, radiation feedback in dense regions needs to be properly addressed, especially when the Stromgren sphere is under-resolved. For example, none of the cloud simulations conducted thus far include \Lya\ feedback, which may disrupt the metal-poor dense clumps near young stars \citep{kimm18}. The inclusion of such strong early feedback leads to the efficient expansion of photoionized bubbles and thus results in a rather monotonic evolution of the escape fractions, as in our fiducial runs where star particles are randomly placed inside the cloud. For these models, we do not expect extremely variable escape fractions even if we resolve the detailed structure of the clouds (see Section~\ref{sec:lya_pressure}), and the fluctuating escape fractions on $10$--$30\,{\rm Myr}$ are likely to persist as a ramification of the sporadic nature of star formation episodes distributed over the galaxy.

\subsection{Properties of \Lya\ photons}

We now turn to the scattering and absorption of \Lya\ photons in the simulated clouds. Note that the propagation of \Lya\ photons in a neutral medium is considerably different from LyC photons in the sense that the interaction with neutral hydrogen does not destroy \Lya\ photons but simply changes their frequency and direction. 

To model the propagation of \Lya\ photons in star-forming clouds, we post-process our simulations using the Monte Carlo \Lya\ radiative transfer code, {\sc rascas} (Michel-Dansac et al. {\sl in prep}). We compute the \Lya\ emissivity by taking into account recombination and collisional radiation, as
\begin{align}
&\epsilon_{\rm Ly\alpha} = \epsilon_{\rm rec}  + \epsilon_{\rm coll} &  \nonumber\\
&\epsilon_{\rm rec} =  P_B (T)\, \alpha_B(T)\, n_e \, n_{\rm HII}  \, e_{\rm Ly\alpha} ~~~~~& \textrm{(recombinative)} \label{eq:rec}  \\
&\epsilon_{\rm coll} =  C(T)  \, n_e  \, n_{\rm HI} \, e_{\rm Ly\alpha}  ~~~~~& \textrm{(collisional)}
\label{eq:coll}
\end{align}
where $e_{\rm Ly\alpha}$ is the energy of \Lya\ photon (10.16 eV), $n_e$ and $n_{\rm HII}$ are the number density of electron and ionized hydrogen. Here, $P_B$ is the probability for an absorbed LyC photon to be re-emitted as a \Lya\ photon \citep{cantalupo08}, 
\begin{align}
&  P_B(T) = 0.686 -0.106 \log T_4 - 0.009 \,T_4^{-0.44},
\end{align}
where $T_4 = T / 10^4\,{\rm K}$, $\alpha_B$ is the case B recombination coefficient \citep{hui97}, 
\begin{align}
& \alpha_B = 2.753\times10^{-14}~{\rm cm^3\,s^{-1}} \frac{\lambda^{1.5}}{\left[ 1+\left(\lambda/2.74 \right)^{0.407} \right]^{2.242}},
\end{align}
with $\lambda = 315 614\,{\rm K} / T$, and $C(T)$ is the coefficient for the cooling radiation \citep[e.g.][]{callaway87}
\begin{align}
& C(T) = \frac{2.41\times10^{-6} \, {\rm cm^3\,s^{-1}}}{T^{0.5}} \, T_4^{0.22} \, \exp\left[ \frac{-1.63\times10^{-11}}{k_B T}\right].
\label{eq:coe_cooling}
\end{align}
Based on the emissivity of each cell, we randomly sample the initial position for $10^4$ \Lya\ photons. Once the position is determined, the initial frequency is drawn randomly from a Gaussian distribution with the thermal Doppler broadening set by the temperature of each cell. The Doppler parameter ($\left<b\right>=\sqrt{2 k_B T / m_{\rm H}}$) at the source position is approximately $13$--$15\,\kms$ for the runs that we examine in this work. We include scattering due to deuterium with a fixed abundance of $3\times10^{-5}$, recoil effects, and the scattering and destruction due to dust based on \citet{laursen09}. The loss of \Lya\ photons due to molecular hydrogen  \citep[][]{shull78,black87} is neglected for simplicity. As such, molecular hydrogen is transparent to \Lya\ photons in this study, and the resulting escape fractions may be slightly over-estimated, especially in the warm ($10^3 \la T\la 10^4 \,{\rm K}$), molecular regions, although it is unlikely to be significant in the typical cold ($T\sim100\,{\rm K}$) star-forming sites \citep[see Figure~20 in][]{neufeld90}. 

\begin{figure}
   \centering
   \includegraphics[width=8.4cm]{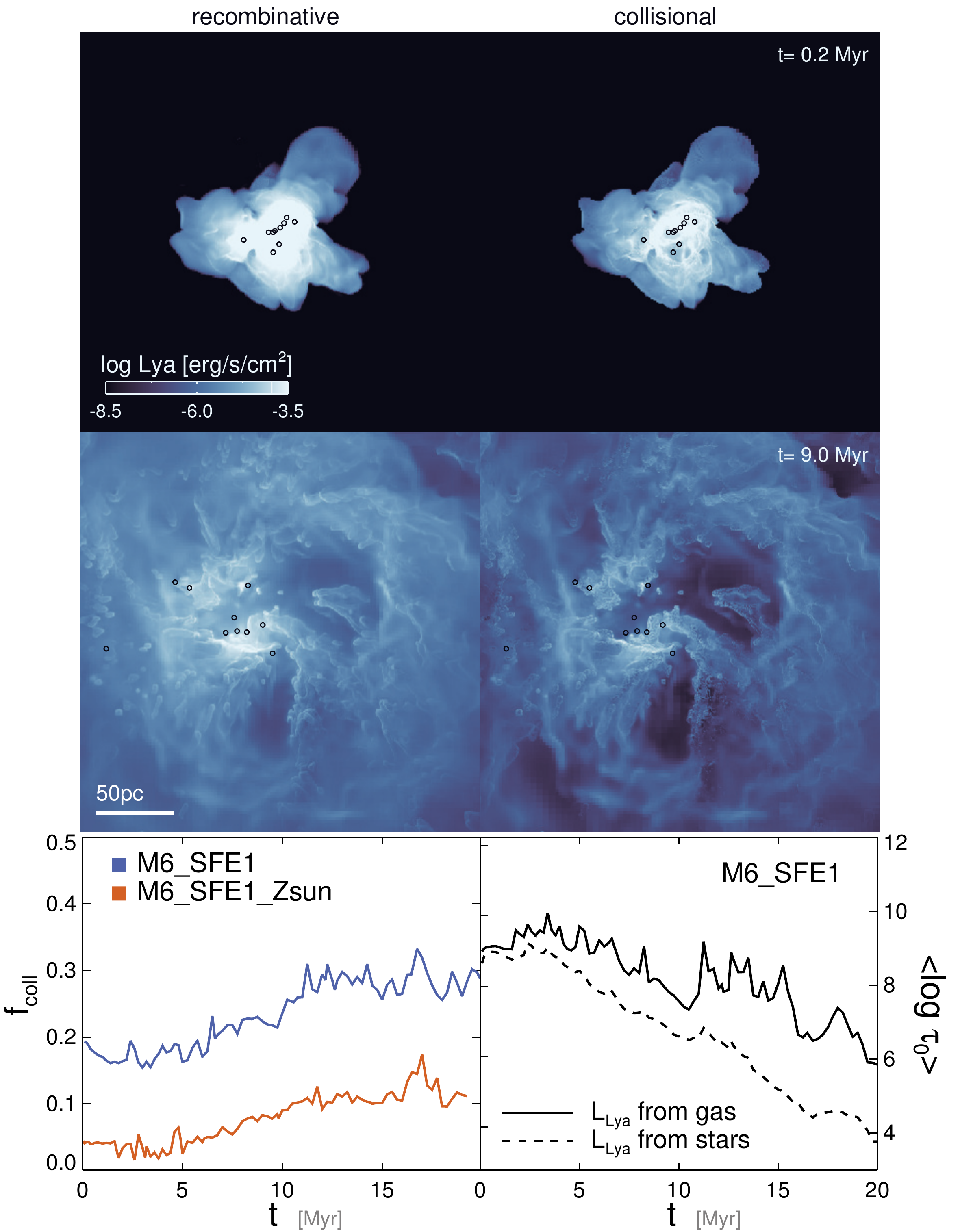}  
   \caption{\Lya\ emissivity map from the \texttt{M6\_SFE1} run at two different epochs, 0.2 and 9.0 Myr. The top and middle left panels show recombinative \Lya\ radiation, while the right panels display the contribution from collisional radiation. Colour codings indicate the strength of \Lya\ surface brightness, as indicated in the legend. The black dots correspond to stellar particles. Note that \Lya\ is emitted by the extended region of the cloud, especially when the cloud gets disrupted. The bottom left panel shows the fraction of collisional radiation to the total \Lya\ radiation from our fiducial runs. The dark blue line corresponds to the metal-poor case ($Z_{\rm gas}=0.002$), while the orange line indicates the runs with solar metallicity ($Z_{\rm gas}=0.02$). In the bottom right panel, we also include the logarithmically averaged optical depth to the \Lya\ line centre depending on the assumption on the source positions (see text).
   }
   \label{fig:lya_origin}
\end{figure}

\begin{figure}
   \centering
   \includegraphics[width=8.4cm]{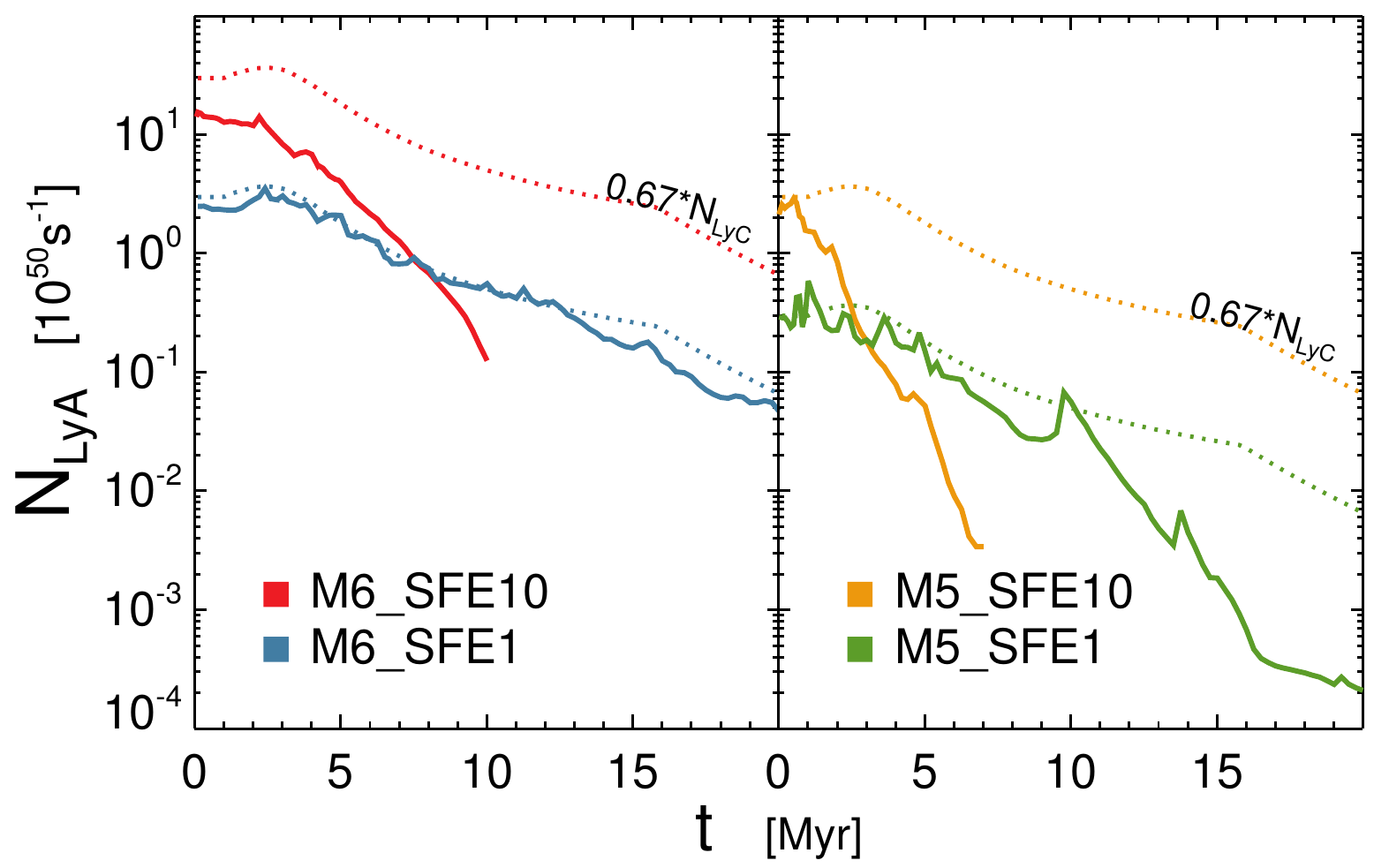} 
   \caption{Rate of \Lya\ photons emitted ($N_{\rm Ly\alpha}$) in metal-poor  ($Z_{\rm gas}=0.002$) turbulent clouds of mass $10^{6} \,\msun$ (left) and $10^{5} \,\msun$ (right). Different color-codings denote the runs with different SFEs, as indicated in the legend. Note that the cloud with a higher SFE produces a smaller number of \Lya\ photons at the late stage of the evolution because a larger fraction of LyC photons escapes from the system.  All models are based on the binary stellar evolution model with the upper mass limit of $100\,\msun$. Also included as dotted lines is the simple case where 67\% of the LyC photons available from stars are assumed to yield \Lya\ photons.
   }
   \label{fig:NLya}
\end{figure}

\subsubsection{Production of \Lya\ radiation}

Figure~\ref{fig:lya_origin} shows an example of \Lya\ emissivity maps from the \texttt{M6\_SFE1} run at two different times (c.f. Figure~\ref{fig:img}). Initially, LyC photons are well confined within the cloud, and \Lya\ emissivity map closely follows the distributions of young stellar particles (i.e., LyC photon density). At this stage, most \Lya\ photons are produced at high densities, $\left<n_{\rm H} \right>\sim100\,\cmq$, mostly via recombinative radiation. Once the pressure from photo-ionization pushes the dense gas away, more diffuse gas in the vicinity of young stars and the irradiated surfaces of the clumps with $\left<n_{\rm H} \right>\sim1-5\,\cmq$ become the main sites of \Lya\ production. The initial position of \Lya\ photons does not match the distribution of stellar particles precisely at the late stage of the cloud evolution, as LyC radiation that provides photoionized electrons is widespread over the cloud. To be more quantitative, we measure the optical depth to the line centre ($\tau_0$) for 3072 sight lines from each star particle and compare them with $\tau_0$ measured from the positions of the actual \Lya\ emitting gas to 30,000 random directions in the bottom panel of Figure~\ref{fig:lya_origin}. The plot shows that $\tau_0$ measured from the stars is systematically lower than $\tau_0$ measured from the gas, as the young stars photo-ionize the surrounding neutral hydrogen and create low density channels via stellar feedback. Note that the difference between the two measurements becomes more prominent in the latter stages of the cloud evolution, indicating that the actual line emitting gas distribution should be used to compute the profile of \Lya\ photons rather than the stellar distribution as a radiation source \citep[e.g.][]{verhamme12,behrens14a} if the structure of the ISM is resolved.

Because collisional radiation (right panels) as well as recombination (left panels) requires electrons to produce \Lya, the emissivity maps from the two different mechanisms appear quite similar, although the former tends to better trace the dense structures and is thus less extended. Approximately 19\% of the total \Lya\ radiation is produced via collisional radiation in the massive cloud with a 1\% SFE. When the metallicity is increased by an order of magnitude (i.e., $Z_{\rm gas}=0.02$), the contribution from collisional radiation decreases to $\sim 5\%$ for a 1\% SFE run. This is mainly because the enhanced metal cooling lowers the temperature of the \Lya-emitting, ionized gas from $\approx 12,700 \, {\rm K}$ to $\approx 11,400\,{\rm K}$, and collisional ionization accordingly becomes less efficient.

Figure~\ref{fig:NLya} shows the rate of \Lya\ photon production inside the cloud. When \fescLyC\ is negligible (\texttt{M6\_SFE1} and \texttt{M6\_SFE1}), the number of \Lya\ photons can be reasonably approximated by the simple calculation that  67\% of all the LyC photons from the stellar population are converted into \Lya\ photons, as is often done in the literature \citep[e.g.][]{verhamme12,dijkstra14}. This is possible i) because LyC photons are efficiently absorbed and ii) because the contribution from cooling radiation to the total \Lya\ photon budget is not very significant ($\sim20\%$). The simple approximation ($N_{\rm Ly\alpha}\approx0.67 N_{\rm LyC}$) is no longer valid in the early phase ($t\la 1\,{\rm Myr}$) of bubble expansion during which a large number of stars shine simultaneously and the local surplus of LyC photons does not directly contribute to recombinative radiation (see \texttt{M6\_SFE10} for an example). In addition, if stellar feedback ejects a large amount of gas from the cloud, LyC photons would leave without interacting with ionizing neutral hydrogen, and a larger number of \Lya\ photons may be produced in the ISM/CGM of the galaxy compared with those produced inside the cloud.

\begin{figure}
   \centering
   \includegraphics[width=8.4cm]{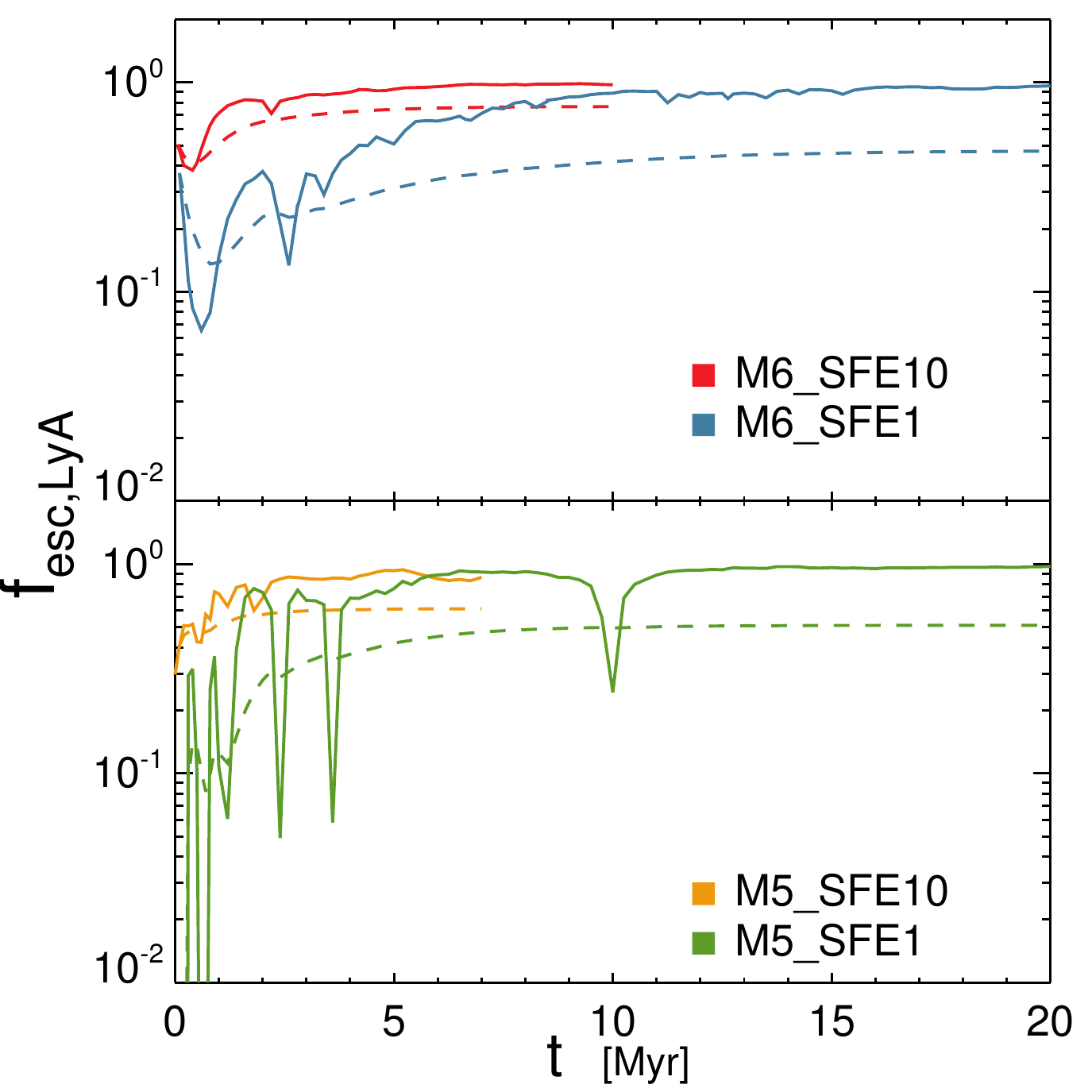} 
   \caption{Escape fractions of the \Lya\ photons (\fescLya) produced inside the metal-poor ($Z_{\rm gas}=0.002$) turbulent clouds with  $10^6\,\msun$ (top) and $10^5\,\msun$ (bottom). Different color-codings represent the runs with different SFEs. Solid lines display the instantaneous fraction ($f_{\rm esc,Ly\alpha}$), while dashed lines show the luminosity-weighted, time-averaged value until time $t$ ($\left<f_{\rm esc}^{\rm Ly\alpha}\right>$). The escape fraction increases with increasing SFE and decreasing cloud mass. The sudden change in \fescLya\ in the case of \texttt{M5\_SFE1} occurs when the only star particle encounters a dense clump.    }
   \label{fig:fescLyA}
\end{figure}

\begin{table}
   \caption{Summary of the simulation results. From left to right, each column represents the name, the final epoch of the simulation, the luminosity-weighted, time-averaged LyC escape fractions, the total number of LyC photons escaped, the luminosity-weighted, time-averaged \Lya\ escape fractions, and the total number of \Lya\ photons that escaped from the cloud. Note that the last two columns do not include the contribution from the \Lya\ photons created by escaping LyC photons. }
   \centering
   \begin{tabular}{lccccc}
   \hline
  Name & $t_{\rm final}$ & $\left<f_{\rm esc}^{\rm LyC}\right>$   & $N_{\rm esc}^{\rm LyC}$ & $\left<f_{\rm esc}^{\rm Ly\alpha}\right>$  & $N_{\rm esc}^{\rm Ly\alpha}$ \\
  & [Myr] & [\%] & [$10^{50}$] & [\%] & [$10^{50}$] \\
     \hline
   M6\_SFE10 & 10  & 45.4 & 132.2 & 76.7 & 41.7 \\
   M6\_SFE1  & 20  &  4.8 &   1.6 & 47.0 & 8.8 \\
   M5\_SFE10 &  7  & 71.6 &  18.7 & 61.1 & 3.9  \\
   M5\_SFE1  & 20  & 23.1 &   0.8 & 51.0 & 1.6\\
     \hline
   M6\_SFE10\_300  & 10  & 53.1 & 240.9 & 82.6 & 51.5 \\
   M6\_SFE10\_sng  & 10  & 30.0 &  60.3 & 76.5  & 39.6 \\
   M6\_SFE10\_Zsun & 10  &  5.2 &  15.1 & 19.9 & 11.4 \\
   M6\_SFE10\_noSN & 10  & 47.8 & 139.1 & 75.2 & 38.1 \\
   \hline
   M6\_SFE1\_300 & 20  &  7.1  &   3.5  & 56.8  & 13.7 \\
   M6\_SFE1\_sng & 20  &  0.4  &  0.08 & 35.7  & 4.4 \\
   M6\_SFE1\_Zsun & 20 &  1.9  &   0.6  & 20.5  & 2.3 \\
   M6\_SFE1\_noTurb & 20  & 8.5  & 2.8  & 31.3  & 5.4 \\
   M6\_SFE1\_noSN & 20 &  3.9  &   1.3  & 53.3  & 9.8 \\
    M6\_SFE1\_dSF & 20  & 1.1  &   0.4  & 26.1  & 5.4 \\
    M6\_SFE1\_dSF\_Lya & 20  & 13.0 &  4.3 & 63.1 & 10.9 \\
        \hline
   \end{tabular}
   \label{tab:result}
\end{table}

\subsubsection{Escape fractions of \Lya\ photons}

Figure~\ref{fig:fescLyA} shows the escape fractions of \Lya\ photons (\fescLya) from the simulated clouds. We find that \fescLya\ generally increases with time as the covering fraction of dense dusty gas diminishes due to stellar feedback (Figure~\ref{fig:img}). Because this process occurs more rapidly in clouds with higher SFEs, more \Lya\ photons escape from the runs with 10\% SFE ($\left<\fescLya\right>=76.7\%$ and $61.1\%$ for $M_{\rm cloud}=10^6\,\msun$ and $10^5\,\msun$) than the runs with 1\% SFE ($\left<\fescLya\right>=47.0\%$ and $51.0\%$, respectively). Similar to the escape of LyC photons (Figure~\ref{fig:fescLyC}), \fescLya\ in the \texttt{M5\_SFE1} run becomes temporarily very low when the only star particle is enshrouded by a dense dusty clump (blue solid line in the bottom panel). This phase does not last long, as the clumps are destroyed by radiation feedback. High escape fractions ($\sim 40-80\%$) are commonly found in the clouds with $Z_{\rm gas}=0.002$, regardless of the choice of the SED (see Table~\ref{tab:result}), demonstrating that in metal-poor environments, most \Lya\ photons from young stellar populations are likely to escape from their birth clouds.

In contrast, \Lya\ photons are more efficiently destroyed in dusty environments. \citet{neufeld90} showed that the escape of \Lya\ photons in a uniform medium may be described as a function of the optical depth to \Lya\ ($\tau_0$) and dust ($\tau_d$), as $\fescLya=1/ \cosh \left[A \left(a\tau_0\right)^{1/6} \tau_d^{1/2} \right]$, where $A\approx2$ is a fitting parameter and $a$ is the Voigt parameter. If we apply this formula to obtain $\left<\fescLya\right>$ in the metal-rich environment by replacing $\tau_d$ with $10\, \tau_d$, it should be $\approx 28\%$ and $9\%$ for the \texttt{M6\_SFE10} and \texttt{M6\_SFE1} run, respectively. However, post-processing of the metal-rich runs with {\sc rascas} yields a somewhat different $\left<\fescLya\right>$ of  $\approx 20\%$ both for 1\% and 10\% SFEs (Table~\ref{tab:result}). This is not unexpected, given that the metals are distributed inhomogeneously in the simulated domain and that the optical depth to individual photons can be different, as they are produced in relatively extended regions of the cloud.

\begin{figure}
   \centering
            \includegraphics[width=7cm]{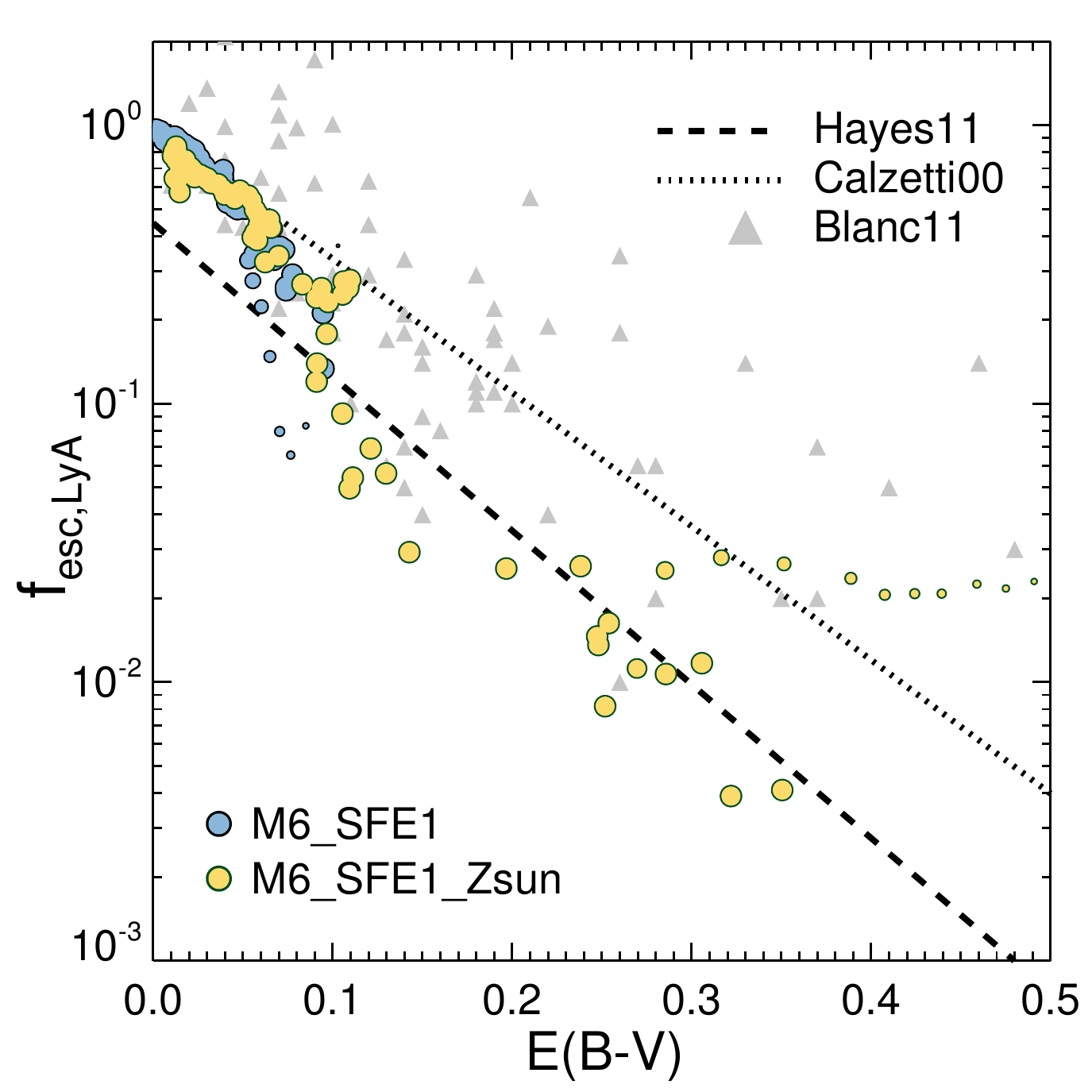} 
   \caption{Relation between the escape fraction of \Lya\ photons and the reddening due to dust, $E(B-V)$, in the metal-poor (blue) and metal-rich (orange) run. Smaller symbols indicate the results at the early stage of the evolution ($t\la 3\,{\rm Myr}$) during which the ionization front is effectively confined within the central region (see the text). The dashed line exhibits the observational trend obtained from a wide redshift range ($0.3 < z < 6$) \citep{hayes11}, while the relation derived from the simple attenuation law by \citet{calzetti00} is shown as the dotted line. Also included as grey triangles are the results from the HETDEX pilot survey \citep{blanc11}.
      }
   \label{fig:EBV}
\end{figure}

 \citet{hayes11} and  \citet{blanc11} find that more dust reddened \Lya\ emitters exhibit lower escape fractions of \Lya\ photons by comparing the observed \Lya\ line with the intrinsic flux derived either from the UV or $H\alpha$. To examine whether the trend is established already on cloud scales, we compute the colour index of the simulated clouds by convolving the angle-averaged, dust-attenuated spectrum with the B and V band filter throughputs. Figure~\ref{fig:EBV} demonstrates that the escape of \Lya\ photons is less efficient in more dust-reddened clouds, largely consistent with the observed trend. However, two interesting differences are found. First, when the radiation field from young stars does not permeate the cloud, dust reddening is very significant, despite that a few percent of \Lya\ photons still emerge from the cloud through low-density channels. As LyC photons ionize the neutral hydrogen around the young stars, dust is destroyed (by construction) while the majority of neutral gas remains intact (see Figure~\ref{fig:img_all}). Consequently, in this early phase ($t\la 3 \,{\rm Myr}$), dust reddening decreases while \fescLya\ is kept nearly fixed, as is shown with smaller yellow symbols in Figure~\ref{fig:EBV}. Second, when dust reddening is negligible, i.e., $E(B-V)\la 0.1$, the majority of the \Lya\ photons escape, which is a factor of two higher than the measurement by \citet{hayes11}. During the transparent phase, our simulated cloud is better represented by \citet{calzetti00}, although attenuation due to the ISM/CGM can shift the sequence to the \citet{hayes11} line. We note that the two differences can potentially contribute to the scatter in the observed $\fescLya-E(B-V)$ relations, which is consistent with the results from the HETDEX pilot survey \citep{blanc11}.

\begin{figure}
   \centering
   \includegraphics[width=8cm]{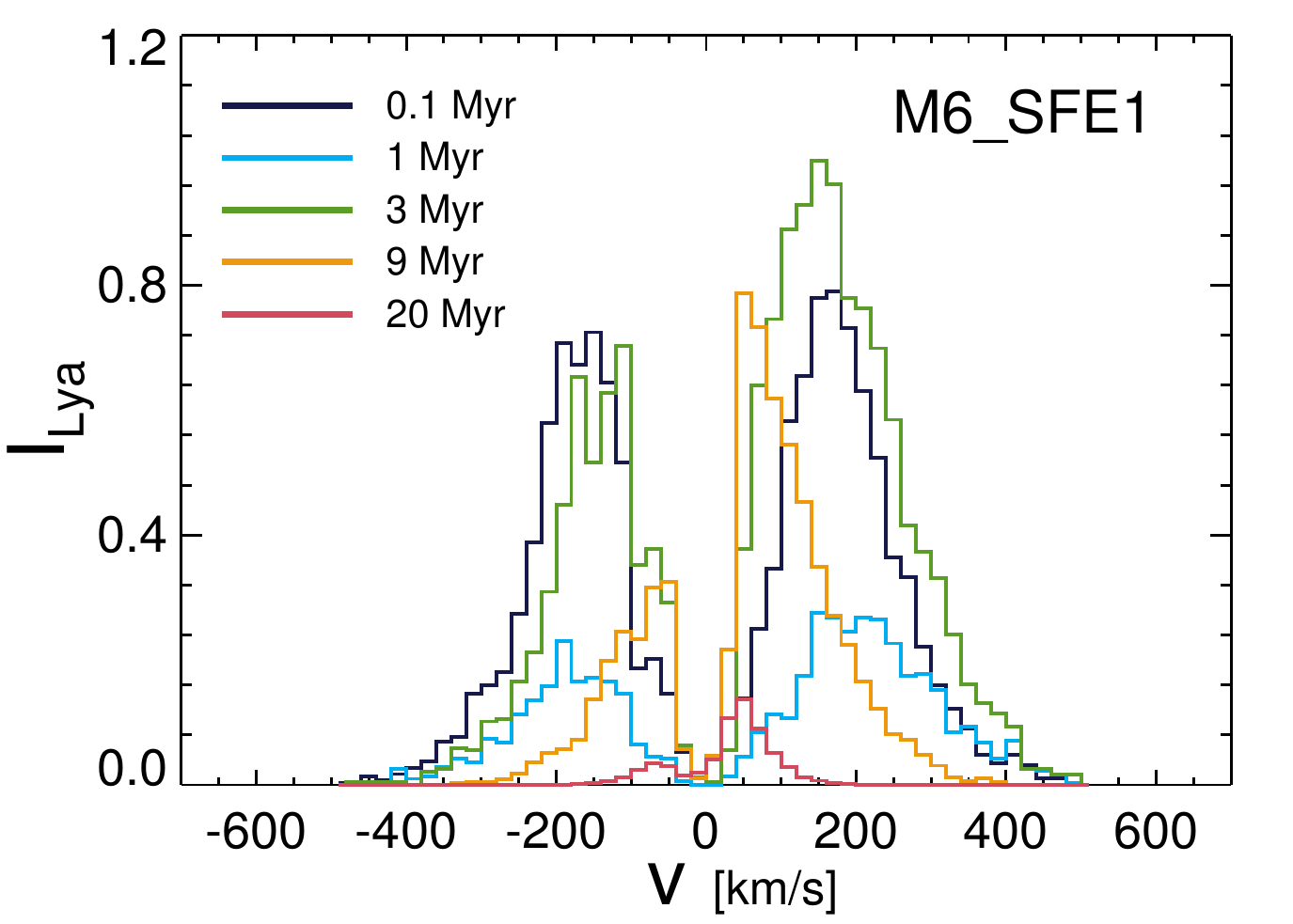} 
      \includegraphics[width=8cm]{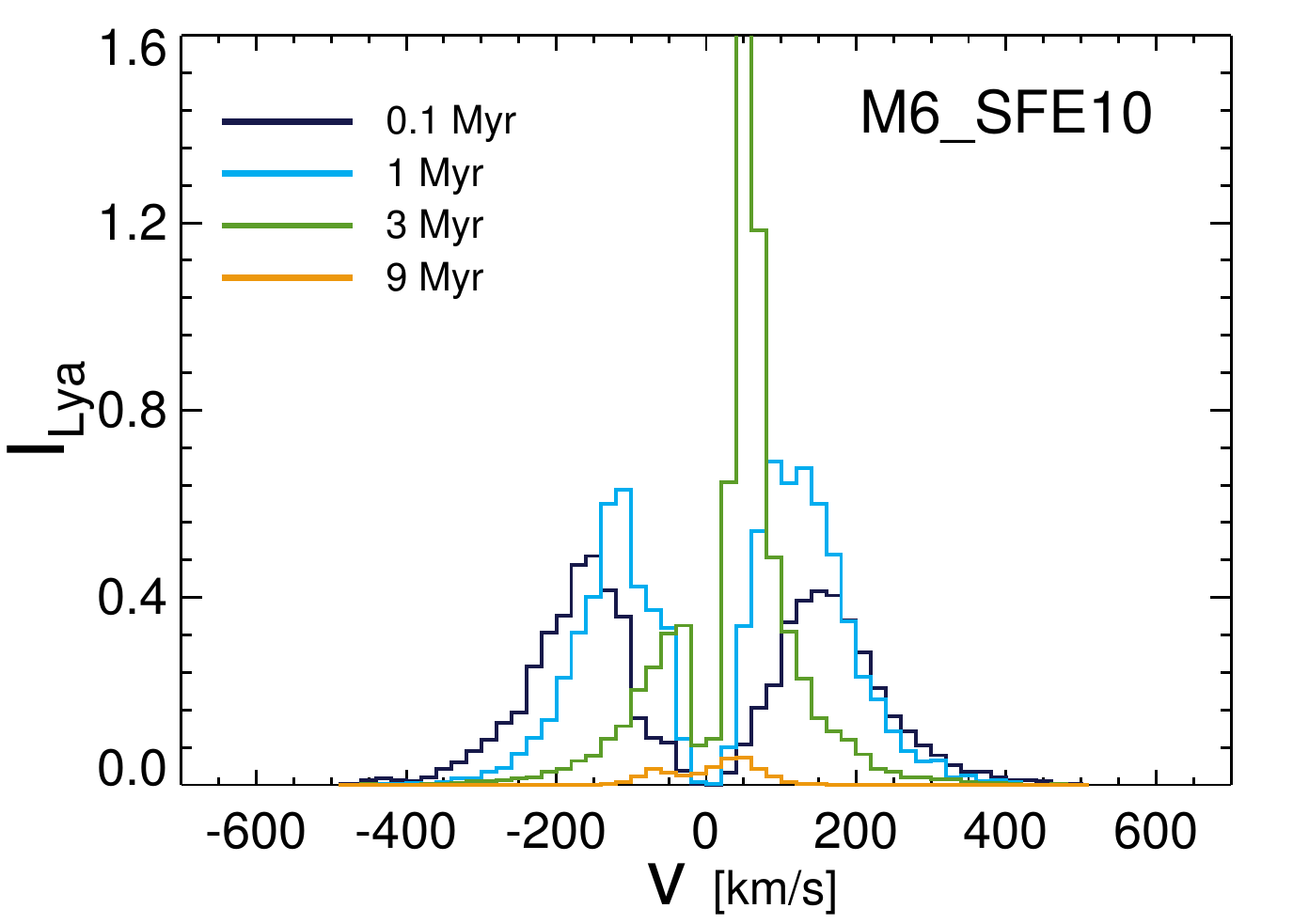} 
   \caption{Emergent spectra of \Lya\ photons generated from partially ionized gas inside the massive cloud of mass $10^6\,\msun$. Note that \Lya\ photons that would be produced outside the cloud from the escaping LyC photons are not included in this plot. The top and bottom panels show the spectra from the run with a 1\% and 10\% SFE, respectively. Different colour codings denote different times of the evolution, as indicated in the legend. The spectra are in arbitrary units and normalized to the maximum intensity. The profiles initially show symmetric double peaks, but the red peak becomes more pronounced as the cloud gets expanded by  stellar feedback.    }
   \label{fig:prof_lya}
\end{figure}

\subsubsection{Line profile of \Lya\ photons}

We find that the velocity profile of \Lya\ photons is already complex on cloud scales. Figure~\ref{fig:prof_lya} presents angle-averaged \Lya\ profiles as a function of time in the massive cloud ($M_{\rm cloud}=10^6\,\msun$) with two different SFEs (1\% and 10\%). In the early phase, during which \fescLyC\ is small and the optical depth is high, the velocity profile exhibits well-known symmetric double peaks \citep{neufeld90,ahn01,verhamme06,dijkstra14}. Once radiation feedback drives outflowing motions, the red peak becomes more pronounced (green line in the \texttt{M6\_SFE1} case) and eventually dominates the velocity profile (orange and pink lines in the \texttt{M6\_SFE1} run or green and orange lines in the \texttt{M6\_SFE10} run).  Note that few photons with zero velocity shift escape from the cloud \citep[c.f.][]{behrens14}.

We also note that the emergent spectrum is quite broad ($\Delta v_{\rm peak} \sim 200-400\,\kms$) in the early phase of the cloud evolution. Because the young stars are enshrouded by a large amount of neutral hydrogen, the initial average column density along the sight lines of star particles is large ($\left< N_{\rm H} \right>\sim 10^{22}\,{\rm cm^{-2}}$), and the line-centre optical depth is $\tau_0 \sim10^8-10^{9}$ (see the bottom right panel of Figure~\ref{fig:lya_origin}). However, since \Lya\ photons preferentially propagate along low-density channels due to their resonant nature, the velocity peaks are not separated as much as the uniform case with the given optical depth ($v_{\rm sep}\equiv v_{\rm peak,red}-v_{\rm peak,blue}\sim1000\,\kms$), but this separation is certainly broader than the thermally broadened spectra by the warm ($T\sim10^4\,{\rm K}$) ISM.  At later stages ($t\ga 10\,{\rm Myr}$), the logarithmic mean of  $\left<\tau_0\right>$ at the \Lya\ production sites becomes smaller ($\log \tau_0\sim 6-8$) in the case of  \texttt{M6\_SFE1}, resulting in the narrower offset of the velocity peak ($\Delta v_{\rm peak} \la 100\,\kms$). It is worth noting that \Lya\ photons scatter significantly at all times, possibly imparting a large amount of momentum to unbind the star-forming gas via resonant scattering \citep[see Section~\ref{sec:lya_pressure},][]{kimm18,dijkstra08,smith17}.

\begin{figure}
   \centering
   \includegraphics[width=8.0cm]{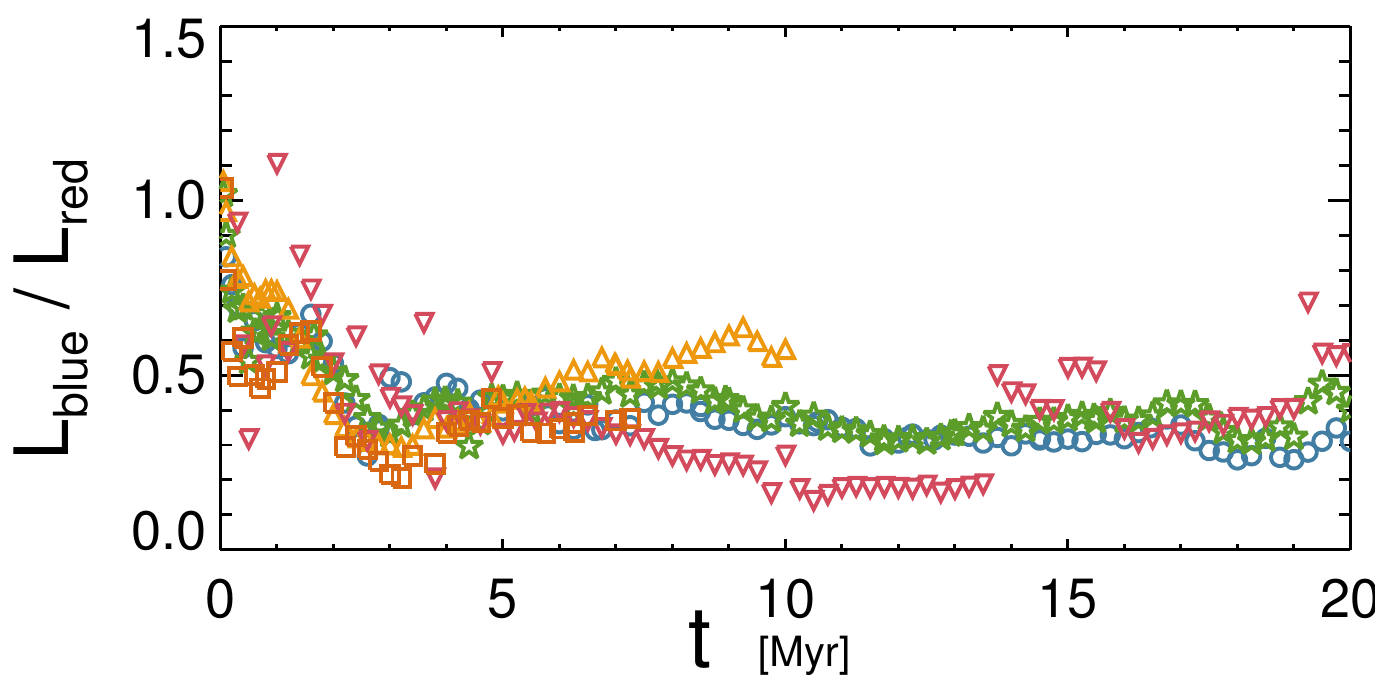} 
      \includegraphics[width=8.0cm]{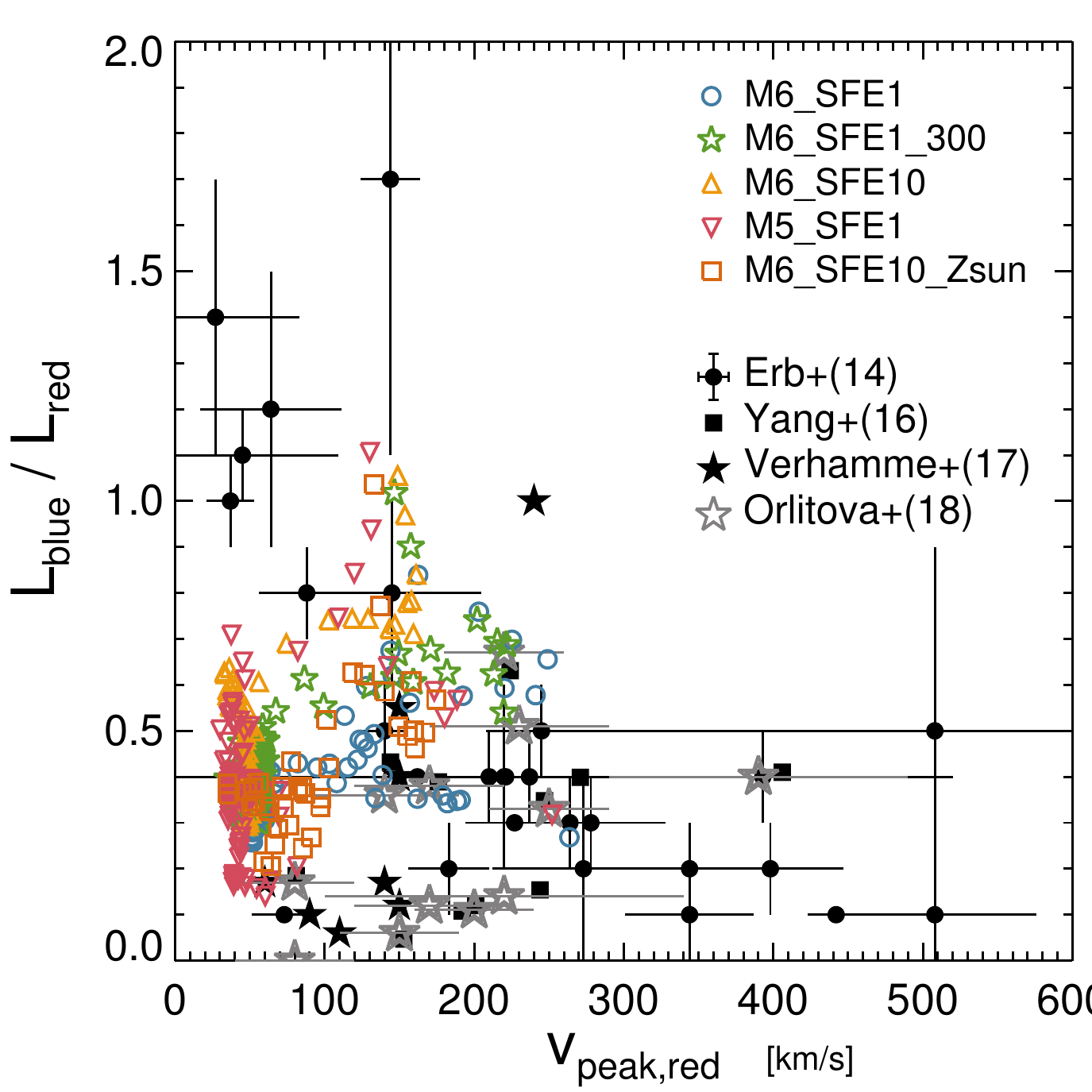} 
   \caption{Ratio of the number of photons blueward of the \Lya\ line centre to the number of photons redward of the line centre ($L_{\rm blue}/L_{\rm red}$). The top panel shows the time evolution of the ratio in different runs, while the bottom panel exhibits the relation between the position of the red peak in velocity and the luminosity ratio. Observational data points by \citet{erb14}, \citet{yang16}, \citet{verhamme17}, and  \citet{orlitova18} are shown as black filled circles, squares, stars, and empty grey stars, respectively.    } 
   \label{fig:Lratio}
\end{figure}

To quantify the asymmetry of the velocity profile \citep[e.g.][]{erb14}, we  present the ratio of the number of photons blueward of the line centre to the number of photons redward of the line centre ($=L_{\rm blue}/L_{\rm red}$) as a function of time in Figure~\ref{fig:Lratio} (top panel). When \Lya\ photons are efficiently trapped, the ratio is close to unity, although no simulated clouds exhibit a large velocity offset of $v_{\rm peak,red} > 300\,\kms$ with $L_{\rm blue}/L_{\rm red}\sim 1$ due to the presence of turbulent structures. Once the gas near the young stars is radially accelerated, the outflowing motion leads to a smaller ratio of $L_{\rm blue}/L_{\rm red}\approx 0.2-0.5$. Because the velocity of the outflows is not very large (Figure~\ref{fig:cloud_mass}) and because the optical depth to the \Lya\ photon decreases (Figure~\ref{fig:lya_origin}), we find that the velocity peak of the spectrum redward of the line centre in the simulated clouds is no greater than 100$-$200\,\kms\ at $t\ga 3\,{\rm Myr}$. This seems inconsistent with the observations that some galaxies show a large velocity offset ($v_{\rm peak,red}\ga 400 \,\kms$) with a pronounced red peak \citep[$L_{\rm blue}/L_{\rm red} \la 0.5$,][]{erb14}. However, the discrepancy may be reconciled if scattering with large-scale galactic outflows is included, as it can shift the frequency of the photons to even redward directions. It is also interesting to point out that the simulated spectra of the clouds cannot account for the small velocity offset with $L_{\rm blue}/L_{\rm red}\ga 1$, suggesting that a large fraction of \Lya\ photons may directly arise from an optically thin ISM with little outflowing motions in some galaxies. Not surprisingly, both cases demonstrate that the modelling of emergent \Lya\ spectra requires the scattering with birth clouds {\it and} the ISM/CGM \citep[e.g.][]{smith19}.

\begin{figure}
   \centering
   \includegraphics[width=8cm]{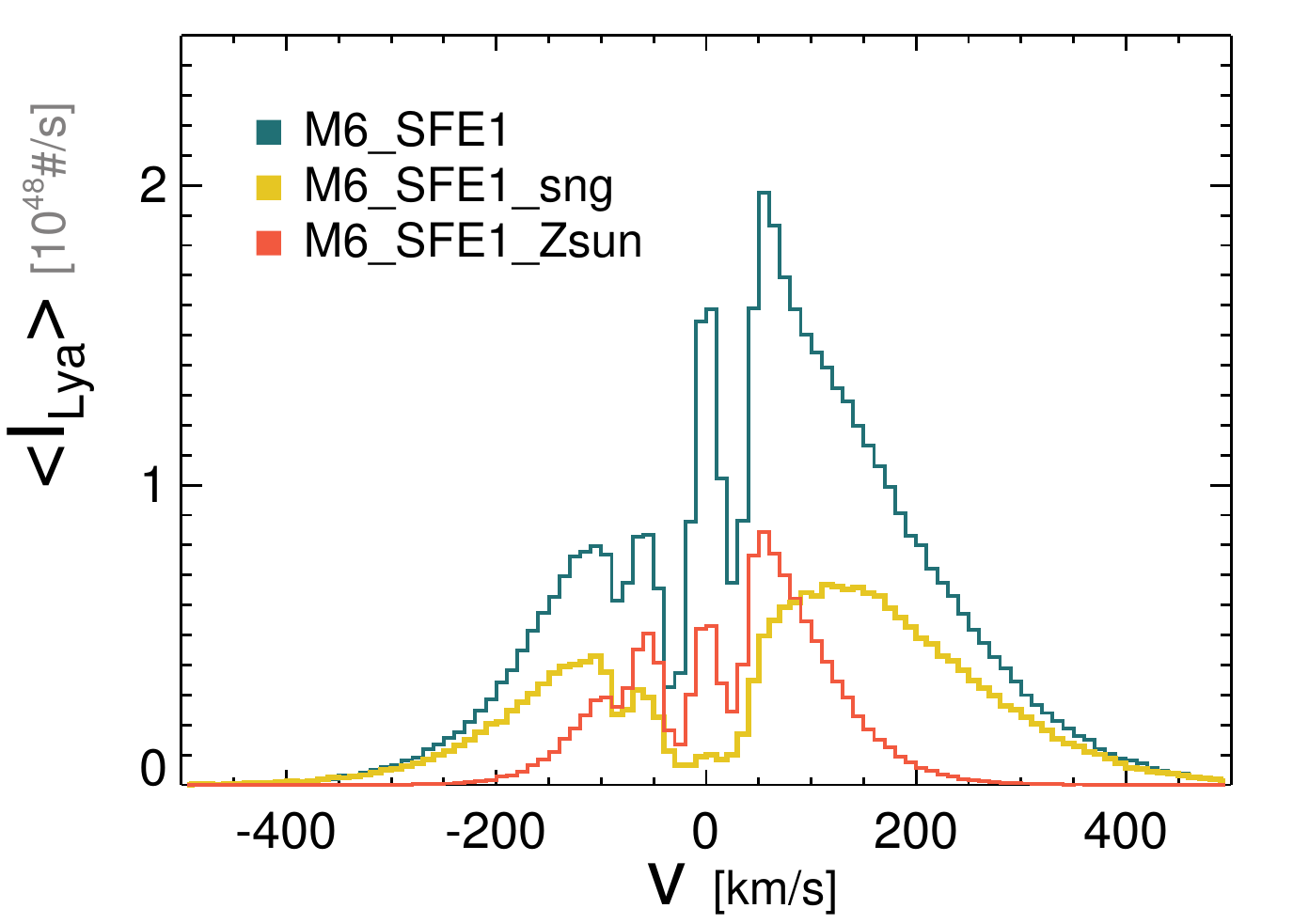} 
      \includegraphics[width=8cm]{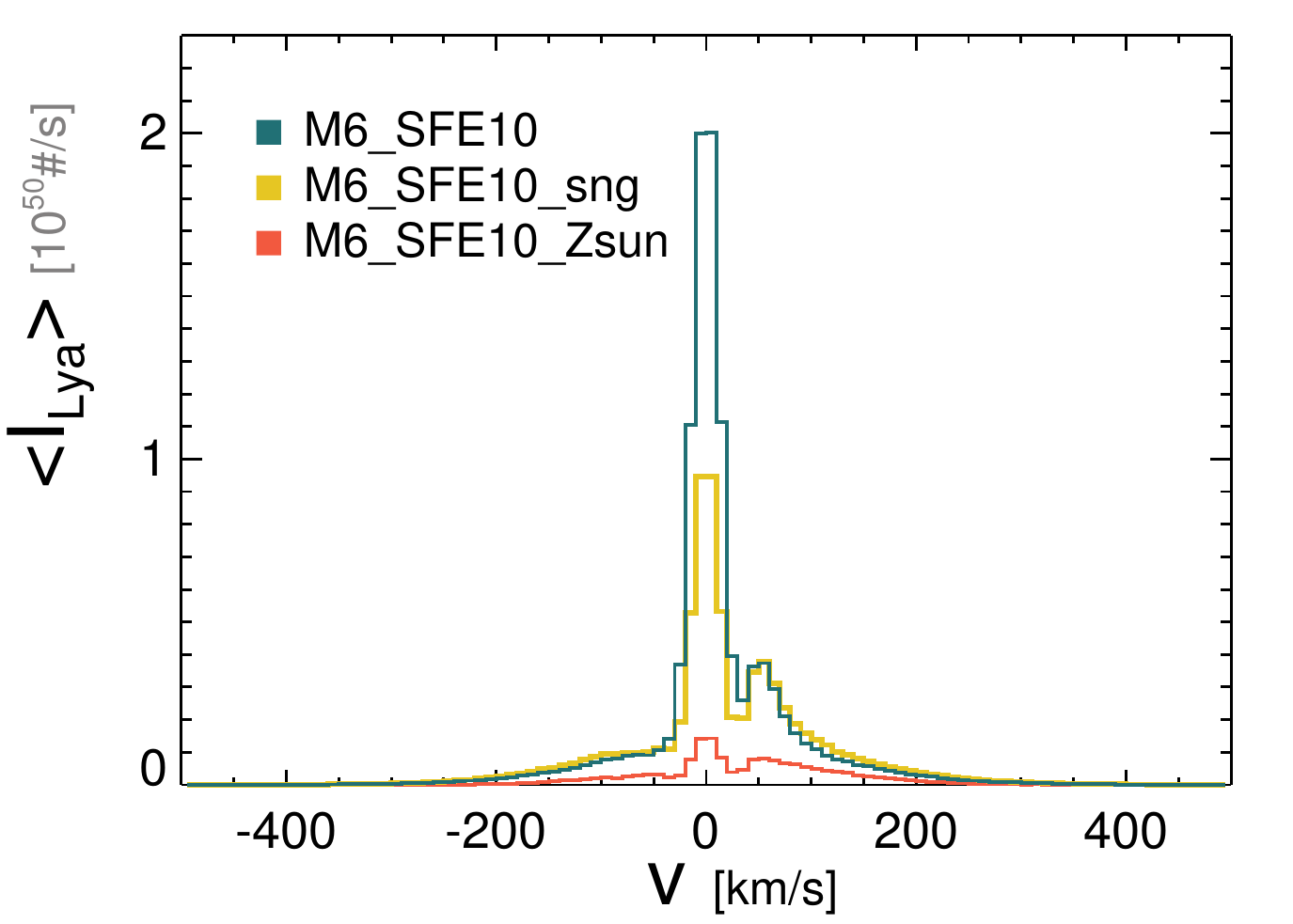} 
   \caption{   The luminosity-weighted, {\em time-averaged} spectra of \Lya\ photons from the massive cloud of mass $10^6\,\msun$, assuming that all of the LyC photons escaping from the cloud are processed to \Lya\ photons by the ISM of temperature $10^4 \, {\rm K}$. When the net escape fractions are reasonably low, the combined spectra show triple peaks, while only two peaks are noticeable in the high SFE case. Note that the profiles are already quite broad on cloud scales and certainly different from the simple Gaussian profiles that are often used as an input for the \Lya\ source in under-resolved simulations.}
   \label{fig:prof_lya_avg}
\end{figure}

Figure~\ref{fig:prof_lya} also shows that the velocity profile of \Lya\ photons is highly time variable, which can potentially be an issue when modelling \Lya\ emission in under-resolved galactic-scale simulations. In this case, a possible option is to use the luminosity-weighted time average of the velocity profile as an input spectrum for star-forming regions and let them propagate from the cloud boundaries. This is certainly an approximation but would be a good alternative to using simple Gaussian profiles in simulations with under-resolved clouds especially if there is a large number of GMCs hosting stellar populations with different ages. To provide an idea of how the  processed spectrum from the cloud would contribute to the total \Lya\ spectrum in a galaxy, we compute the luminosity-weighted, time average spectrum of \Lya\ photons in different clouds in Figure~\ref{fig:prof_lya_avg}. Note that we assume 100\% of escaping LyC photons are absorbed and re-processed into \Lya\ photons outside the cloud, i.e., in a volume-filling warm-ionized ISM with $T\sim10^4\,{\rm K}$ \citep[e.g.][]{kim13,kimm18}. For simplicity, we do not take into account any existing macroscopic motions, such as outflows, which can make the spectrum broader (see the discussion section). The plot shows that when some fraction of the LyC photons leak out from the system (i.e., \texttt{M6\_SFE1} series), the average spectrum tends to have triple peaks. If we fit the velocity distributions with the single Gaussian profile, the FWHM would be $\approx 330$ and $410\, \kms$ for the binary and single stellar evolution SEDs, respectively, indicating that \Lya\ photons scatter more in the cloud with weaker radiation feedback (\texttt{M6\_SFE1\_sng}). When the absorption by dust increases ($\texttt{M6\_SFE1\_Zsun}$), photons that travel a longer distance are preferentially destroyed by dust, resulting in a narrower FWHM of $\approx220\,\kms$ \citep[see also][]{laursen09}. Indeed, we confirm that if we artificially lower the gas metallicity by an order of magnitude ($Z_{\rm gas}=0.002$) before post-processing with {\sc rascas}, we recover the large FWHM ($\sim 400\,\kms$) found in the  \texttt{M6\_SFE1} run.

In contrast, if the SFE is high enough to disrupt the cloud very quickly (i.e. \texttt{M6\_SFE10} series), we find little dependence of the FWHM on the cloud properties. The velocity profiles of the runs with the binary star and single star SEDs look very similar essentially because both SEDs produce a similar number of LyC photons before the cloud is destroyed. An exception is the metal-rich case in which the cloud is not efficiently destroyed, as opposed to other high SFE runs. The different amplitude at the velocity centre between  \texttt{M6\_SFE10} and  \texttt{M6\_SFE10\_sng} is simply due to the fact that more ionizing photons are emitted in the binary SED than in the single SED. At the late stage of the evolution, the profiles become asymmetric and display signs of outflows (red peak), even though these weak signals are likely to be modified by the scattering due to neutral hydrogen in the ISM/CGM.

\begin{figure}
   \centering
   \includegraphics[width=8.4cm]{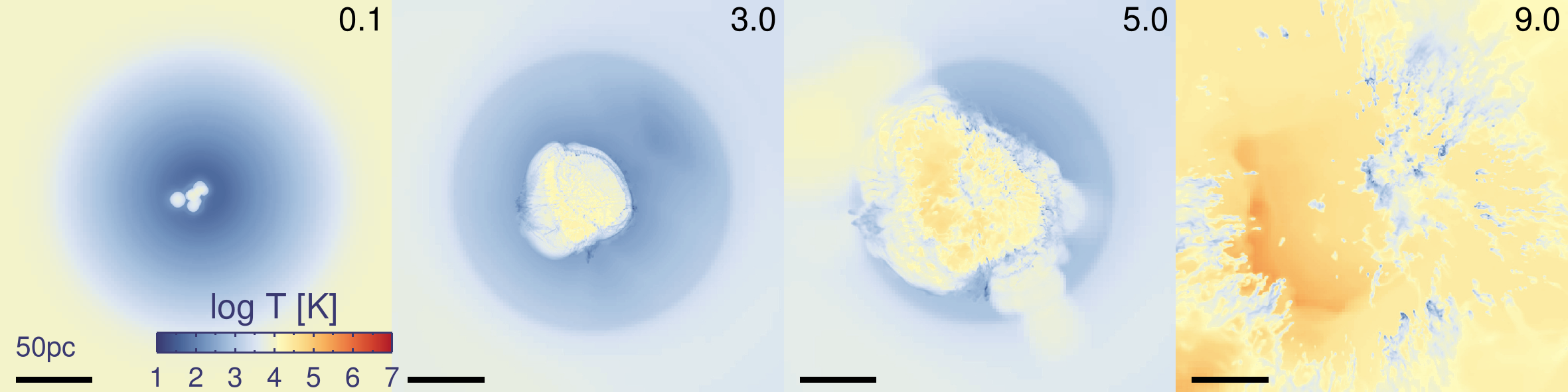} 
      \includegraphics[width=4.0cm]{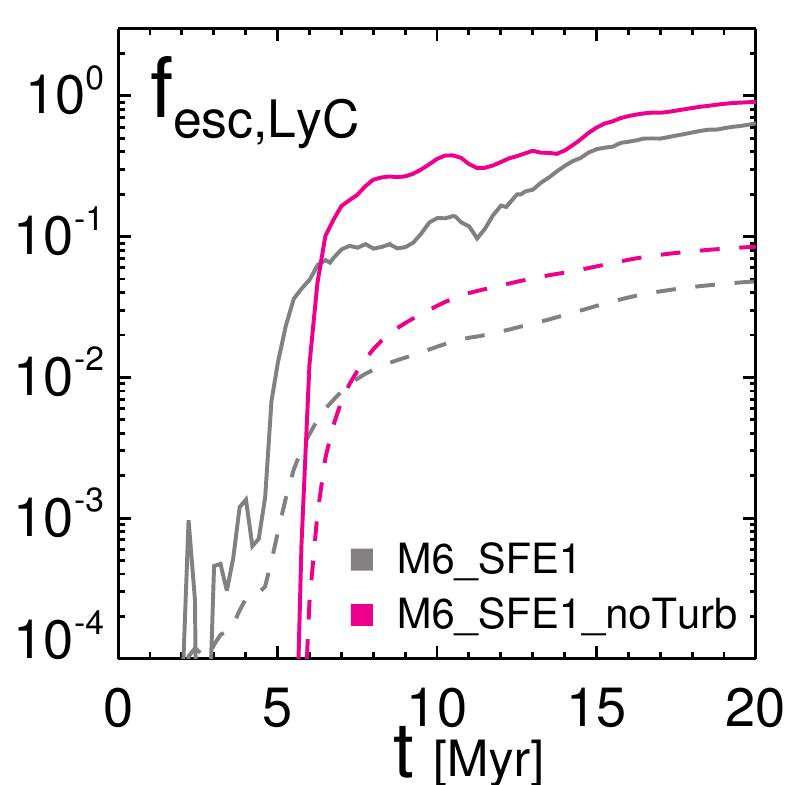} 
      \includegraphics[width=4.0cm]{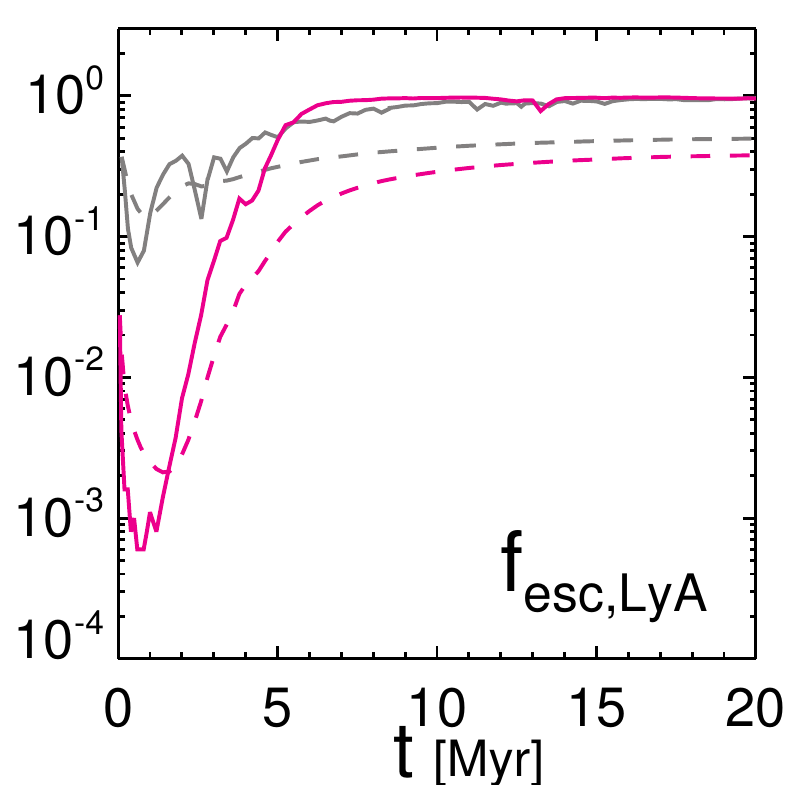} 
      \includegraphics[width=8.4cm]{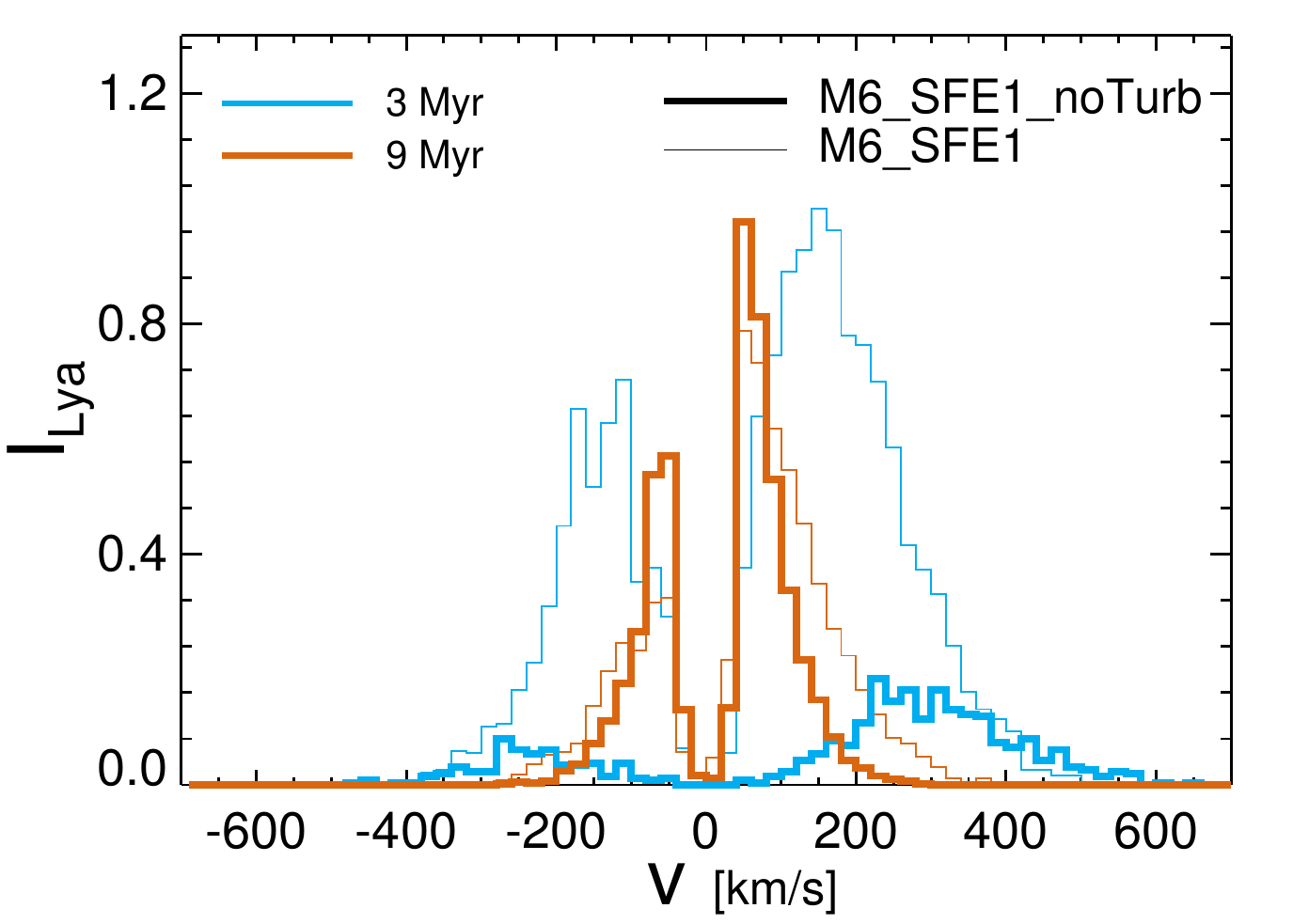} 
   \caption{ Effects of initial turbulence on the escape of LyC and \Lya\ photons. The top panels show the projected temperature distributions of the cloud at different times, as displayed in the top right corner in units of Myr. We show the corresponding escape fraction of LyC (middle left) and \Lya\ photons (middle right) as pink lines. For comparison, we also include the escape fractions from our fiducial run as grey colours (\texttt{M6\_SFE1}). The solid lines represent the instantaneous escape fraction, whereas the luminosity-weighted average is shown as dashed lines. The bottom panel displays the emergent spectra from the clouds without (thick solid lines) and with (thin solid lines) initial turbulence at two different times, as indicated in the legend.}
   \label{fig:noTurb}
\end{figure}

\section{Discussion}
In this section, we discuss how the presence of turbulent structures affects the escape of LyC and \Lya\ photons, and how these photons are related by comparison with observations. We also discuss the implication of the escape fractions measured from turbulent clouds to the reionization history of the Universe, and show possible effects derived from modelling of star formation and feedback. 
\subsection{Effects of turbulent structures}

As mentioned previously, the typical resolution of galactic-scale simulations ($\Delta x \ga 10 \,{\rm pc}$) is insufficient to capture complex turbulent structures in star-forming regions. In this case, low-density channels through which LyC and \Lya\ photons propagate may be under-resolved, leading to an underestimation in the true escape fractions. To examine the effects of turbulent structures, we run a control simulation without initial turbulence while keeping other parameters including positions, metallicity, age, and mass of the star particles fixed, as in \texttt{M6\_SFE1}. This is done by fitting the radial gas distribution of the initial turbulent cloud to the analytic form of $n_{\rm H} = n_{\rm H,0} \left[ 1 + \left(r / r_s\right)^\alpha \right]^{-\beta}$. We use the following set of parameters that best matches the gas distributions: $n_{\rm H,0}=499.4 \,\cmq$, $r_s=40.1\,{\rm pc}$, $\alpha=1.343$, and $\beta=6.509$.

Figure~\ref{fig:noTurb} (top panel) shows that the propagation of the LyC photons is efficiently confined within the cloud in the early phase ($t \la 6\,{\rm Myr}$) in the absence of initial turbulence. Although SN explosions drive turbulence inside the cloud, the outer neutral gas shells that block LyC photons are relatively well maintained until the ionization front breaks out of the cloud. Later, the shell experiences hydrodynamic instabilities \citep[e.g.][]{elmegreen94}, breaking into smaller clumps, similar to the final stage of the cloud evolution in the turbulent case (c.f. Figure~\ref{fig:img}). As such, the escape of LyC photons is efficiently suppressed until $\sim6$ Myr and increases rapidly afterwards. The resulting luminosity-weighted escape fraction ($\left<\fescLyC\right>=8.5\%$) is found to be even higher than the turbulent case ($\left<\fescLyC\right>=4.8\%$), as the propagation of the ionization front is faster in the case with no turbulence where the mean density around young stellar populations drops steeply at large radii \citep[c.f.][]{safarzadeh16}. This indicates that the porous structure inside the cloud does not necessarily lead to the significantly enhanced escape fractions, as speculated by previous studies as a possible solution to reproduce an early reionization of the Universe \citep[e.g.][]{kimm14}. 

Interestingly, we find that the presence of a turbulent structure leads to more significant escape of \Lya\ photons, as suggested by the observations of 14 galaxies with strong \Lya\ emission \citep{herenz16}. The middle right panel of Figure~\ref{fig:noTurb} shows that the run without the initial turbulence begins with a high  \fescLya\ value at $t\approx0$, as the cloud is assumed to be initially half molecular and transparent to \Lya\ photons. However, molecular hydrogen is quickly dissociated by Lyman-Werner radiation from stars on a timescale of  $\approx 1\,{\rm Myr}$, and the scattering of \Lya\ photons becomes significant again. Because there are no low-density channels due to the lack of a turbulent structure, \Lya\ photons are efficiently trapped inside the HI shells, resulting in larger velocity offsets compared to those in the turbulent clouds (see blue lines in the bottom panel). Furthermore, \fescLya\ is reduced as \Lya\ photons have a high probability of encountering dust and being destroyed. This phase stops and the high escape fractions observed in the turbulent case are recovered once the gas cloud fragments into smaller clumps ($t> 5\,{\rm Myr}$). Because the majority of the \Lya\ photons are created in the early phase, the resulting net \Lya\ escape fractions become lower in the absence of turbulence ($\left< \fescLya \right> \approx 31\%$ vs. $47\%$).

\begin{figure*}
   \centering
         \includegraphics[width=8.cm]{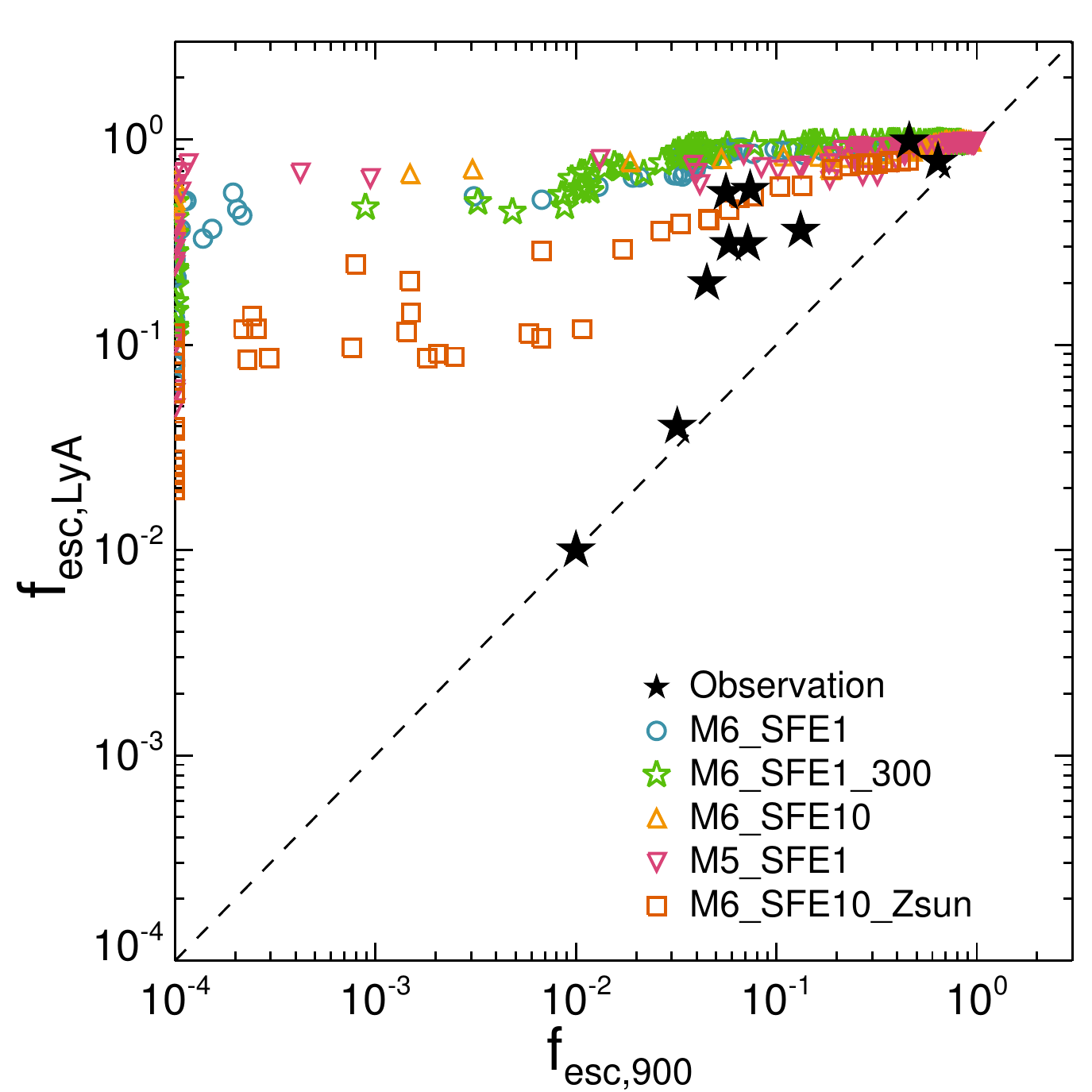} 
      \includegraphics[width=8.cm]{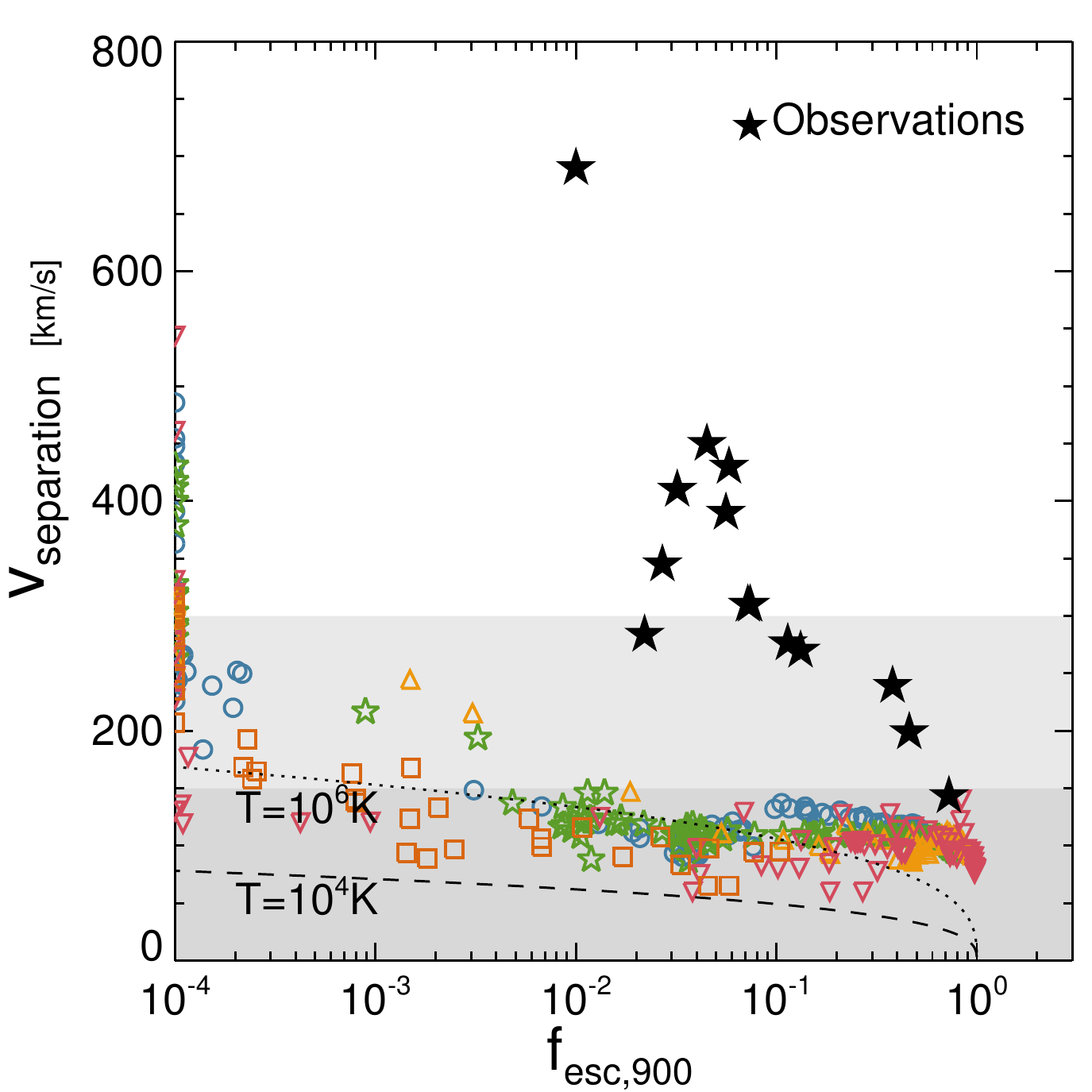} 
   \caption{Left: relation between the \Lya\ and LyC escape fractions  measured around 900 \AA\ in turbulent clouds.  Different symbols denote the results from the runs with different conditions, as indicated in the legend. \fescLya\ is always larger than $f_{\rm esc}^{\rm 900}$, and the clouds where LyC photons escape more efficiently tend to show higher \fescLya. Note that we have displayed the results with $f_{\rm esc}^{\rm 900}<10^{-4}$ as $f_{\rm esc}^{\rm 900}=10^{-4}$ for better readability. Also included as black star symbols is the {\em galactic} escape fractions from observations \citep{vanzella15,izotov16a,izotov16b,verhamme17,vanzella18,izotov18}. Right: the separation of the peaks ($v_{\rm sep}$) in the \Lya\ velocity profile. The grey lines are the analytic predictions for uniform media with different gas temperatures (Eq.~\ref{eq:d14}). Black stars indicate the data obtained from low-redshift galaxies \citep{verhamme17,izotov18,izotov18b}. Note that the clouds with moderate to high $f_{\rm esc}^{\rm 900}$ ($\ga 1\%$) show a peak separation of less than $150\,\kms$.
     }
   \label{fig:vsepa}
\end{figure*}

\subsection{Connection between LyC and \Lya\ photons}

In Figure~\ref{fig:vsepa}, we compare the escape fractions of \LyC\ around 900\AA and \Lya\ photons from different environments. We confirm the previous finding that \Lya\ photons leave the cloud more efficiently than ionizing photons \citep{yajima14,verhamme15,dijkstra16,verhamme17}. A large fraction ($\sim 20-80\%$) of \Lya\ photons leave the cloud regardless of the input stellar spectrum and size of the cloud, even when almost all ($f_{\rm esc}^{\rm 900}\la10^{-3}$) of the \LyC\ photons are absorbed by neutral hydrogen. \citet{dijkstra16} show that the escape fractions of \Lya\ photons tend to be low ($\la 10\%$) if there are more than $\sim 5$ clumps along the sight line from the centre to the edge of the cloud. In light of this, the high \fescLya\ observed in our turbulent cloud simulations suggest that \Lya\ photons do not interact with dense clumps and filaments very often but rather propagate through low-density channels. We also find that \fescLya\ decreases down to $\sim10\%$ when the amount of dust increases (i.e., $Z_{\rm gas}=0.02$), although the difference is negligible once the cloud becomes optically thin to both \Lya\ and \LyC\ photons. As a result, the positive correlation between $f_{\rm esc}^{\rm 900}$ and $\fescLya$ becomes more pronounced in the metal-rich case.

\citet{verhamme15} argue that \LyC\ leaking candidates may be pre-selected by one of the following two features. First, if the intervening medium is optically thin to \LyC\ photons, the separation of double peaks in the \Lya\ profile should be smaller than $ v_{\rm sep} \la 300\,\kms$. Second, if the ISM is clumpy, \Lya\ photons propagating through low-density channels will be seen at line centre, while some photons will escape on the blue side of the profile. The latter point is straightforward to understand, and indeed can be inferred from the top panel of Figure~\ref{fig:prof_lya_avg}. Although the precise determination of the amplitude of the central peak would depend on the distribution of the ISM, the clouds with $\left<\fescLyC\right>\ga 2\%$ exhibit some photons near the line centre (e.g., \texttt{M6\_SFE1} and \texttt{M6\_SFE1\_Zsun}), while they nearly disappear in the cloud with a lower $\left<\fescLyC\right>=0.3\%$ (\texttt{M6\_SFE1\_sng}).

To explore the first possibility, we measure the peak separation ($v_{\rm sep}$) by fitting the \Lya\ line profile with two different components, i.e., two skewed Gaussian profiles on the left and right side of the line centre in the right panel of Figure~\ref{fig:vsepa}. Our simulated velocity separation of the two peaks generally follows the  simple analytic trends expected in a uniform medium \citep[e.g.][]{neufeld90,dijkstra17},
\begin{gather}
v_{\rm sep} \approx 320\, \kms \left( \frac{N_{\rm HI}}{10^{20}\,{\rm cm^{-2}}} \right)^{1/3} \left( \frac{T}{10^4\,{\rm K}} \right)^{1/6},
\label{eq:d14}
\end{gather}
but two differences are found.  First, the predicted $v_{\rm sep}$ appears to be larger than the analytic calculation with $T\sim10^4\,{\rm K}$, appropriate for our simulated clouds, which is due to the fact that LyC photons can escape through low-density channels, while the majority of the \Lya\ photons still scatter inside dense regions of the cloud. Second, $v_{\rm sep}$ does not approach zero even when the escape fraction becomes very high ($f_{\rm esc}^{\rm 900}\ga 50\%$), as most \Lya\ photons are generated near (a few surviving) relatively dense regions where the optical depth to \Lya\ is still high. More importantly, we find that when $f_{\rm esc}^{\rm 900}$ is intermediate or high ($\ga 1\%$),  the separation of the peaks is small ($v_{\rm sep} \la 200 \,\kms$), supporting the criterion proposed by \citet{verhamme15}. However, the small separation does not guarantee pre-selection of the efficient LyC leakers on cloud scales \citep[as pointed out by][]{verhamme15} because the turbulent structure inside the cloud allows the \Lya\ photons to propagate more efficiently than the uniform or shell case. As a result, some clouds with low LyC escape fractions ($f_{\rm esc}^{\rm 900}\la0.1\%$) show $v_{\rm sep} \la 300\,\kms$.

Figure~\ref{fig:vsepa} also shows that $\fescLya$ from the simulated clouds has a weaker correlation with $f_{\rm esc}^{\rm 900}$ compared to that obtained from metal-poor ($\sim0.1$--$0.2\,Z_\odot$) compact starburst galaxies \citep{vanzella15,izotov16a,izotov16b,verhamme17,vanzella18,izotov18}. The two significant LyC emitters, {\it Ion2} \citep{vanzella15} and J1154+2443 \citep{izotov18}, exhibit escape of the order of unity for both \Lya\ and LyC photons, similar to our findings. However, other observed galaxies with lower $f_{\rm esc}^{\rm 900}$ tend to have a smaller \fescLya, but with a much more broadened \Lya\ spectrum \citep[the right panel,][see also the equation 2 of \citealt{izotov18b}]{verhamme17,izotov18,izotov18b}. This suggests that the interaction of \Lya\ photons with neutral hydrogen and dust inside the star-forming clouds does not fully account for the observed features of the UV spectrum and that scattering with the ISM may significantly change the propagation of \Lya\ photons in galaxies. It may also be possible that multiple clouds in different stages of cloud evolution contribute in a complex way to reproduce the observations, but in this case a large degree of scatter may be present in the low \fescLya\ regime.

Recently, \citet{vanzella18,rivera-thorsen17,izotov18} observed LyC leakers with triple \Lya\ peaks. They claim that this is consistent with a simple model in which LyC and \Lya\ photons escape through the same cavity \citep{behrens14,verhamme15}. Our simulations also support this picture in the sense that the central peaks do not appear to arise from inside of the clouds, but only seem possible if we include \Lya\ photons that would be generated by LyC photons leaking into an ISM (Figure~\ref{fig:prof_lya_avg}). An interesting difference from the results of {\it Ion3} \citep{vanzella18} is that the central peak in  {\it Ion3}  extends to $v\sim \pm 100 \, \kms$, whereas our thermally and turbulently broadened spectra do not reach more than $v\sim \pm 50 \,\kms$. This indicates that the motions of \Lya\ emitting gas in the ISM from this compact starburst galaxy are likely to be dominated by strong turbulence corresponding to the Doppler parameter of $b\ga 20\,\kms$ (see Figure~1 of \citealt{verhamme15}).

\subsection{Implications for reionization of the Universe}

In a dwarf galaxy-driven scenario for reionization, the key quantity that governs the expansion of ionized bubbles is the number of escaping photons \citep[$N_{\rm esc}^{\rm LyC}$, e.g.,][]{wise14,kimm17,koh18,rosdahl18}. However, making theoretical predictions for $N_{\rm esc}^{\rm LyC}$ is not trivial, not only because the turbulent structure for high-$z$ galaxies is unknown, but also because the input stellar spectra are not well constrained in the early Universe. We find that in the massive cloud with a 10\% SFE, the runs adopting a single, binary, and binary SED with larger cut-off mass (i.e., $M_{\rm upper}=300\,\msun$) yield total $0.6\times10^{52}$, $1.3\times10^{52}$, and $2.4\times10^{52}$ number of escaping LyC photons during the first 10 Myr (Table~\ref{tab:result}). In the case of a lower SFE (1\%), the total $N_{\rm esc}^{\rm LyC}$ within $20\,{\rm Myr}$ is found to be $0.08\times10^{50}$,  $1.6\times10^{50}$, and  $3.5\times10^{50}$ for the single, binary, and binary SED with the larger cut-off mass, respectively. These results indicate that the total $N_{\rm esc}^{\rm LyC}$ can vary by a factor of $\sim4-40$ depending on the choice of the stellar spectra, and that one must be careful with interpreting the results from previous simulations where the single stellar evolution is adopted for the photon production rate. Indeed, using high-resolution ($\sim$10 pc), cosmological radiation-hydrodynamic simulations, \citet{rosdahl18} show that the simulated volume of $(10\,{\rm Mpc})^3$ is fully ionized by $z_{\rm reion}\approx7$ when the binary stellar evolution model is used, while single stars cannot ionize the volume in time ($z_{\rm reion}<6$) \citep[see also][]{ma16}. Given that a large fraction of massive stars do live in binaries in the local environments \citep{sana12}, it is encouraging that the binary model manages to ionize the simulated universe early enough. But the question remains how efficiently star formation needs to be suppressed at high redshifts in order to precisely match the end of reionization (i.e. $z_{\rm reion}\approx6$), as the maximum stellar mass of low-metallicity stars is not well constrained.

\begin{figure}
   \centering
         \includegraphics[width=8.cm]{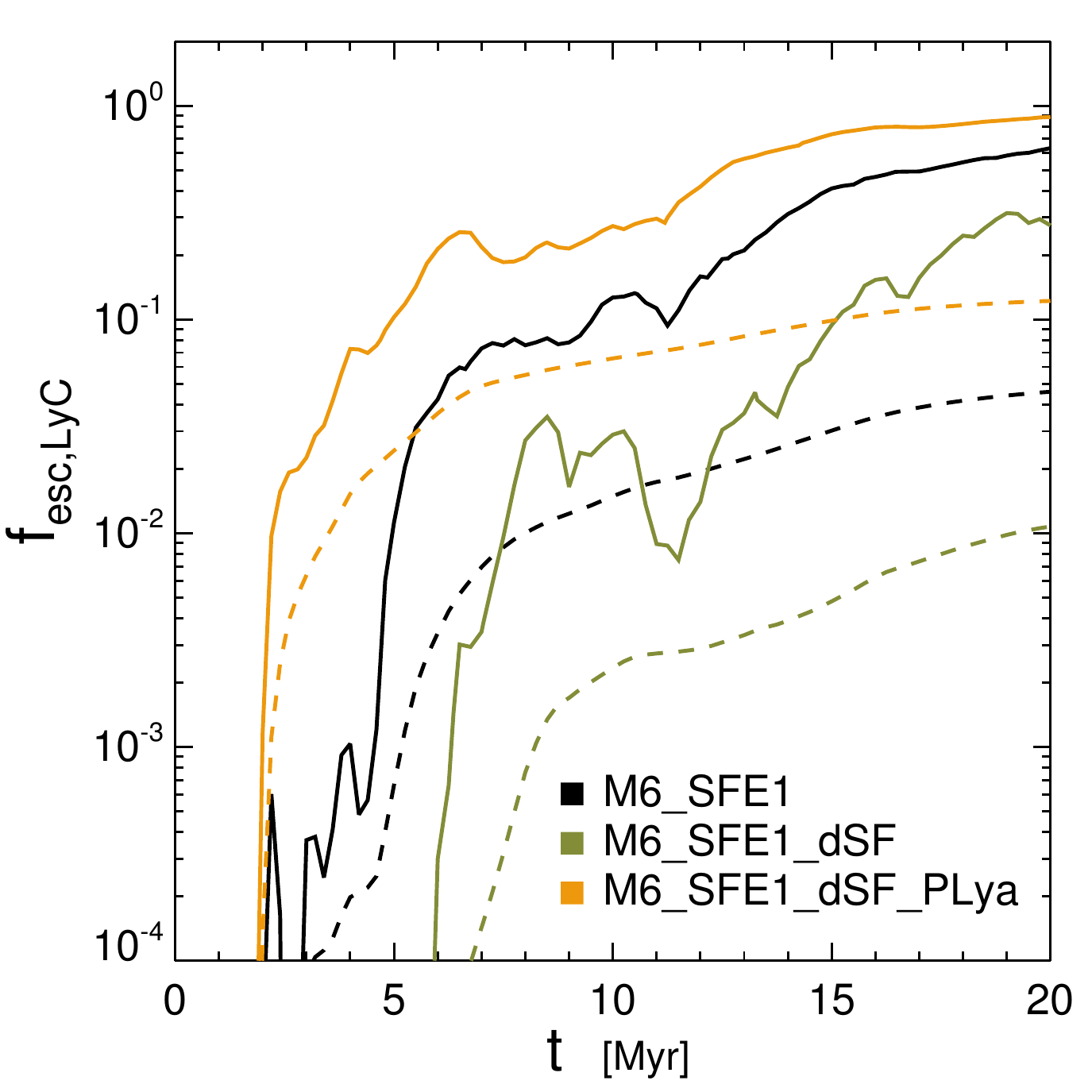} 
   \caption{ Effects of star formation and strong \Lya\ pressure on the escape fractions of LyC photons ($\fescLyC$). The dark green lines indicate $\fescLyC$ from the run where the star particles are initially placed in gravitationally well bound, dense regions, based on the thermo-turbulent star formation model \citep{kimm17}. Our fiducial run (\texttt{M6\_SFE1}) where star particles are placed randomly in space is shown in black. The orange lines show the results when \Lya\ pressure is included on top of photo-ionization heating, radiation pressure, and SNe. The solid lines indicate the instantaneous escape fractions, while the dashed lines show the luminosity-weighted ones. Note that the inclusion of the strong radiation feedback can elevate $\left<\fescLyC\right>$ by an order of magnitude compared to the run in which stars cannot destroy their birth clumps early (\texttt{M6\_SFE1\_dSF}).
     }
   \label{fig:fesc_fb}
\end{figure}

\subsection{Impact of star formation and feedback schemes}
\label{sec:lya_pressure}
In our simulations, we place star particles randomly inside the clouds instead of directly modelling star formation \citep[e.g.][]{bate95,gong13,hubber13,bleuler14} so that we can control the SFE per cloud  while resolving the initial Stromgren sphere. Nevertheless, it is true that stars form in dense environments, and our experiments may under-estimate the interaction of LyC and \Lya\ photons with neutral hydrogen patches. 

In order to understand the difference of adopting a more realistic star formation model, we run additional simulations by placing star particles preferentially in dense pockets of gas. This is done by computing the SFE per free-fall time based on the local thermo-turbulent conditions \citep{federrath12}, as described in \citet[][Equation~2]{kimm17}. We then assign star particles based on the probability of forming a star particle in each cell. The resulting average density of the host cell is $\left<\nH\right>\sim2\times10^5\,\cmq$ (\texttt{M6\_SFE1\_dSF}), which is $10^4$ times denser than that of the runs with a random sampling (i.e., \texttt{M6\_SFE1}). Because young stars are deeply embedded in dense, high-pressure regions, LyC photons are efficiently absorbed and ionization fronts stall at smaller radii. Although some star particles destroy their local clump through a combination of radiation and supernova explosions, the ionization fronts tend to propagate slowly, as the HII bubble develops from over-dense regions. As a result, only $\approx1\%$ of LyC photons escape from the cloud in the \texttt{M6\_SFE1\_dSF} run (Figure~\ref{fig:fesc_fb}), which is a factor of $\sim5$ lower than that of the fiducial model (\texttt{M6\_SFE1}), demonstrating the importance of self-consistent modelling of star formation in simulations.

However, we find that the previous results rely sensitively on the presence of strong feedback processes, such as \Lya\ pressure.  To examine the possible impact of \Lya\ pressure, we adopt the simple model developed in \citet{kimm18}. Briefly, we compute the number of \Lya\ photons produced in each cell using Equations~\ref{eq:rec}--\ref{eq:coe_cooling} and estimate the total neutral hydrogen column density by summing the neutral hydrogen from the host and the adjacent cell along the direction of the propagation. The direction of the momentum input is taken as the direction of the LyC flux, as it allows us to trace the position of the ionizing source. Note that this method neglects the long-range force due to \Lya\ that escapes from the cell of interest and propagates to the neighbouring cells \citep[see][for example]{smith19}. Also, we do not use the Sobolev approximation to conservatively estimate \Lya\ pressure. In this regard, the actual impact from \Lya\ could be even more significant. Despite these simplifications, we find that the average density of the host cell of the young star particles is reduced to $\left<\nH\right>\sim0.1-1\,\cmq$ in about 0.1 Myr. Note that this is four orders of magnitude lower than that of the \texttt{M6\_SFE1\_dSF}  run ($\left<\nH\right>\sim4000\,\cmq$). Consequently, a large fraction of LyC photons leave from the cloud even from $t\sim2\,{\rm Myr}$, leading to $\left<\fescLyC\right>\approx  13\%$ (Figure~\ref{fig:fesc_fb}, orange lines). This suggests that predictions from the simulations without strong radiation pressure should be taken with caution and that even our results based on the randomly distributed star particles without \Lya\ pressure are likely to be a lower limit of the true $\left<\fescLyC\right>$. As a first step, we do not include \Lya\ pressure in our fiducial set of runs, but this issue needs to be addressed for a variety of conditions in the near future.

\section{Conclusions}
 Motivated by the recent results that a large fraction of LyC photons are absorbed on small scales \citep{kimjh13,kimm14,paardekooper15,trebitsch17} and to give useful insights into how we should interpret the emergent \Lya\ spectra, we perform a suite of RHD simulations of turbulent star-forming clouds with stellar feedback processes, including direct radiation pressure, photo-ionization heating, photo-electric heating on dust, non-thermal pressure due to multiply scattering infrared photons, and Type II supernova explosions. By randomly placing star particles around the central region of the cloud, we follow the evolution of $\fescLyC$, $\fescLya$, and the spectral shape of the \Lya\ photons from the runs adopting different cloud masses, SED shapes, gas metallicity, and turbulence.  Our findings can be summarized as follows.

\begin{itemize}
\item[(i)] The escape fractions of \LyC\ tend to increase rapidly and rather monotonically over the cloud lifetime (Figure~\ref{fig:fescLyC}). Although the porous structures inside the turbulent cloud allow for LyC photons to propagate locally, the optically thin channels are not necessarily well aligned, and all of the LyC photons are absorbed by neighbouring gas in the early stage of the evolution (Figure~\ref{fig:img}). Once radiation feedback clears away the neighbouring neutral regions and blows out the dense clumps, the HII bubble expands and the covering fraction of optically thick regions becomes smaller, elevating the escape fractions. In the case of the runs with less efficient radiation feedback (i.e., \texttt{M6\_SFE1\_sng}), SN explosions help to disrupt the cloud, although the escape fractions are not as high as in the fiducial runs (Figure~\ref{fig:fescLyCextra}).

\item[(ii)] We find that the luminosity-weighted, time-averaged escape fractions of LyC photons ($\left< \fescLyC\right>$) are relatively low from a massive cloud ($M_{\rm cloud}=10^6\,\msun$) with a 1\% SFE, which is the typical value derived in the local GMCs \citep[e.g.][]{heyer09} or simulations \citep[e.g.][]{grisdale18}.  With binary star SEDs, the metal-poor clouds with $Z_{\rm gas}=0.002$ show $\left< \fescLyC\right>\sim5\%$. For the metal-rich run ($Z_{\rm gas}=0.02$), radiative cooling enhances the recombination and significantly reduces $\left< \fescLyC\right>$ to $\sim1\%$ (Figures~\ref{fig:fescLyC} and \ref{fig:fescLyCextra}). In contrast, when the SFE is higher or when the cloud mass is smaller, overpressure due to photo-ionization heating and SN explosions blow away the cloud more rapidly (Figure~\ref{fig:img}), leading to a very large $\left< \fescLyC\right>$ of $30$--$70\%$.

\item[(iii)] The runs with binary star SEDs or a higher stellar mass upper limit in the IMF show a higher $\left< \fescLyC\right>$, as radiation feedback is enhanced due to the larger number of LyC photons produced via binary interactions or by very massive stars ($100 < M/\msun < 300$) compared to the SED with single star evolution (Figure~\ref{fig:fescLyCextra}). As a result, the number of escaping LyC photons can easily be different by a factor of $\sim$4 depending on the choice of the SED (Table~\ref{tab:result}).

\item[(iv)] The majority of the absorption takes place on small scales especially when the SFE is low (1\%). We find that 50\%, 90\%, and 99\% of the LyC photons are absorbed within a distance of 8, 33, and 83 pc from each star particles, respectively. The scale becomes larger (49, 140, and 191~pc) if the SFE is higher (10\%) as the cloud is dispersed due to stellar feedback.

\item[(v)] In the run with a $1\%$ SFE, most of the \Lya\ photons are generated via the recombinative process, while collisional recombination contributes to $\sim20$--$30\%$ of the total \Lya\ (Figure~\ref{fig:lya_origin}). The latter fraction becomes lower ($\sim10\%$) in the metal-rich cloud, as the \Lya-emitting gas becomes cooler. The resulting number of \Lya\ photons produced inside the cloud reasonably matches the simple estimate based on the assumption that 67\% of the LyC photons available from young stars produce \Lya\ photons, provided that $\left< \fescLyC\right>$ is low. 

\item[(vi)] Our simple experiment without initial turbulence shows that fewer \Lya\ escape from the cloud until it becomes disrupted, while the majority ($\sim 40$--$80\%$) of the \Lya\ photons escape from the cloud with turbulent structures in the early phase of the evolution (Figure~\ref{fig:noTurb}). Even clouds with a large amount of dust (i.e. $Z_{\rm gas}=0.02$) show a slightly lower  $\fescLya$ of $\approx 20\%$, which is systemically larger than $\fescLyC$ (Figure~\ref{fig:vsepa}). This suggests that the low $\fescLya$ observed in local and high redshift galaxies \citep[][]{deharveng08,cowie10,ono10,hayes10} may be largely due to the substantial absorption occurring in the ISM.

\item[(vii)]  We find that emergent \Lya\ spectra can be broad even on cloud scales. When ionizing radiation is effectively confined (i.e., $\fescLyC \approx 0$), the velocity spectrum from the $10^6\,\msun$ cloud shows symmetric double peaks separated by $\sim 400 \, \kms$ (Figure~\ref{fig:prof_lya}). The peak separation becomes smaller ($v_{\rm sep} \sim 100$--$300\,\kms$) if $\fescLyC \approx 0.1-1\%$, but does not become closer to $\Delta v \la 50\,\kms$ even when $\fescLyC$ becomes very high ($\ga 30\%$), as there exists residual neutral hydrogen in the relatively dense regions around which \Lya\ photons are produced (Figure~\ref{fig:vsepa}). Consequently, the luminosity-weighted \Lya\ profiles over the cloud lifetime are found to be more complex than the simple Gaussian profile that is often used as an input spectrum for \Lya\ photons in the galactic scales simulation (Figure~\ref{fig:prof_lya_avg}).

\item[(viii)] Finally, LyC leaking clouds ( $\fescLyC \ga 1\%$) show the separation of peaks less than $v_{\rm sep} \approx 150\,\kms$, consistent with \citet{verhamme15}, as the turbulent structure allows \Lya\ photons to escape more efficiently (Figure~\ref{fig:vsepa}). However, we find that the predicted $v_{\rm sep}$ for a given \fescLyC\ is a factor of two smaller than the observed in compact metal-poor systems \citep[e.g.][]{vanzella18,izotov18,verhamme17}, again suggesting that the interaction of \Lya\ photons with the ISM is likely to be crucial to determine the emergent spectrum.
\end{itemize}

We note that the number of simulations performed in this study is limited, hence it may be difficult to generalize our results to the GMCs under a wide variety of conditions \citep[e.g.][]{heyer09}. However, these experiments clearly demonstrate that the escape fractions of LyC photons are driven by radiation and SN feedback to steadily establish the low-density channels, although this process is highly dependent on the small-scale physics, such as star formation and input SEDs. Our work also emphasizes the complexity in predicting the \Lya\ line profiles, necessitating a more comprehensive understanding of the dynamics in the star-forming clouds and the ISM. Future simulations that can resolve these processes in galactic scales will be the natural step forward to make firm predictions on the escape of LyC and \Lya\ photons in the high-redshift Universe.

\appendix

\section{Analytic calculations of the escape of ionizing radiation}

\begin{figure}
   \centering
   \includegraphics[width=8.6cm]{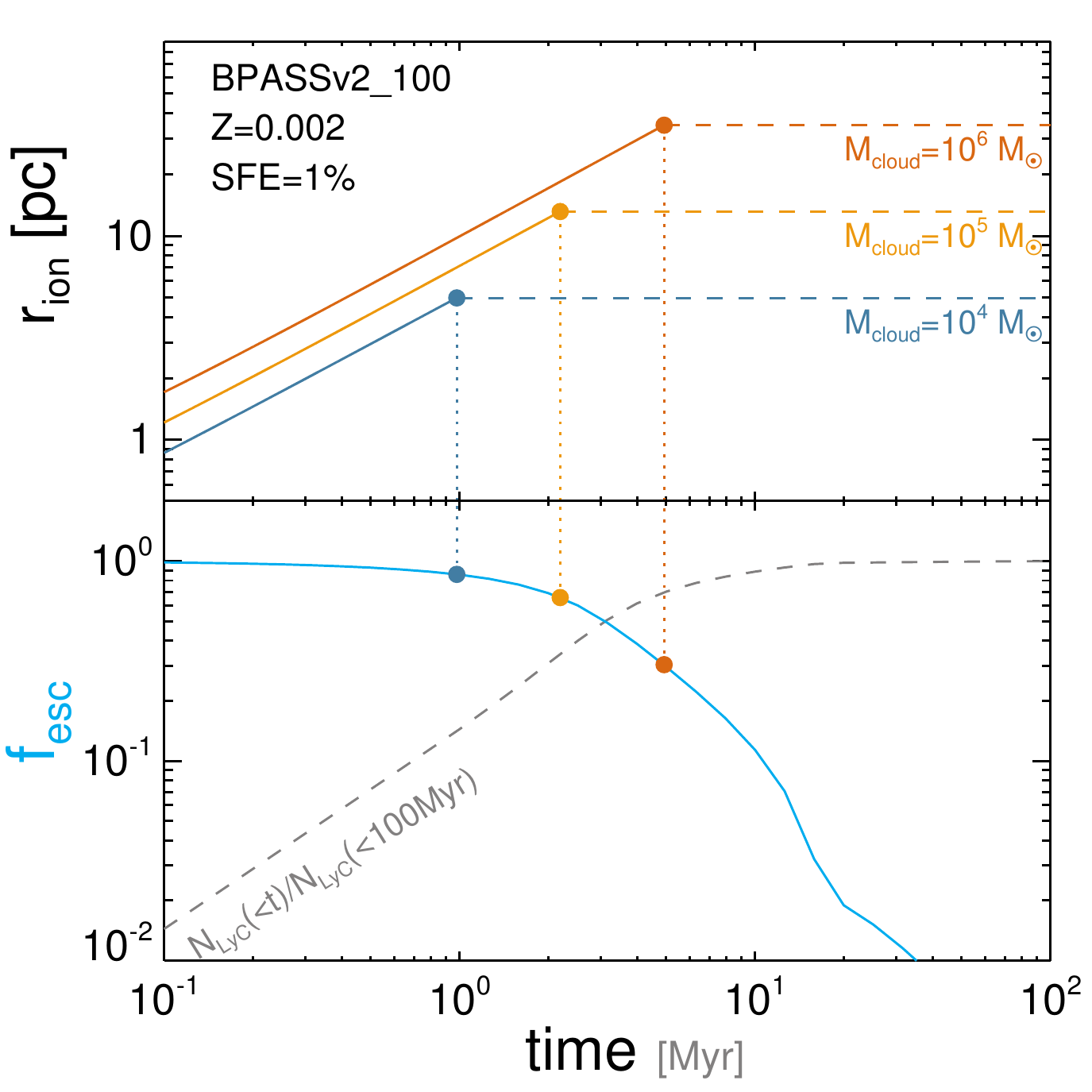} 
   \caption{Simple calculations of the propagation of the ionization front in a uniform cloud with a metal-poor binary stellar evolution model. We assume that the total SFE is 1\%.  The top panel shows the size of the ionized bubbles in clouds of different masses, as indicated in the legend. The dashed lines mark the typical radius of the GMC observed in the local Universe. The bottom panel shows the cumulative fraction of LyC photons generated until some time $t$ with respect to the total number of LyC photons emitted until 100 Myr (dashed grey line). The escape fraction is computed by subtracting this fraction from unity (cyan lines). 
   }
   \label{fig:analytic_ex}
\end{figure}

In this section, we present a simple model for the propagation of an ionization front in a spherically symmetric cloud and compute the timescale for which the ionization front reaches the edge of the GMC. One may then calculate the photon number-weighted average escape fractions, assuming that all of the ionizing radiation leaves the cloud once the ionization front reaches the edge. Note that this experiment is highly idealized but provides useful insights into understanding the dependence of the escape fraction on the basic properties of simulated clouds.  

\citet[][Appendix A]{geen15b} derive the propagation of the ionization front \citep[see also][]{raga12} 
in a cloud with a power-law density profile,
\begin{equation}
n_{\rm ext} (r)=n_0 \, \left(r/r_0\right)^{-w},
\end{equation}  
as
\begin{equation}
\frac{1}{c_{\rm s,i} } \frac{d r_{\rm i}(t)}{dt} = F(r,t) - \left(\frac{c_{\rm ext}}{c_{\rm s,i}}\right)^2 \frac{1}{F(r,t)} + \frac{v_{\rm ext}(r,t)}{ c_{\rm s,i}},
\label{eq:hii_another}
\end{equation}
where $r$ is the radius, $t$ is the time, $c_{s,i}$ is the sound speed of the ionized medium, $v_{\rm ext}$ is the infall velocity of the ambient gas, $c_{\rm ext}$ represents the velocity term due to the thermal and turbulent pressure, and 
\begin{align} 
F(r,t) & \equiv \sqrt{\frac{n_i(t)}{n_{\rm ext}(r)}} = \left( \frac{r_{\rm Strom}}{r_{\rm i}(t)} \right)^{3/4} \, \left( \frac{n_i(t=0)}{n_{\rm ext}(r,t)}\right)^{1/2} \nonumber \\
& = \left( \frac{r_{\rm Strom}}{r_i (t)}\right)^{\frac{3}{4} - \frac{w}{2}}.
\end{align}
Here we assume that $r_i (t=0) = r_{\rm Strom}$ and that the density profile is quasi-static. This neglects the timescale necessary to develop the Stromgren sphere, i.e., a roughly recombination timescale ($\tau < {\rm 0.01\, Myr}$), but this is usually negligible compared to the lifetime of a GMC ($\sim \rm 1 - 10\, Myr$). The density of the ionized region is related to the source as 
\begin{gather}
n_i = \left( \frac{3 \dot{N}_{\rm ph} }{4 \pi r_{\rm Strom}^3 \alpha_B} \right)^{1/2}.
\end{gather}
The ionization front stalls at the radius where $\dot{r}_{\rm i} = 0$, and for a pressure-dominated region without infalling motion, one finds
\begin{equation}
r_{\rm stall} = r_{\rm Strom} \left( \frac{c_{\rm s,i}}{c_{\rm ext}} \right)^{\frac{4}{3-2w}}.
\end{equation}
This indicates that in a dynamically cold medium ($c_{\rm ext} < c_{\rm s,i}$), the over-pressure due to photo-ionization heating is significant only if the cloud is not too compact ($w < 3/2$). If this condition is met, the ionized bubble would expand with time and can reach the edge of the cloud even though it may take a long time. 

For the uniform profile ($w=0$), Equation~\ref{eq:hii_another} can be written as
\begin{equation}
\frac{d y}{dt} =  \frac{c_{\rm s,i} }{r_{\rm Strom}}\,\left(  y^{-3/4}- \left(\frac{c_{\rm s,ext}}{c_{\rm s,i}}\right)^2 y^{3/4} \right)
\label{eq:hii2}
\end{equation}
where $y\equiv r_{\rm i} (t) / r_{\rm Strom}$. The time required for the ionization front to propagate to some radius $r$ in a pressure-dominated region (i.e., $v_{\rm ext}\sim 0$) can then be computed as
\begin{equation}
\tau_{\rm ion} (r)= \int_1^{r/r_{\rm Strom}} \, \frac{r_{\rm Strom} / c_{\rm s,i}}{  y^{-3/4}- \left(c_{\rm ext}/c_{\rm s,i}\right)^2 y^{3/4}  } dy .
\end{equation} 
Note that $\fescLyC$ is zero if  $r_{\rm stall} < r_{\rm cloud}$.

The radius of a cloud is taken from the empirical relation between the mass and radius \citep{roman-duval10}, as
\begin{gather}
M_{\rm cloud} \approx 228\, \msun \, (R_{\rm cloud} / {\rm pc} )^{2.36}.
\end{gather} 
Figure~\ref{fig:analytic_ex} shows that the ionization front increases steadily, as the LyC photon production rates are nearly constant until $5\,{\rm Myr}$. The ionization front can in principle shrink back if the external pressure is strong enough to counterbalance the pressure due to photo-ionization heating. This occurs mostly after $\sim$ 10 Myr when the emissivity drops rapidly, which is well after the ionization propagates to the outer region in this simple setup. Although  massive clouds are more compact and that recombination is more efficient, the resulting escape fractions are higher because they are smaller in size.

\section{Resolution test and temperature distributions of the turbulent clouds}
\begin{figure}
   \centering
   \includegraphics[width=8cm]{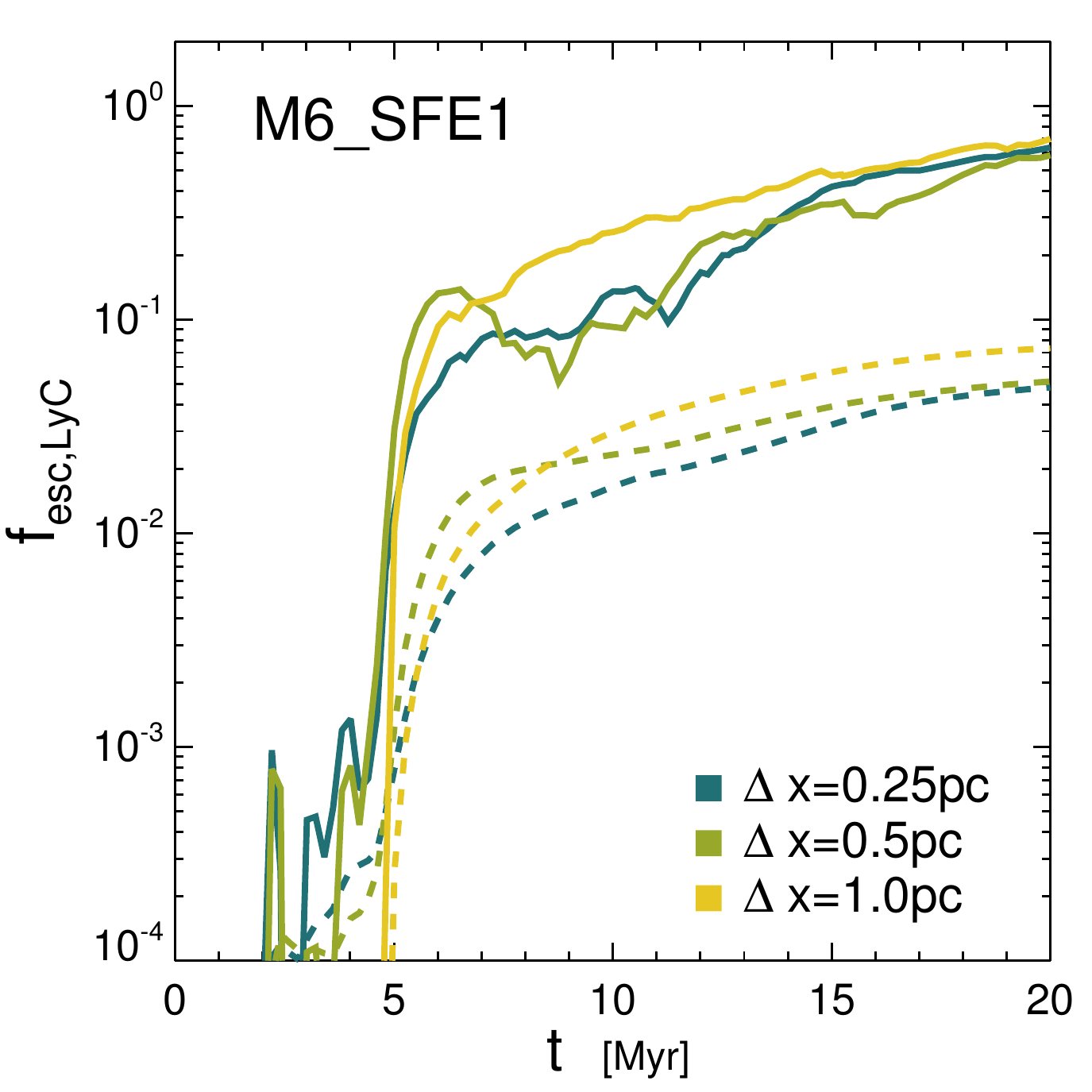}
   \caption{Resolution test for the escape of LyC photons in the fiducial case with $M_{\rm cloud}=10^6\,\msun$ and $M_{\rm star}=10^4\,\msun$. The dark green lines indicate the escape fractions in the runs with our fiducial resolution ($\Delta x_{\rm min}=0.25\,{\rm pc}$), while the light green lines correspond to the results with one or two fewer levels of refinement, as indicated in the legend. The instantaneous escape fractions are shown as solid lines, while the time averaged values are displayed as dashed lines.  
   }
   \label{fig:res_test}
\end{figure}

Figure~\ref{fig:res_test} shows the escape fractions of LyC photons in our fiducial run with different resolutions. Although the Stromgren sphere around individual star particle is initially well resolved, the propagation of the LyC photons are affected by the maximum AMR resolution, as star particles encounter dense clumps and become enshrouded by neutral hydrogen before stellar feedback entirely destroys the cloud. As a result, 4.8, 5.1, and 7.3\% of the total LyC photons escape from the cloud with 0.25, 0.5, and 1.0 pc resolution. However, the general trends in the evolution of the clouds are very similar, indicating that our conclusions are little affected by the resolution. 

\begin{figure*}
   \centering
   \includegraphics[width=16cm]{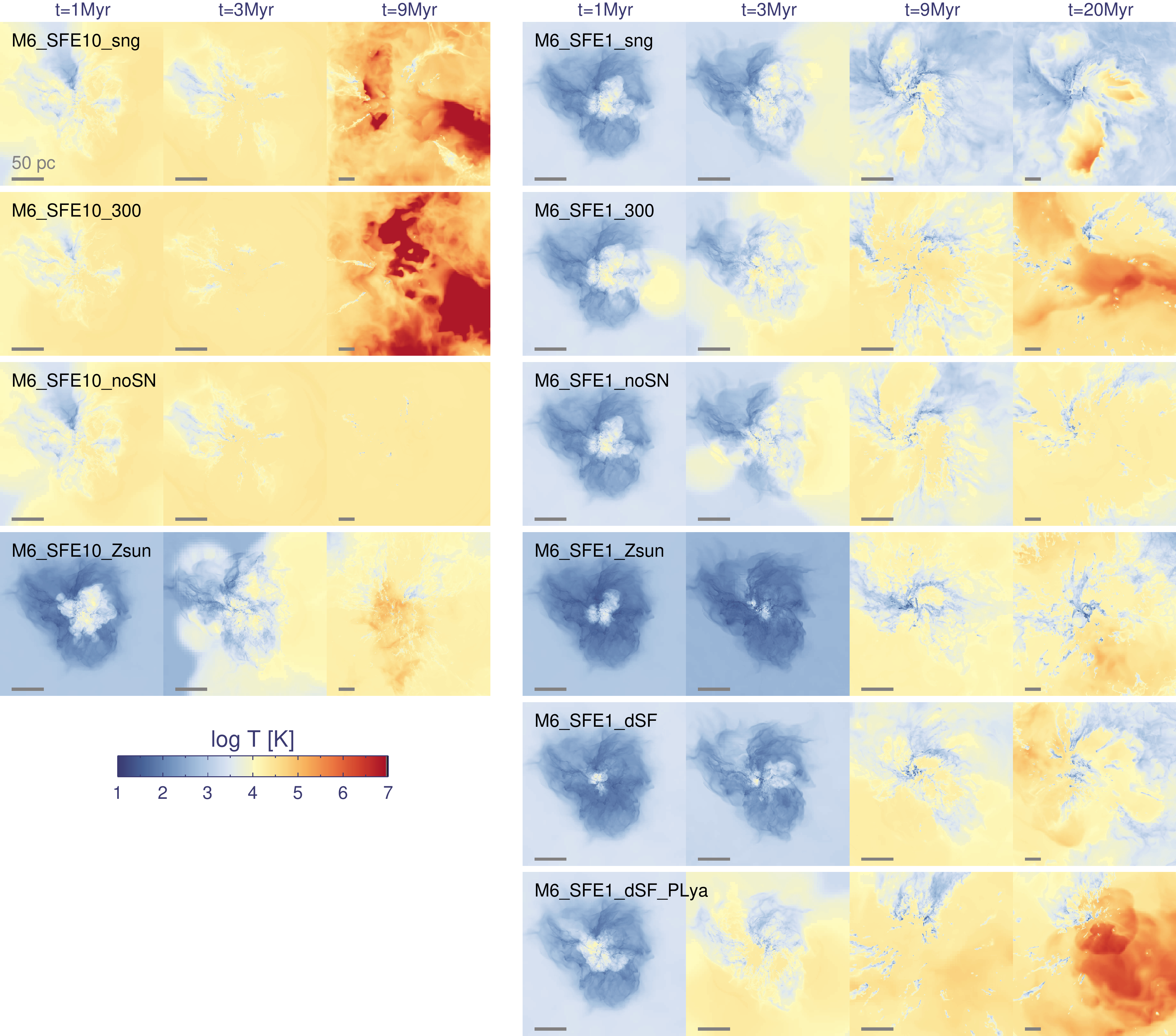} 
   \caption{Temperature evolution of the turbulent clouds as a function of time. The left panels show the clouds with a 10\% SFE, while the evolution of clouds with a 1\% SFE is shown in the right. The grey scale bar displays 50 pc. Note that the last column of each run shows a region that is twice the size of that in other panels.
   }
   \label{fig:img_all}
\end{figure*}

In Figure~\ref{fig:img_all}, we show the projected temperature distributions of turbulent clouds with different SED, metallicity, and input physics.

\section*{Acknowledgements}
 We thank Julien Devriendt, Kwang-Il Seon, and an anonymous referee for helpful comments. This work was supported by the Supercomputing Center/Korea Institute of Science and Technology Information with supercomputing resources including technical support (KSC-2017-C3-0073) and in part by the ERC Advanced Grant 320596 ``The Emergence of Structure during the Epoch of Reionization". This study was performed under the umbrella of the joint collaboration between Yonsei University Observatory and the Korean Astronomy and Space Science Institute. TK was supported in part by the Yonsei University Future-leading Research Initiative  (RMS2-2018-22-0183) and in part by the National Research Foundation of Korea (No. 2017R1A5A1070354 and No. 2018036146). HK thanks Brasenose College and the support of the Nicholas Kurti Junior Fellowship as well as the Beecroft Fellowship. JR and JB acknowledge support from the ORAGE project from the Agence Nationale de la Recherche under grand ANR-14-CE33-0016-03. TG is grateful to the LABEX Lyon Institute of Origins (ANR-10-LABX-0066) of the Universit\'e de Lyon for its financial support within the programme `Investissements d'Avenir' (ANR-11-IDEX-0007) of the French government operated by the National Research Agency (ANR). TG acknowledges support from the European Research Council under grant agreement ERC-stg-757258 (TRIPLE). This work made use of the following DiRAC facilities (www.dirac.ac.uk): the Data Analytic system at the University of Cambridge (funded by BIS National E-infrastructure capital grant ST/K001590/1, STFC capital grants ST/H008861/1 and ST/H00887X/1, and STFC DiRAC Operations grant ST/K00333X/1) and the Complexity system at the University of Leicester (funded by BIS National E-infrastructure capital grant ST/K000373/1 and STFC DiRAC Operations grant ST/K0003259/1). DiRAC is part of the National E-Infrastructure. 
\small
\bibliographystyle{mnras}

\begin{thebibliography}{}
\makeatletter
\relax
\def\mn@urlcharsother{\let\do\@makeother \do\$\do\&\do\#\do\^\do\_\do\%\do\~}
\def\mn@doi{\begingroup\mn@urlcharsother \@ifnextchar [ {\mn@doi@}
  {\mn@doi@[]}}
\def\mn@doi@[#1]#2{\def\@tempa{#1}\ifx\@tempa\@empty \href
  {http://dx.doi.org/#2} {doi:#2}\else \href {http://dx.doi.org/#2} {#1}\fi
  \endgroup}
\def\mn@eprint#1#2{\mn@eprint@#1:#2::\@nil}
\def\mn@eprint@arXiv#1{\href {http://arxiv.org/abs/#1} {{\tt arXiv:#1}}}
\def\mn@eprint@dblp#1{\href {http://dblp.uni-trier.de/rec/bibtex/#1.xml}
  {dblp:#1}}
\def\mn@eprint@#1:#2:#3:#4\@nil{\def\@tempa {#1}\def\@tempb {#2}\def\@tempc
  {#3}\ifx \@tempc \@empty \let \@tempc \@tempb \let \@tempb \@tempa \fi \ifx
  \@tempb \@empty \def\@tempb {arXiv}\fi \@ifundefined
  {mn@eprint@\@tempb}{\@tempb:\@tempc}{\expandafter \expandafter \csname
  mn@eprint@\@tempb\endcsname \expandafter{\@tempc}}}

\bibitem[\protect\citeauthoryear{{Ahn}, {Lee}  \& {Lee}}{{Ahn}
  et~al.}{2001}]{ahn01}
{Ahn} S.-H.,  {Lee} H.-W.,   {Lee} H.~M.,  2001, \mn@doi [\apj]
  {10.1086/321374}, \href {http://adsabs.harvard.edu/abs/2001ApJ...554..604A}
  {554, 604}

\bibitem[\protect\citeauthoryear{{Ahn}, {Lee}  \& {Lee}}{{Ahn}
  et~al.}{2003}]{ahn03}
{Ahn} S.-H.,  {Lee} H.-W.,   {Lee} H.~M.,  2003, \mn@doi [\mnras]
  {10.1046/j.1365-8711.2003.06353.x}, \href
  {http://adsabs.harvard.edu/abs/2003MNRAS.340..863A} {340, 863}

\bibitem[\protect\citeauthoryear{{Baczynski}, {Glover}  \&
  {Klessen}}{{Baczynski} et~al.}{2015}]{baczynski15}
{Baczynski} C.,  {Glover} S.~C.~O.,   {Klessen} R.~S.,  2015, \mn@doi [\mnras]
  {10.1093/mnras/stv1906}, \href
  {http://adsabs.harvard.edu/abs/2015MNRAS.454..380B} {454, 380}

\bibitem[\protect\citeauthoryear{{Barnes}, {Haehnelt}, {Tescari}  \&
  {Viel}}{{Barnes} et~al.}{2011}]{barnes11}
{Barnes} L.~A.,  {Haehnelt} M.~G.,  {Tescari} E.,   {Viel} M.,  2011, \mn@doi
  [\mnras] {10.1111/j.1365-2966.2011.18789.x}, \href
  {http://adsabs.harvard.edu/abs/2011MNRAS.416.1723B} {416, 1723}

\bibitem[\protect\citeauthoryear{{Bate}, {Bonnell}  \& {Price}}{{Bate}
  et~al.}{1995}]{bate95}
{Bate} M.~R.,  {Bonnell} I.~A.,   {Price} N.~M.,  1995, \mn@doi [\mnras]
  {10.1093/mnras/277.2.362}, \href
  {http://adsabs.harvard.edu/abs/1995MNRAS.277..362B} {277, 362}

\bibitem[\protect\citeauthoryear{{Behrens} \& {Braun}}{{Behrens} \&
  {Braun}}{2014}]{behrens14a}
{Behrens} C.,  {Braun} H.,  2014, \mn@doi [\aap] {10.1051/0004-6361/201424755},
  \href {http://adsabs.harvard.edu/abs/2014A%26A...572A..74B} {572, A74}

\bibitem[\protect\citeauthoryear{{Behrens}, {Dijkstra}  \&
  {Niemeyer}}{{Behrens} et~al.}{2014}]{behrens14}
{Behrens} C.,  {Dijkstra} M.,   {Niemeyer} J.~C.,  2014, \mn@doi [\aap]
  {10.1051/0004-6361/201322949}, \href
  {http://adsabs.harvard.edu/abs/2014A%26A...563A..77B} {563, A77}

\bibitem[\protect\citeauthoryear{{Bian}, {Fan}, {McGreer}, {Cai}  \&
  {Jiang}}{{Bian} et~al.}{2017}]{bian17}
{Bian} F.,  {Fan} X.,  {McGreer} I.,  {Cai} Z.,   {Jiang} L.,  2017, \mn@doi
  [\apjl] {10.3847/2041-8213/aa5ff7}, \href
  {http://adsabs.harvard.edu/abs/2017ApJ...837L..12B} {837, L12}

\bibitem[\protect\citeauthoryear{{Black} \& {van Dishoeck}}{{Black} \& {van
  Dishoeck}}{1987}]{black87}
{Black} J.~H.,  {van Dishoeck} E.~F.,  1987, \mn@doi [\apj] {10.1086/165740},
  \href {http://adsabs.harvard.edu/abs/1987ApJ...322..412B} {322, 412}

\bibitem[\protect\citeauthoryear{{Blanc} et~al.,}{{Blanc}
  et~al.}{2011}]{blanc11}
{Blanc} G.~A.,  et~al., 2011, \mn@doi [\apj] {10.1088/0004-637X/736/1/31},
  \href {http://adsabs.harvard.edu/abs/2011ApJ...736...31B} {736, 31}

\bibitem[\protect\citeauthoryear{{Bleuler} \& {Teyssier}}{{Bleuler} \&
  {Teyssier}}{2014}]{bleuler14}
{Bleuler} A.,  {Teyssier} R.,  2014, \mn@doi [\mnras] {10.1093/mnras/stu2005},
  \href {http://adsabs.harvard.edu/abs/2014MNRAS.445.4015B} {445, 4015}

\bibitem[\protect\citeauthoryear{{Blondin}, {Wright}, {Borkowski}  \&
  {Reynolds}}{{Blondin} et~al.}{1998}]{blondin98}
{Blondin} J.~M.,  {Wright} E.~B.,  {Borkowski} K.~J.,   {Reynolds} S.~P.,
  1998, \mn@doi [\apj] {10.1086/305708}, \href
  {http://adsabs.harvard.edu/abs/1998ApJ...500..342B} {500, 342}

\bibitem[\protect\citeauthoryear{{Bouwens} et~al.,}{{Bouwens}
  et~al.}{2016}]{bouwens16}
{Bouwens} R.~J.,  et~al., 2016, \mn@doi [\apj] {10.3847/1538-4357/833/1/72},
  \href {http://adsabs.harvard.edu/abs/2016ApJ...833...72B} {833, 72}

\bibitem[\protect\citeauthoryear{{Bruzual} \& {Charlot}}{{Bruzual} \&
  {Charlot}}{2003}]{bruzual03}
{Bruzual} G.,  {Charlot} S.,  2003, \mn@doi [\mnras]
  {10.1046/j.1365-8711.2003.06897.x}, \href
  {http://adsabs.harvard.edu/abs/2003MNRAS.344.1000B} {344, 1000}

\bibitem[\protect\citeauthoryear{{Callaway}, {Unnikrishnan}  \&
  {Oza}}{{Callaway} et~al.}{1987}]{callaway87}
{Callaway} J.,  {Unnikrishnan} K.,   {Oza} D.~H.,  1987, \mn@doi [\pra]
  {10.1103/PhysRevA.36.2576}, \href
  {http://adsabs.harvard.edu/abs/1987PhRvA..36.2576C} {36, 2576}

\bibitem[\protect\citeauthoryear{{Calzetti}, {Armus}, {Bohlin}, {Kinney},
  {Koornneef}  \& {Storchi-Bergmann}}{{Calzetti} et~al.}{2000}]{calzetti00}
{Calzetti} D.,  {Armus} L.,  {Bohlin} R.~C.,  {Kinney} A.~L.,  {Koornneef} J.,
   {Storchi-Bergmann} T.,  2000, \mn@doi [\apj] {10.1086/308692}, \href
  {http://adsabs.harvard.edu/abs/2000ApJ...533..682C} {533, 682}

\bibitem[\protect\citeauthoryear{{Cantalupo}, {Porciani}  \&
  {Lilly}}{{Cantalupo} et~al.}{2008}]{cantalupo08}
{Cantalupo} S.,  {Porciani} C.,   {Lilly} S.~J.,  2008, \mn@doi [\apj]
  {10.1086/523298}, \href {http://adsabs.harvard.edu/abs/2008ApJ...672...48C}
  {672, 48}

\bibitem[\protect\citeauthoryear{{Chardin}, {Kulkarni}  \&
  {Haehnelt}}{{Chardin} et~al.}{2018}]{chardin18}
{Chardin} J.,  {Kulkarni} G.,   {Haehnelt} M.~G.,  2018, \mn@doi [\mnras]
  {10.1093/mnras/sty992}, \href
  {http://adsabs.harvard.edu/abs/2018MNRAS.478.1065C} {478, 1065}

\bibitem[\protect\citeauthoryear{{Choudhury}, {Puchwein}, {Haehnelt}  \&
  {Bolton}}{{Choudhury} et~al.}{2015}]{choudhury15}
{Choudhury} T.~R.,  {Puchwein} E.,  {Haehnelt} M.~G.,   {Bolton} J.~S.,  2015,
  \mn@doi [\mnras] {10.1093/mnras/stv1250}, \href
  {http://adsabs.harvard.edu/abs/2015MNRAS.452..261C} {452, 261}

\bibitem[\protect\citeauthoryear{{Cowie}, {Barger}  \& {Hu}}{{Cowie}
  et~al.}{2010}]{cowie10}
{Cowie} L.~L.,  {Barger} A.~J.,   {Hu} E.~M.,  2010, \mn@doi [\apj]
  {10.1088/0004-637X/711/2/928}, \href
  {http://adsabs.harvard.edu/abs/2010ApJ...711..928C} {711, 928}

\bibitem[\protect\citeauthoryear{{Crowther}, {Schnurr}, {Hirschi}, {Yusof},
  {Parker}, {Goodwin}  \& {Kassim}}{{Crowther} et~al.}{2010}]{crowther10}
{Crowther} P.~A.,  {Schnurr} O.,  {Hirschi} R.,  {Yusof} N.,  {Parker} R.~J.,
  {Goodwin} S.~P.,   {Kassim} H.~A.,  2010, \mn@doi [\mnras]
  {10.1111/j.1365-2966.2010.17167.x}, \href
  {http://adsabs.harvard.edu/abs/2010MNRAS.408..731C} {408, 731}

\bibitem[\protect\citeauthoryear{{Dale}, {Ercolano}  \& {Bonnell}}{{Dale}
  et~al.}{2012}]{dale12}
{Dale} J.~E.,  {Ercolano} B.,   {Bonnell} I.~A.,  2012, \mn@doi [\mnras]
  {10.1111/j.1365-2966.2012.21205.x}, \href
  {http://adsabs.harvard.edu/abs/2012MNRAS.424..377D} {424, 377}

\bibitem[\protect\citeauthoryear{{Dale}, {Ercolano}  \& {Bonnell}}{{Dale}
  et~al.}{2013}]{dale13}
{Dale} J.~E.,  {Ercolano} B.,   {Bonnell} I.~A.,  2013, \mn@doi [\mnras]
  {10.1093/mnras/sts592}, \href
  {http://adsabs.harvard.edu/abs/2013MNRAS.430..234D} {430, 234}

\bibitem[\protect\citeauthoryear{{Dale}, {Ngoumou}, {Ercolano}  \&
  {Bonnell}}{{Dale} et~al.}{2014}]{dale14}
{Dale} J.~E.,  {Ngoumou} J.,  {Ercolano} B.,   {Bonnell} I.~A.,  2014, \mn@doi
  [\mnras] {10.1093/mnras/stu816}, \href
  {http://adsabs.harvard.edu/abs/2014MNRAS.442..694D} {442, 694}

\bibitem[\protect\citeauthoryear{{Deharveng} et~al.,}{{Deharveng}
  et~al.}{2008}]{deharveng08}
{Deharveng} J.-M.,  et~al., 2008, \mn@doi [\apj] {10.1086/587953}, \href
  {http://adsabs.harvard.edu/abs/2008ApJ...680.1072D} {680, 1072}

\bibitem[\protect\citeauthoryear{{Dijkstra}}{{Dijkstra}}{2014}]{dijkstra14}
{Dijkstra} M.,  2014, \mn@doi [\pasa] {10.1017/pasa.2014.33}, \href
  {http://adsabs.harvard.edu/abs/2014PASA...31...40D} {31, e040}

\bibitem[\protect\citeauthoryear{{Dijkstra}}{{Dijkstra}}{2017}]{dijkstra17}
{Dijkstra} M.,  2017, arXiv e-prints, \href
  {http://adsabs.harvard.edu/abs/2017arXiv170403416D} {}

\bibitem[\protect\citeauthoryear{{Dijkstra} \& {Loeb}}{{Dijkstra} \&
  {Loeb}}{2008}]{dijkstra08}
{Dijkstra} M.,  {Loeb} A.,  2008, \mn@doi [\mnras]
  {10.1111/j.1365-2966.2008.13920.x}, \href
  {http://adsabs.harvard.edu/abs/2008MNRAS.391..457D} {391, 457}

\bibitem[\protect\citeauthoryear{{Dijkstra}, {Haiman}  \& {Spaans}}{{Dijkstra}
  et~al.}{2006}]{dijkstra06}
{Dijkstra} M.,  {Haiman} Z.,   {Spaans} M.,  2006, \mn@doi [\apj]
  {10.1086/506243}, \href {http://adsabs.harvard.edu/abs/2006ApJ...649...14D}
  {649, 14}

\bibitem[\protect\citeauthoryear{{Dijkstra}, {Gronke}  \&
  {Venkatesan}}{{Dijkstra} et~al.}{2016}]{dijkstra16}
{Dijkstra} M.,  {Gronke} M.,   {Venkatesan} A.,  2016, \mn@doi [\apj]
  {10.3847/0004-637X/828/2/71}, \href
  {http://adsabs.harvard.edu/abs/2016ApJ...828...71D} {828, 71}

\bibitem[\protect\citeauthoryear{{Dove} \& {Shull}}{{Dove} \&
  {Shull}}{1994}]{dove94}
{Dove} J.~B.,  {Shull} J.~M.,  1994, \mn@doi [\apj] {10.1086/174397}, \href
  {http://adsabs.harvard.edu/abs/1994ApJ...430..222D} {430, 222}

\bibitem[\protect\citeauthoryear{{Dove}, {Shull}  \& {Ferrara}}{{Dove}
  et~al.}{2000}]{dove00}
{Dove} J.~B.,  {Shull} J.~M.,   {Ferrara} A.,  2000, \mn@doi [\apj]
  {10.1086/308481}, \href {http://adsabs.harvard.edu/abs/2000ApJ...531..846D}
  {531, 846}

\bibitem[\protect\citeauthoryear{{Draine} \& {Li}}{{Draine} \&
  {Li}}{2007}]{draine07}
{Draine} B.~T.,  {Li} A.,  2007, \mn@doi [\apj] {10.1086/511055}, \href
  {http://adsabs.harvard.edu/abs/2007ApJ...657..810D} {657, 810}

\bibitem[\protect\citeauthoryear{{Draine} et~al.,}{{Draine}
  et~al.}{2007}]{draine07b}
{Draine} B.~T.,  et~al., 2007, \mn@doi [\apj] {10.1086/518306}, \href
  {http://adsabs.harvard.edu/abs/2007ApJ...663..866D} {663, 866}

\bibitem[\protect\citeauthoryear{{Dwek}}{{Dwek}}{1998}]{dwek98}
{Dwek} E.,  1998, \mn@doi [\apj] {10.1086/305829}, \href
  {http://adsabs.harvard.edu/abs/1998ApJ...501..643D} {501, 643}

\bibitem[\protect\citeauthoryear{{Efstathiou}}{{Efstathiou}}{1992}]{efstathiou92}
{Efstathiou} G.,  1992, \mn@doi [\mnras] {10.1093/mnras/256.1.43P}, \href
  {http://adsabs.harvard.edu/abs/1992MNRAS.256P..43E} {256, 43P}

\bibitem[\protect\citeauthoryear{{Elmegreen}}{{Elmegreen}}{1994}]{elmegreen94}
{Elmegreen} B.~G.,  1994, \mn@doi [\apj] {10.1086/174147}, \href
  {http://adsabs.harvard.edu/abs/1994ApJ...427..384E} {427, 384}

\bibitem[\protect\citeauthoryear{{Erb} et~al.,}{{Erb} et~al.}{2014}]{erb14}
{Erb} D.~K.,  et~al., 2014, \mn@doi [\apj] {10.1088/0004-637X/795/1/33}, \href
  {http://adsabs.harvard.edu/abs/2014ApJ...795...33E} {795, 33}

\bibitem[\protect\citeauthoryear{{Evans}, {Heiderman}  \&
  {Vutisalchavakul}}{{Evans} et~al.}{2014}]{evans14}
{Evans} II N.~J.,  {Heiderman} A.,   {Vutisalchavakul} N.,  2014, \mn@doi
  [\apj] {10.1088/0004-637X/782/2/114}, \href
  {http://adsabs.harvard.edu/abs/2014ApJ...782..114E} {782, 114}

\bibitem[\protect\citeauthoryear{{Fan} et~al.,}{{Fan} et~al.}{2001}]{fan01}
{Fan} X.,  et~al., 2001, \mn@doi [\aj] {10.1086/324111}, \href
  {http://adsabs.harvard.edu/abs/2001AJ....122.2833F} {122, 2833}

\bibitem[\protect\citeauthoryear{{Fan} et~al.,}{{Fan} et~al.}{2006}]{fan06}
{Fan} X.,  et~al., 2006, \mn@doi [\aj] {10.1086/504836}, \href
  {http://adsabs.harvard.edu/abs/2006AJ....132..117F} {132, 117}

\bibitem[\protect\citeauthoryear{{Faucher-Gigu{\`e}re}, {Kere{\v s}},
  {Dijkstra}, {Hernquist}  \& {Zaldarriaga}}{{Faucher-Gigu{\`e}re}
  et~al.}{2010}]{faucher-giguere10}
{Faucher-Gigu{\`e}re} C.-A.,  {Kere{\v s}} D.,  {Dijkstra} M.,  {Hernquist} L.,
    {Zaldarriaga} M.,  2010, \mn@doi [\apj] {10.1088/0004-637X/725/1/633},
  \href {http://adsabs.harvard.edu/abs/2010ApJ...725..633F} {725, 633}

\bibitem[\protect\citeauthoryear{{Federrath} \& {Klessen}}{{Federrath} \&
  {Klessen}}{2012}]{federrath12}
{Federrath} C.,  {Klessen} R.~S.,  2012, \mn@doi [\apj]
  {10.1088/0004-637X/761/2/156}, \href
  {http://adsabs.harvard.edu/abs/2012ApJ...761..156F} {761, 156}

\bibitem[\protect\citeauthoryear{{Finlator}, {Keating}, {Oppenheimer},
  {Dav{\'e}}  \& {Zackrisson}}{{Finlator} et~al.}{2018}]{finlator18}
{Finlator} K.,  {Keating} L.,  {Oppenheimer} B.~D.,  {Dav{\'e}} R.,
  {Zackrisson} E.,  2018, \mn@doi [\mnras] {10.1093/mnras/sty1949}, \href
  {http://adsabs.harvard.edu/abs/2018MNRAS.480.2628F} {480, 2628}

\bibitem[\protect\citeauthoryear{{Gavagnin}, {Bleuler}, {Rosdahl}  \&
  {Teyssier}}{{Gavagnin} et~al.}{2017}]{gavagnin17}
{Gavagnin} E.,  {Bleuler} A.,  {Rosdahl} J.,   {Teyssier} R.,  2017, \mn@doi
  [\mnras] {10.1093/mnras/stx2222}, \href
  {http://adsabs.harvard.edu/abs/2017MNRAS.472.4155G} {472, 4155}

\bibitem[\protect\citeauthoryear{{Geen}, {Slyz}  \& {Devriendt}}{{Geen}
  et~al.}{2013}]{geen13}
{Geen} S.,  {Slyz} A.,   {Devriendt} J.,  2013, \mn@doi [\mnras]
  {10.1093/mnras/sts364}, \href
  {http://adsabs.harvard.edu/abs/2013MNRAS.429..633G} {429, 633}

\bibitem[\protect\citeauthoryear{{Geen}, {Rosdahl}, {Blaizot}, {Devriendt}  \&
  {Slyz}}{{Geen} et~al.}{2015a}]{geen15}
{Geen} S.,  {Rosdahl} J.,  {Blaizot} J.,  {Devriendt} J.,   {Slyz} A.,  2015a,
  \mn@doi [\mnras] {10.1093/mnras/stv251}, \href
  {http://adsabs.harvard.edu/abs/2015MNRAS.448.3248G} {448, 3248}

\bibitem[\protect\citeauthoryear{{Geen}, {Hennebelle}, {Tremblin}  \&
  {Rosdahl}}{{Geen} et~al.}{2015b}]{geen15b}
{Geen} S.,  {Hennebelle} P.,  {Tremblin} P.,   {Rosdahl} J.,  2015b, \mn@doi
  [\mnras] {10.1093/mnras/stv2272}, \href
  {http://adsabs.harvard.edu/abs/2015MNRAS.454.4484G} {454, 4484}

\bibitem[\protect\citeauthoryear{{Geen}, {Hennebelle}, {Tremblin}  \&
  {Rosdahl}}{{Geen} et~al.}{2016}]{geen16}
{Geen} S.,  {Hennebelle} P.,  {Tremblin} P.,   {Rosdahl} J.,  2016, \mn@doi
  [\mnras] {10.1093/mnras/stw2235}, \href
  {http://adsabs.harvard.edu/abs/2016MNRAS.463.3129G} {463, 3129}

\bibitem[\protect\citeauthoryear{{Geen}, {Watson}, {Rosdahl}, {Bieri},
  {Klessen}  \& {Hennebelle}}{{Geen} et~al.}{2018}]{geen18}
{Geen} S.,  {Watson} S.~K.,  {Rosdahl} J.,  {Bieri} R.,  {Klessen} R.~S.,
  {Hennebelle} P.,  2018, preprint, \href
  {http://adsabs.harvard.edu/abs/2018arXiv180610575G} {} (\mn@eprint {arXiv}
  {1806.10575})

\bibitem[\protect\citeauthoryear{{Gnedin}}{{Gnedin}}{2000}]{gnedin00}
{Gnedin} N.~Y.,  2000, \mn@doi [\apj] {10.1086/308876}, \href
  {http://adsabs.harvard.edu/abs/2000ApJ...535..530G} {535, 530}

\bibitem[\protect\citeauthoryear{{Gnedin} \& {Kaurov}}{{Gnedin} \&
  {Kaurov}}{2014}]{gnedin14}
{Gnedin} N.~Y.,  {Kaurov} A.~A.,  2014, \mn@doi [\apj]
  {10.1088/0004-637X/793/1/30}, \href
  {http://adsabs.harvard.edu/abs/2014ApJ...793...30G} {793, 30}

\bibitem[\protect\citeauthoryear{{Gnedin}, {Kravtsov}  \& {Chen}}{{Gnedin}
  et~al.}{2008}]{gnedin08}
{Gnedin} N.~Y.,  {Kravtsov} A.~V.,   {Chen} H.-W.,  2008, \mn@doi [\apj]
  {10.1086/524007}, \href {http://adsabs.harvard.edu/abs/2008ApJ...672..765G}
  {672, 765}

\bibitem[\protect\citeauthoryear{{Gong} \& {Ostriker}}{{Gong} \&
  {Ostriker}}{2013}]{gong13}
{Gong} H.,  {Ostriker} E.~C.,  2013, \mn@doi [\apjs]
  {10.1088/0067-0049/204/1/8}, \href
  {http://adsabs.harvard.edu/abs/2013ApJS..204....8G} {204, 8}

\bibitem[\protect\citeauthoryear{{G{\'o}rski}, {Hivon}, {Banday}, {Wandelt},
  {Hansen}, {Reinecke}  \& {Bartelmann}}{{G{\'o}rski} et~al.}{2005}]{gorski05}
{G{\'o}rski} K.~M.,  {Hivon} E.,  {Banday} A.~J.,  {Wandelt} B.~D.,  {Hansen}
  F.~K.,  {Reinecke} M.,   {Bartelmann} M.,  2005, \mn@doi [\apj]
  {10.1086/427976}, \href {http://adsabs.harvard.edu/abs/2005ApJ...622..759G}
  {622, 759}

\bibitem[\protect\citeauthoryear{{Grazian} et~al.,}{{Grazian}
  et~al.}{2016}]{grazian16}
{Grazian} A.,  et~al., 2016, \mn@doi [\aap] {10.1051/0004-6361/201526396},
  \href {http://adsabs.harvard.edu/abs/2016A%26A...585A..48G} {585, A48}

\bibitem[\protect\citeauthoryear{{Grisdale}, {Agertz}, {Renaud}  \&
  {Romeo}}{{Grisdale} et~al.}{2018}]{grisdale18}
{Grisdale} K.,  {Agertz} O.,  {Renaud} F.,   {Romeo} A.~B.,  2018, \mn@doi
  [\mnras] {10.1093/mnras/sty1595}, \href
  {http://adsabs.harvard.edu/abs/2018MNRAS.479.3167G} {479, 3167}

\bibitem[\protect\citeauthoryear{{Gronke}}{{Gronke}}{2017}]{gronke17}
{Gronke} M.,  2017, \mn@doi [\aap] {10.1051/0004-6361/201731791}, \href
  {http://adsabs.harvard.edu/abs/2017A%26A...608A.139G} {608, A139}

\bibitem[\protect\citeauthoryear{{Guillet} \& {Teyssier}}{{Guillet} \&
  {Teyssier}}{2011}]{guillet11}
{Guillet} T.,  {Teyssier} R.,  2011, \mn@doi [Journal of Computational Physics]
  {10.1016/j.jcp.2011.02.044}, \href
  {http://adsabs.harvard.edu/abs/2011JCoPh.230.4756G} {230, 4756}

\bibitem[\protect\citeauthoryear{{Hayes} et~al.,}{{Hayes}
  et~al.}{2010}]{hayes10}
{Hayes} M.,  et~al., 2010, \mn@doi [\nat] {10.1038/nature08881}, \href
  {http://adsabs.harvard.edu/abs/2010Natur.464..562H} {464, 562}

\bibitem[\protect\citeauthoryear{{Hayes}, {Schaerer}, {{\"O}stlin},
  {Mas-Hesse}, {Atek}  \& {Kunth}}{{Hayes} et~al.}{2011}]{hayes11}
{Hayes} M.,  {Schaerer} D.,  {{\"O}stlin} G.,  {Mas-Hesse} J.~M.,  {Atek} H.,
  {Kunth} D.,  2011, \mn@doi [\apj] {10.1088/0004-637X/730/1/8}, \href
  {http://adsabs.harvard.edu/abs/2011ApJ...730....8H} {730, 8}

\bibitem[\protect\citeauthoryear{{Herenz} et~al.,}{{Herenz}
  et~al.}{2016}]{herenz16}
{Herenz} E.~C.,  et~al., 2016, \mn@doi [\aap] {10.1051/0004-6361/201527373},
  \href {http://adsabs.harvard.edu/abs/2016A%26A...587A..78H} {587, A78}

\bibitem[\protect\citeauthoryear{{Heyer} \& {Brunt}}{{Heyer} \&
  {Brunt}}{2004}]{heyer04}
{Heyer} M.~H.,  {Brunt} C.~M.,  2004, \mn@doi [\apjl] {10.1086/425978}, \href
  {http://adsabs.harvard.edu/abs/2004ApJ...615L..45H} {615, L45}

\bibitem[\protect\citeauthoryear{{Heyer}, {Krawczyk}, {Duval}  \&
  {Jackson}}{{Heyer} et~al.}{2009}]{heyer09}
{Heyer} M.,  {Krawczyk} C.,  {Duval} J.,   {Jackson} J.~M.,  2009, \mn@doi
  [\apj] {10.1088/0004-637X/699/2/1092}, \href
  {http://adsabs.harvard.edu/abs/2009ApJ...699.1092H} {699, 1092}

\bibitem[\protect\citeauthoryear{{Hopkins} \& {Grudic}}{{Hopkins} \&
  {Grudic}}{2018}]{hopkins18}
{Hopkins} P.~F.,  {Grudic} M.~Y.,  2018, preprint, \href
  {http://adsabs.harvard.edu/abs/2018arXiv180307573H} {} (\mn@eprint {arXiv}
  {1803.07573})

\bibitem[\protect\citeauthoryear{{Hopkins}, {Quataert}  \& {Murray}}{{Hopkins}
  et~al.}{2012a}]{hopkins12b}
{Hopkins} P.~F.,  {Quataert} E.,   {Murray} N.,  2012a, \mn@doi [\mnras]
  {10.1111/j.1365-2966.2012.20578.x}, \href
  {http://adsabs.harvard.edu/abs/2012MNRAS.421.3488H} {421, 3488}

\bibitem[\protect\citeauthoryear{{Hopkins}, {Quataert}  \& {Murray}}{{Hopkins}
  et~al.}{2012b}]{hopkins12a}
{Hopkins} P.~F.,  {Quataert} E.,   {Murray} N.,  2012b, \mn@doi [\mnras]
  {10.1111/j.1365-2966.2012.20593.x}, \href
  {http://adsabs.harvard.edu/abs/2012MNRAS.421.3522H} {421, 3522}

\bibitem[\protect\citeauthoryear{{Hopkins}, {Kere{\v s}}, {O{\~n}orbe},
  {Faucher-Gigu{\`e}re}, {Quataert}, {Murray}  \& {Bullock}}{{Hopkins}
  et~al.}{2014}]{hopkins14}
{Hopkins} P.~F.,  {Kere{\v s}} D.,  {O{\~n}orbe} J.,  {Faucher-Gigu{\`e}re}
  C.-A.,  {Quataert} E.,  {Murray} N.,   {Bullock} J.~S.,  2014, \mn@doi
  [\mnras] {10.1093/mnras/stu1738}, \href
  {http://adsabs.harvard.edu/abs/2014MNRAS.445..581H} {445, 581}

\bibitem[\protect\citeauthoryear{{Howard}, {Pudritz}, {Harris}  \&
  {Klessen}}{{Howard} et~al.}{2018}]{howard18}
{Howard} C.~S.,  {Pudritz} R.~E.,  {Harris} W.~E.,   {Klessen} R.~S.,  2018,
  \mn@doi [\mnras] {10.1093/mnras/stx3276}, \href
  {http://adsabs.harvard.edu/abs/2018MNRAS.475.3121H} {475, 3121}

\bibitem[\protect\citeauthoryear{{Hubber}, {Walch}  \& {Whitworth}}{{Hubber}
  et~al.}{2013}]{hubber13}
{Hubber} D.~A.,  {Walch} S.,   {Whitworth} A.~P.,  2013, \mn@doi [\mnras]
  {10.1093/mnras/stt128}, \href
  {http://adsabs.harvard.edu/abs/2013MNRAS.430.3261H} {430, 3261}

\bibitem[\protect\citeauthoryear{{Hui} \& {Gnedin}}{{Hui} \&
  {Gnedin}}{1997}]{hui97}
{Hui} L.,  {Gnedin} N.~Y.,  1997, \mn@doi [\mnras] {10.1093/mnras/292.1.27},
  \href {http://adsabs.harvard.edu/abs/1997MNRAS.292...27H} {292, 27}

\bibitem[\protect\citeauthoryear{{Izotov}, {Schaerer}, {Thuan}, {Worseck},
  {Guseva}, {Orlitov{\'a}}  \& {Verhamme}}{{Izotov} et~al.}{2016a}]{izotov16b}
{Izotov} Y.~I.,  {Schaerer} D.,  {Thuan} T.~X.,  {Worseck} G.,  {Guseva} N.~G.,
   {Orlitov{\'a}} I.,   {Verhamme} A.,  2016a, \mn@doi [\mnras]
  {10.1093/mnras/stw1205}, \href
  {http://adsabs.harvard.edu/abs/2016MNRAS.461.3683I} {461, 3683}

\bibitem[\protect\citeauthoryear{{Izotov}, {Orlitov{\'a}}, {Schaerer}, {Thuan},
  {Verhamme}, {Guseva}  \& {Worseck}}{{Izotov} et~al.}{2016b}]{izotov16a}
{Izotov} Y.~I.,  {Orlitov{\'a}} I.,  {Schaerer} D.,  {Thuan} T.~X.,  {Verhamme}
  A.,  {Guseva} N.~G.,   {Worseck} G.,  2016b, \mn@doi [\nat]
  {10.1038/nature16456}, \href
  {http://adsabs.harvard.edu/abs/2016Natur.529..178I} {529, 178}

\bibitem[\protect\citeauthoryear{{Izotov}, {Schaerer}, {Worseck}, {Guseva},
  {Thuan}, {Verhamme}, {Orlitov{\'a}}  \& {Fricke}}{{Izotov}
  et~al.}{2018a}]{izotov18}
{Izotov} Y.~I.,  {Schaerer} D.,  {Worseck} G.,  {Guseva} N.~G.,  {Thuan} T.~X.,
   {Verhamme} A.,  {Orlitov{\'a}} I.,   {Fricke} K.~J.,  2018a, \mn@doi
  [\mnras] {10.1093/mnras/stx3115}, \href
  {http://adsabs.harvard.edu/abs/2018MNRAS.474.4514I} {474, 4514}

\bibitem[\protect\citeauthoryear{{Izotov}, {Worseck}, {Schaerer}, {Guseva},
  {Thuan}, {Fricke}  \& {Orlitov{\'a}}}{{Izotov} et~al.}{2018b}]{izotov18b}
{Izotov} Y.~I.,  {Worseck} G.,  {Schaerer} D.,  {Guseva} N.~G.,  {Thuan} T.~X.,
   {Fricke} A. V.,   {Orlitov{\'a}} I.,  2018b, \mn@doi [\mnras]
  {10.1093/mnras/sty1378}, \href
  {http://adsabs.harvard.edu/abs/2018MNRAS.478.4851I} {478, 4851}

\bibitem[\protect\citeauthoryear{{Kannan}, {Marinacci}, {Simpson}, {Glover}  \&
  {Hernquist}}{{Kannan} et~al.}{2018}]{kannan18}
{Kannan} R.,  {Marinacci} F.,  {Simpson} C.~M.,  {Glover} S.~C.~O.,
  {Hernquist} L.,  2018, astro-ph/1812.01614, \href
  {http://adsabs.harvard.edu/abs/2018arXiv181201614K} {}

\bibitem[\protect\citeauthoryear{{Katz}, {Kimm}, {Sijacki}  \&
  {Haehnelt}}{{Katz} et~al.}{2017}]{katz17}
{Katz} H.,  {Kimm} T.,  {Sijacki} D.,   {Haehnelt} M.~G.,  2017, \mn@doi
  [\mnras] {10.1093/mnras/stx608}, \href
  {http://adsabs.harvard.edu/abs/2017MNRAS.468.4831K} {468, 4831}

\bibitem[\protect\citeauthoryear{{Katz}, {Kimm}, {Haehnelt}, {Sijacki},
  {Rosdahl}  \& {Blaizot}}{{Katz} et~al.}{2018}]{katz18a}
{Katz} H.,  {Kimm} T.,  {Haehnelt} M.,  {Sijacki} D.,  {Rosdahl} J.,
  {Blaizot} J.,  2018, \mn@doi [\mnras] {10.1093/mnras/sty1225}, \href
  {http://adsabs.harvard.edu/abs/2018MNRAS.478.4986K} {478, 4986}

\bibitem[\protect\citeauthoryear{{Katz}, {Kimm}, {Haehnelt}, {Sijacki},
  {Rosdahl}  \& {Blaizot}}{{Katz} et~al.}{2019}]{katz19}
{Katz} H.,  {Kimm} T.,  {Haehnelt} M.~G.,  {Sijacki} D.,  {Rosdahl} J.,
  {Blaizot} J.,  2019, \mn@doi [\mnras] {10.1093/mnras/sty3154}, \href
  {http://adsabs.harvard.edu/abs/2019MNRAS.483.1029K} {483, 1029}

\bibitem[\protect\citeauthoryear{{Kim}, {Krumholz}, {Wise}, {Turk}, {Goldbaum}
  \& {Abel}}{{Kim} et~al.}{2013a}]{kimjh13}
{Kim} J.-h.,  {Krumholz} M.~R.,  {Wise} J.~H.,  {Turk} M.~J.,  {Goldbaum}
  N.~J.,   {Abel} T.,  2013a, \mn@doi [\apj] {10.1088/0004-637X/775/2/109},
  \href {http://adsabs.harvard.edu/abs/2013ApJ...775..109K} {775, 109}

\bibitem[\protect\citeauthoryear{{Kim}, {Ostriker}  \& {Kim}}{{Kim}
  et~al.}{2013b}]{kim13}
{Kim} C.-G.,  {Ostriker} E.~C.,   {Kim} W.-T.,  2013b, \mn@doi [\apj]
  {10.1088/0004-637X/776/1/1}, \href
  {http://adsabs.harvard.edu/abs/2013ApJ...776....1K} {776, 1}

\bibitem[\protect\citeauthoryear{{Kim}, {Kim}, {Ostriker}  \& {Skinner}}{{Kim}
  et~al.}{2017}]{kim17}
{Kim} J.-G.,  {Kim} W.-T.,  {Ostriker} E.~C.,   {Skinner} M.~A.,  2017, \mn@doi
  [\apj] {10.3847/1538-4357/aa9b80}, \href
  {http://adsabs.harvard.edu/abs/2017ApJ...851...93K} {851, 93}

\bibitem[\protect\citeauthoryear{{Kim}, {Kim}  \& {Ostriker}}{{Kim}
  et~al.}{2018}]{kimjg18}
{Kim} J.-G.,  {Kim} W.-T.,   {Ostriker} E.~C.,  2018, \mn@doi [\apj]
  {10.3847/1538-4357/aabe27}, \href
  {http://adsabs.harvard.edu/abs/2018ApJ...859...68K} {859, 68}

\bibitem[\protect\citeauthoryear{{Kimm} \& {Cen}}{{Kimm} \&
  {Cen}}{2014}]{kimm14}
{Kimm} T.,  {Cen} R.,  2014, \mn@doi [\apj] {10.1088/0004-637X/788/2/121},
  \href {http://adsabs.harvard.edu/abs/2014ApJ...788..121K} {788, 121}

\bibitem[\protect\citeauthoryear{{Kimm}, {Cen}, {Devriendt}, {Dubois}  \&
  {Slyz}}{{Kimm} et~al.}{2015}]{kimm15}
{Kimm} T.,  {Cen} R.,  {Devriendt} J.,  {Dubois} Y.,   {Slyz} A.,  2015,
  \mn@doi [\mnras] {10.1093/mnras/stv1211}, \href
  {http://adsabs.harvard.edu/abs/2015MNRAS.451.2900K} {451, 2900}

\bibitem[\protect\citeauthoryear{{Kimm}, {Katz}, {Haehnelt}, {Rosdahl},
  {Devriendt}  \& {Slyz}}{{Kimm} et~al.}{2017}]{kimm17}
{Kimm} T.,  {Katz} H.,  {Haehnelt} M.,  {Rosdahl} J.,  {Devriendt} J.,   {Slyz}
  A.,  2017, \mn@doi [\mnras] {10.1093/mnras/stx052}, \href
  {http://adsabs.harvard.edu/abs/2017MNRAS.466.4826K} {466, 4826}

\bibitem[\protect\citeauthoryear{{Kimm}, {Haehnelt}, {Blaizot}, {Katz},
  {Michel-Dansac}, {Garel}, {Rosdahl}  \& {Teyssier}}{{Kimm}
  et~al.}{2018}]{kimm18}
{Kimm} T.,  {Haehnelt} M.,  {Blaizot} J.,  {Katz} H.,  {Michel-Dansac} L.,
  {Garel} T.,  {Rosdahl} J.,   {Teyssier} R.,  2018, \mn@doi [\mnras]
  {10.1093/mnras/sty126}, \href
  {http://adsabs.harvard.edu/abs/2018MNRAS.475.4617K} {475, 4617}

\bibitem[\protect\citeauthoryear{{Koh} \& {Wise}}{{Koh} \&
  {Wise}}{2018}]{koh18}
{Koh} D.,  {Wise} J.~H.,  2018, \mn@doi [\mnras] {10.1093/mnras/stx3018}, \href
  {http://adsabs.harvard.edu/abs/2018MNRAS.474.3817K} {474, 3817}

\bibitem[\protect\citeauthoryear{{Kornei}, {Shapley}, {Erb}, {Steidel},
  {Reddy}, {Pettini}  \& {Bogosavljevi{\'c}}}{{Kornei} et~al.}{2010}]{kornei10}
{Kornei} K.~A.,  {Shapley} A.~E.,  {Erb} D.~K.,  {Steidel} C.~C.,  {Reddy}
  N.~A.,  {Pettini} M.,   {Bogosavljevi{\'c}} M.,  2010, \mn@doi [\apj]
  {10.1088/0004-637X/711/2/693}, \href
  {http://adsabs.harvard.edu/abs/2010ApJ...711..693K} {711, 693}

\bibitem[\protect\citeauthoryear{{Kroupa}}{{Kroupa}}{2001}]{kroupa01}
{Kroupa} P.,  2001, \mn@doi [\mnras] {10.1046/j.1365-8711.2001.04022.x}, \href
  {http://adsabs.harvard.edu/abs/2001MNRAS.322..231K} {322, 231}

\bibitem[\protect\citeauthoryear{{Kroupa}, {Weidner}, {Pflamm-Altenburg},
  {Thies}, {Dabringhausen}, {Marks}  \& {Maschberger}}{{Kroupa}
  et~al.}{2013}]{kroupa13}
{Kroupa} P.,  {Weidner} C.,  {Pflamm-Altenburg} J.,  {Thies} I.,
  {Dabringhausen} J.,  {Marks} M.,   {Maschberger} T.,  2013, {The Stellar and
  Sub-Stellar Initial Mass Function of Simple and Composite Populations}.
p.~115, \mn@doi{10.1007/978-94-007-5612-0_4}

\bibitem[\protect\citeauthoryear{{Krumholz}}{{Krumholz}}{2015}]{krumholz15}
{Krumholz} M.~R.,  2015, preprint, \href
  {http://adsabs.harvard.edu/abs/2015arXiv151103457K} {} (\mn@eprint {arXiv}
  {1511.03457})

\bibitem[\protect\citeauthoryear{{Krumholz}}{{Krumholz}}{2018}]{krumholz18}
{Krumholz} M.~R.,  2018, \mn@doi [\mnras] {10.1093/mnras/sty2105}, \href
  {http://adsabs.harvard.edu/abs/2018MNRAS.480.3468K} {480, 3468}

\bibitem[\protect\citeauthoryear{{Krumholz} \& {Thompson}}{{Krumholz} \&
  {Thompson}}{2012}]{krumholz12}
{Krumholz} M.~R.,  {Thompson} T.~A.,  2012, \mn@doi [\apj]
  {10.1088/0004-637X/760/2/155}, \href
  {http://adsabs.harvard.edu/abs/2012ApJ...760..155K} {760, 155}

\bibitem[\protect\citeauthoryear{{Krumholz}, {Stone}  \& {Gardiner}}{{Krumholz}
  et~al.}{2007}]{krumholz07b}
{Krumholz} M.~R.,  {Stone} J.~M.,   {Gardiner} T.~A.,  2007, \mn@doi [\apj]
  {10.1086/522665}, \href {http://adsabs.harvard.edu/abs/2007ApJ...671..518K}
  {671, 518}

\bibitem[\protect\citeauthoryear{{Lada}, {Lombardi}  \& {Alves}}{{Lada}
  et~al.}{2010}]{lada10}
{Lada} C.~J.,  {Lombardi} M.,   {Alves} J.~F.,  2010, \mn@doi [\apj]
  {10.1088/0004-637X/724/1/687}, \href
  {http://adsabs.harvard.edu/abs/2010ApJ...724..687L} {724, 687}

\bibitem[\protect\citeauthoryear{{Larson}}{{Larson}}{1981}]{larson81}
{Larson} R.~B.,  1981, \mn@doi [\mnras] {10.1093/mnras/194.4.809}, \href
  {http://adsabs.harvard.edu/abs/1981MNRAS.194..809L} {194, 809}

\bibitem[\protect\citeauthoryear{{Laursen}, {Sommer-Larsen}  \&
  {Andersen}}{{Laursen} et~al.}{2009}]{laursen09}
{Laursen} P.,  {Sommer-Larsen} J.,   {Andersen} A.~C.,  2009, \mn@doi [\apj]
  {10.1088/0004-637X/704/2/1640}, \href
  {http://adsabs.harvard.edu/abs/2009ApJ...704.1640L} {704, 1640}

\bibitem[\protect\citeauthoryear{{Lee}, {Miville-Desch{\^e}nes}  \&
  {Murray}}{{Lee} et~al.}{2016}]{lee_eve16}
{Lee} E.~J.,  {Miville-Desch{\^e}nes} M.-A.,   {Murray} N.~W.,  2016, \mn@doi
  [\apj] {10.3847/1538-4357/833/2/229}, \href
  {http://adsabs.harvard.edu/abs/2016ApJ...833..229L} {833, 229}

\bibitem[\protect\citeauthoryear{{Leitet}, {Bergvall}, {Hayes}, {Linn{\'e}}  \&
  {Zackrisson}}{{Leitet} et~al.}{2013}]{leitet13}
{Leitet} E.,  {Bergvall} N.,  {Hayes} M.,  {Linn{\'e}} S.,   {Zackrisson} E.,
  2013, \mn@doi [\aap] {10.1051/0004-6361/201118370}, \href
  {http://adsabs.harvard.edu/abs/2013A%26A...553A.106L} {553, A106}

\bibitem[\protect\citeauthoryear{{Leitherer} et~al.,}{{Leitherer}
  et~al.}{1999}]{leitherer99}
{Leitherer} C.,  et~al., 1999, \mn@doi [\apjs] {10.1086/313233}, \href
  {http://adsabs.harvard.edu/abs/1999ApJS..123....3L} {123, 3}

\bibitem[\protect\citeauthoryear{{Leitherer}, {Hernandez}, {Lee}  \&
  {Oey}}{{Leitherer} et~al.}{2016}]{leitherer16}
{Leitherer} C.,  {Hernandez} S.,  {Lee} J.~C.,   {Oey} M.~S.,  2016, \mn@doi
  [\apj] {10.3847/0004-637X/823/1/64}, \href
  {http://adsabs.harvard.edu/abs/2016ApJ...823...64L} {823, 64}

\bibitem[\protect\citeauthoryear{{Leroy} et~al.,}{{Leroy}
  et~al.}{2017}]{leroy17}
{Leroy} A.~K.,  et~al., 2017, \mn@doi [\apj] {10.3847/1538-4357/aa7fef}, \href
  {http://adsabs.harvard.edu/abs/2017ApJ...846...71L} {846, 71}

\bibitem[\protect\citeauthoryear{{Lupi}}{{Lupi}}{2019}]{lupi19}
{Lupi} A.,  2019, \mn@doi [\mnras] {10.1093/mnras/stz100}, \href
  {http://adsabs.harvard.edu/abs/2019MNRAS.484.1687L} {484, 1687}

\bibitem[\protect\citeauthoryear{{Ma}, {Hopkins}, {Kasen}, {Quataert},
  {Faucher-Gigu{\`e}re}, {Kere{\v s}}, {Murray}  \& {Strom}}{{Ma}
  et~al.}{2016}]{ma16}
{Ma} X.,  {Hopkins} P.~F.,  {Kasen} D.,  {Quataert} E.,  {Faucher-Gigu{\`e}re}
  C.-A.,  {Kere{\v s}} D.,  {Murray} N.,   {Strom} A.,  2016, \mn@doi [\mnras]
  {10.1093/mnras/stw941}, \href
  {http://adsabs.harvard.edu/abs/2016MNRAS.459.3614M} {459, 3614}

\bibitem[\protect\citeauthoryear{{Madau}, {Haardt}  \& {Rees}}{{Madau}
  et~al.}{1999}]{madau99}
{Madau} P.,  {Haardt} F.,   {Rees} M.~J.,  1999, \mn@doi [\apj]
  {10.1086/306975}, \href {http://adsabs.harvard.edu/abs/1999ApJ...514..648M}
  {514, 648}

\bibitem[\protect\citeauthoryear{{Marchi} et~al.,}{{Marchi}
  et~al.}{2017}]{marchi17}
{Marchi} F.,  et~al., 2017, \mn@doi [\aap] {10.1051/0004-6361/201630054}, \href
  {http://adsabs.harvard.edu/abs/2017A%26A...601A..73M} {601, A73}

\bibitem[\protect\citeauthoryear{{Massey} \& {Hunter}}{{Massey} \&
  {Hunter}}{1998}]{massey98}
{Massey} P.,  {Hunter} D.~A.,  1998, \mn@doi [\apj] {10.1086/305126}, \href
  {http://adsabs.harvard.edu/abs/1998ApJ...493..180M} {493, 180}

\bibitem[\protect\citeauthoryear{{Matzner}}{{Matzner}}{2002}]{matzner02}
{Matzner} C.~D.,  2002, \mn@doi [\apj] {10.1086/338030}, \href
  {http://adsabs.harvard.edu/abs/2002ApJ...566..302M} {566, 302}

\bibitem[\protect\citeauthoryear{{Mesinger}, {Aykutalp}, {Vanzella},
  {Pentericci}, {Ferrara}  \& {Dijkstra}}{{Mesinger} et~al.}{2015}]{mesinger15}
{Mesinger} A.,  {Aykutalp} A.,  {Vanzella} E.,  {Pentericci} L.,  {Ferrara} A.,
    {Dijkstra} M.,  2015, \mn@doi [\mnras] {10.1093/mnras/stu2089}, \href
  {http://adsabs.harvard.edu/abs/2015MNRAS.446..566M} {446, 566}

\bibitem[\protect\citeauthoryear{{Mostardi}, {Shapley}, {Steidel}, {Trainor},
  {Reddy}  \& {Siana}}{{Mostardi} et~al.}{2015}]{mostardi15}
{Mostardi} R.~E.,  {Shapley} A.~E.,  {Steidel} C.~C.,  {Trainor} R.~F.,
  {Reddy} N.~A.,   {Siana} B.,  2015, \mn@doi [\apj]
  {10.1088/0004-637X/810/2/107}, \href
  {http://adsabs.harvard.edu/abs/2015ApJ...810..107M} {810, 107}

\bibitem[\protect\citeauthoryear{{Neufeld}}{{Neufeld}}{1990}]{neufeld90}
{Neufeld} D.~A.,  1990, \mn@doi [\apj] {10.1086/168375}, \href
  {http://adsabs.harvard.edu/abs/1990ApJ...350..216N} {350, 216}

\bibitem[\protect\citeauthoryear{{Ocvirk} et~al.,}{{Ocvirk}
  et~al.}{2016}]{ocvirk16}
{Ocvirk} P.,  et~al., 2016, \mn@doi [\mnras] {10.1093/mnras/stw2036}, \href
  {http://adsabs.harvard.edu/abs/2016MNRAS.463.1462O} {463, 1462}

\bibitem[\protect\citeauthoryear{{Oey} \& {Clarke}}{{Oey} \&
  {Clarke}}{2005}]{oey05}
{Oey} M.~S.,  {Clarke} C.~J.,  2005, \mn@doi [\apjl] {10.1086/428396}, \href
  {http://adsabs.harvard.edu/abs/2005ApJ...620L..43O} {620, L43}

\bibitem[\protect\citeauthoryear{{Okamoto}, {Gao}  \& {Theuns}}{{Okamoto}
  et~al.}{2008}]{okamoto08}
{Okamoto} T.,  {Gao} L.,   {Theuns} T.,  2008, \mn@doi [\mnras]
  {10.1111/j.1365-2966.2008.13830.x}, \href
  {http://adsabs.harvard.edu/abs/2008MNRAS.390..920O} {390, 920}

\bibitem[\protect\citeauthoryear{{Ono}, {Ouchi}, {Shimasaku}, {Dunlop},
  {Farrah}, {McLure}  \& {Okamura}}{{Ono} et~al.}{2010}]{ono10}
{Ono} Y.,  {Ouchi} M.,  {Shimasaku} K.,  {Dunlop} J.,  {Farrah} D.,  {McLure}
  R.,   {Okamura} S.,  2010, \mn@doi [\apj] {10.1088/0004-637X/724/2/1524},
  \href {http://adsabs.harvard.edu/abs/2010ApJ...724.1524O} {724, 1524}

\bibitem[\protect\citeauthoryear{{Orlitov{\'a}}, {Verhamme}, {Henry},
  {Scarlata}, {Jaskot}, {Oey}  \& {Schaerer}}{{Orlitov{\'a}}
  et~al.}{2018}]{orlitova18}
{Orlitov{\'a}} I.,  {Verhamme} A.,  {Henry} A.,  {Scarlata} C.,  {Jaskot} A.,
  {Oey} M.~S.,   {Schaerer} D.,  2018, \mn@doi [\aap]
  {10.1051/0004-6361/201732478}, \href
  {http://adsabs.harvard.edu/abs/2018A%26A...616A..60O} {616, A60}

\bibitem[\protect\citeauthoryear{{Osterbrock} \& {Ferland}}{{Osterbrock} \&
  {Ferland}}{2006}]{osterbrock06}
{Osterbrock} D.~E.,  {Ferland} G.~J.,  2006, {Astrophysics of gaseous nebulae
  and active galactic nuclei}

\bibitem[\protect\citeauthoryear{{Paardekooper}, {Khochfar}  \& {Dalla
  Vecchia}}{{Paardekooper} et~al.}{2015}]{paardekooper15}
{Paardekooper} J.-P.,  {Khochfar} S.,   {Dalla Vecchia} C.,  2015, \mn@doi
  [\mnras] {10.1093/mnras/stv1114}, \href
  {http://adsabs.harvard.edu/abs/2015MNRAS.451.2544P} {451, 2544}

\bibitem[\protect\citeauthoryear{{Pawlik}, {Schaye}  \& {Dalla
  Vecchia}}{{Pawlik} et~al.}{2015}]{pawlik15}
{Pawlik} A.~H.,  {Schaye} J.,   {Dalla Vecchia} C.,  2015, \mn@doi [\mnras]
  {10.1093/mnras/stv976}, \href
  {http://adsabs.harvard.edu/abs/2015MNRAS.451.1586P} {451, 1586}

\bibitem[\protect\citeauthoryear{{Peters} et~al.,}{{Peters}
  et~al.}{2017}]{peters17}
{Peters} T.,  et~al., 2017, \mn@doi [\mnras] {10.1093/mnras/stw3216}, \href
  {http://adsabs.harvard.edu/abs/2017MNRAS.466.3293P} {466, 3293}

\bibitem[\protect\citeauthoryear{{Pettini}, {Steidel}, {Adelberger},
  {Dickinson}  \& {Giavalisco}}{{Pettini} et~al.}{2000}]{pettini00}
{Pettini} M.,  {Steidel} C.~C.,  {Adelberger} K.~L.,  {Dickinson} M.,
  {Giavalisco} M.,  2000, \mn@doi [\apj] {10.1086/308176}, \href
  {http://adsabs.harvard.edu/abs/2000ApJ...528...96P} {528, 96}

\bibitem[\protect\citeauthoryear{{Puschnig} et~al.,}{{Puschnig}
  et~al.}{2017}]{puschnig17}
{Puschnig} J.,  et~al., 2017, \mn@doi [\mnras] {10.1093/mnras/stx951}, \href
  {http://adsabs.harvard.edu/abs/2017MNRAS.469.3252P} {469, 3252}

\bibitem[\protect\citeauthoryear{{Raga}, {Cant{\'o}}  \&
  {Rodr{\'{\i}}guez}}{{Raga} et~al.}{2012}]{raga12}
{Raga} A.~C.,  {Cant{\'o}} J.,   {Rodr{\'{\i}}guez} L.~F.,  2012, \mn@doi
  [\mnras] {10.1111/j.1745-3933.2011.01173.x}, \href
  {http://adsabs.harvard.edu/abs/2012MNRAS.419L..39R} {419, L39}

\bibitem[\protect\citeauthoryear{{Rivera-Thorsen} et~al.,}{{Rivera-Thorsen}
  et~al.}{2017}]{rivera-thorsen17}
{Rivera-Thorsen} T.~E.,  et~al., 2017, \mn@doi [\aap]
  {10.1051/0004-6361/201732173}, \href
  {http://adsabs.harvard.edu/abs/2017A%26A...608L...4R} {608, L4}

\bibitem[\protect\citeauthoryear{{Roman-Duval}, {Jackson}, {Heyer}, {Rathborne}
   \& {Simon}}{{Roman-Duval} et~al.}{2010}]{roman-duval10}
{Roman-Duval} J.,  {Jackson} J.~M.,  {Heyer} M.,  {Rathborne} J.,   {Simon} R.,
   2010, \mn@doi [\apj] {10.1088/0004-637X/723/1/492}, \href
  {http://adsabs.harvard.edu/abs/2010ApJ...723..492R} {723, 492}

\bibitem[\protect\citeauthoryear{{Rosdahl} \& {Blaizot}}{{Rosdahl} \&
  {Blaizot}}{2012}]{rosdahl12}
{Rosdahl} J.,  {Blaizot} J.,  2012, \mn@doi [\mnras]
  {10.1111/j.1365-2966.2012.20883.x}, \href
  {http://adsabs.harvard.edu/abs/2012MNRAS.423..344R} {423, 344}

\bibitem[\protect\citeauthoryear{{Rosdahl} \& {Teyssier}}{{Rosdahl} \&
  {Teyssier}}{2015}]{rosdahl15}
{Rosdahl} J.,  {Teyssier} R.,  2015, \mn@doi [\mnras] {10.1093/mnras/stv567},
  \href {http://adsabs.harvard.edu/abs/2015MNRAS.449.4380R} {449, 4380}

\bibitem[\protect\citeauthoryear{{Rosdahl}, {Blaizot}, {Aubert}, {Stranex}  \&
  {Teyssier}}{{Rosdahl} et~al.}{2013}]{rosdahl13}
{Rosdahl} J.,  {Blaizot} J.,  {Aubert} D.,  {Stranex} T.,   {Teyssier} R.,
  2013, \mn@doi [\mnras] {10.1093/mnras/stt1722}, \href
  {http://adsabs.harvard.edu/abs/2013MNRAS.436.2188R} {436, 2188}

\bibitem[\protect\citeauthoryear{{Rosdahl}, {Schaye}, {Teyssier}  \&
  {Agertz}}{{Rosdahl} et~al.}{2015}]{rosdahl15b}
{Rosdahl} J.,  {Schaye} J.,  {Teyssier} R.,   {Agertz} O.,  2015,
  astro-ph/1501.04632, \href
  {http://adsabs.harvard.edu/abs/2015arXiv150104632R} {}

\bibitem[\protect\citeauthoryear{{Rosdahl} et~al.,}{{Rosdahl}
  et~al.}{2018}]{rosdahl18}
{Rosdahl} J.,  et~al., 2018, \mn@doi [\mnras] {10.1093/mnras/sty1655}, \href
  {http://adsabs.harvard.edu/abs/2018MNRAS.479..994R} {479, 994}

\bibitem[\protect\citeauthoryear{{Safarzadeh} \& {Scannapieco}}{{Safarzadeh} \&
  {Scannapieco}}{2016}]{safarzadeh16}
{Safarzadeh} M.,  {Scannapieco} E.,  2016, \mn@doi [\apjl]
  {10.3847/2041-8205/832/1/L9}, \href
  {http://adsabs.harvard.edu/abs/2016ApJ...832L...9S} {832, L9}

\bibitem[\protect\citeauthoryear{{Sana} et~al.,}{{Sana} et~al.}{2012}]{sana12}
{Sana} H.,  et~al., 2012, \mn@doi [Science] {10.1126/science.1223344}, \href
  {http://adsabs.harvard.edu/abs/2012Sci...337..444S} {337, 444}

\bibitem[\protect\citeauthoryear{{Schenker}, {Ellis}, {Konidaris}  \&
  {Stark}}{{Schenker} et~al.}{2014}]{schenker14}
{Schenker} M.~A.,  {Ellis} R.~S.,  {Konidaris} N.~P.,   {Stark} D.~P.,  2014,
  \mn@doi [\apj] {10.1088/0004-637X/795/1/20}, \href
  {http://adsabs.harvard.edu/abs/2014ApJ...795...20S} {795, 20}

\bibitem[\protect\citeauthoryear{{Shapley}, {Steidel}, {Pettini}  \&
  {Adelberger}}{{Shapley} et~al.}{2003}]{shapley03}
{Shapley} A.~E.,  {Steidel} C.~C.,  {Pettini} M.,   {Adelberger} K.~L.,  2003,
  \mn@doi [\apj] {10.1086/373922}, \href
  {http://adsabs.harvard.edu/abs/2003ApJ...588...65S} {588, 65}

\bibitem[\protect\citeauthoryear{{Shapley}, {Steidel}, {Strom},
  {Bogosavljevi{\'c}}, {Reddy}, {Siana}, {Mostardi}  \& {Rudie}}{{Shapley}
  et~al.}{2016}]{shapley16}
{Shapley} A.~E.,  {Steidel} C.~C.,  {Strom} A.~L.,  {Bogosavljevi{\'c}} M.,
  {Reddy} N.~A.,  {Siana} B.,  {Mostardi} R.~E.,   {Rudie} G.~C.,  2016,
  \mn@doi [\apjl] {10.3847/2041-8205/826/2/L24}, \href
  {http://adsabs.harvard.edu/abs/2016ApJ...826L..24S} {826, L24}

\bibitem[\protect\citeauthoryear{{Shull}}{{Shull}}{1978}]{shull78}
{Shull} J.~M.,  1978, \mn@doi [\apj] {10.1086/156433}, \href
  {http://adsabs.harvard.edu/abs/1978ApJ...224..841S} {224, 841}

\bibitem[\protect\citeauthoryear{{Siana} et~al.,}{{Siana}
  et~al.}{2010}]{siana10}
{Siana} B.,  et~al., 2010, \mn@doi [\apj] {10.1088/0004-637X/723/1/241}, \href
  {http://adsabs.harvard.edu/abs/2010ApJ...723..241S} {723, 241}

\bibitem[\protect\citeauthoryear{{Skinner} \& {Ostriker}}{{Skinner} \&
  {Ostriker}}{2015}]{skinner15}
{Skinner} M.~A.,  {Ostriker} E.~C.,  2015, \mn@doi [\apj]
  {10.1088/0004-637X/809/2/187}, \href
  {http://adsabs.harvard.edu/abs/2015ApJ...809..187S} {809, 187}

\bibitem[\protect\citeauthoryear{{Smith}, {Bromm}  \& {Loeb}}{{Smith}
  et~al.}{2017}]{smith17}
{Smith} A.,  {Bromm} V.,   {Loeb} A.,  2017, \mn@doi [\mnras]
  {10.1093/mnras/stw2591}, \href
  {http://adsabs.harvard.edu/abs/2017MNRAS.464.2963S} {464, 2963}

\bibitem[\protect\citeauthoryear{{Smith}, {Sijacki}  \& {Shen}}{{Smith}
  et~al.}{2018}]{smith18m}
{Smith} M.~C.,  {Sijacki} D.,   {Shen} S.,  2018, \mn@doi [\mnras]
  {10.1093/mnras/sty994}, \href
  {http://adsabs.harvard.edu/abs/2018MNRAS.478..302S} {478, 302}

\bibitem[\protect\citeauthoryear{{Smith}, {Ma}, {Bromm}, {Finkelstein},
  {Hopkins}, {Faucher-Gigu{\`e}re}  \& {Kere{\v s}}}{{Smith}
  et~al.}{2019}]{smith19}
{Smith} A.,  {Ma} X.,  {Bromm} V.,  {Finkelstein} S.~L.,  {Hopkins} P.~F.,
  {Faucher-Gigu{\`e}re} C.-A.,   {Kere{\v s}} D.,  2019, \mn@doi [\mnras]
  {10.1093/mnras/sty3483}, \href
  {http://adsabs.harvard.edu/abs/2019MNRAS.484...39S} {484, 39}

\bibitem[\protect\citeauthoryear{{Somerville}}{{Somerville}}{2002}]{somerville02}
{Somerville} R.~S.,  2002, \mn@doi [\apjl] {10.1086/341444}, \href
  {http://adsabs.harvard.edu/abs/2002ApJ...572L..23S} {572, L23}

\bibitem[\protect\citeauthoryear{{Song} et~al.,}{{Song} et~al.}{2014}]{song14}
{Song} M.,  et~al., 2014, \mn@doi [\apj] {10.1088/0004-637X/791/1/3}, \href
  {http://adsabs.harvard.edu/abs/2014ApJ...791....3S} {791, 3}

\bibitem[\protect\citeauthoryear{{Stanway}, {Eldridge}  \& {Becker}}{{Stanway}
  et~al.}{2016}]{stanway16}
{Stanway} E.~R.,  {Eldridge} J.~J.,   {Becker} G.~D.,  2016, \mn@doi [\mnras]
  {10.1093/mnras/stv2661}, \href
  {http://adsabs.harvard.edu/abs/2016MNRAS.456..485S} {456, 485}

\bibitem[\protect\citeauthoryear{{Stark}, {Ellis}, {Chiu}, {Ouchi}  \&
  {Bunker}}{{Stark} et~al.}{2010}]{stark10}
{Stark} D.~P.,  {Ellis} R.~S.,  {Chiu} K.,  {Ouchi} M.,   {Bunker} A.,  2010,
  \mn@doi [\mnras] {10.1111/j.1365-2966.2010.17227.x}, \href
  {http://adsabs.harvard.edu/abs/2010MNRAS.408.1628S} {408, 1628}

\bibitem[\protect\citeauthoryear{{Steidel}, {Pettini}  \&
  {Adelberger}}{{Steidel} et~al.}{2001}]{steidel01}
{Steidel} C.~C.,  {Pettini} M.,   {Adelberger} K.~L.,  2001, \mn@doi [\apj]
  {10.1086/318323}, \href {http://adsabs.harvard.edu/abs/2001ApJ...546..665S}
  {546, 665}

\bibitem[\protect\citeauthoryear{{Steidel}, {Erb}, {Shapley}, {Pettini},
  {Reddy}, {Bogosavljevi{\'c}}, {Rudie}  \& {Rakic}}{{Steidel}
  et~al.}{2010}]{steidel10}
{Steidel} C.~C.,  {Erb} D.~K.,  {Shapley} A.~E.,  {Pettini} M.,  {Reddy} N.,
  {Bogosavljevi{\'c}} M.,  {Rudie} G.~C.,   {Rakic} O.,  2010, \mn@doi [\apj]
  {10.1088/0004-637X/717/1/289}, \href
  {http://adsabs.harvard.edu/abs/2010ApJ...717..289S} {717, 289}

\bibitem[\protect\citeauthoryear{{Steidel}, {Bogosavljevi{\'c}}, {Shapley},
  {Reddy}, {Rudie}, {Pettini}, {Trainor}  \& {Strom}}{{Steidel}
  et~al.}{2018}]{steidel18}
{Steidel} C.~C.,  {Bogosavljevi{\'c}} M.,  {Shapley} A.~E.,  {Reddy} N.~A.,
  {Rudie} G.~C.,  {Pettini} M.,  {Trainor} R.~F.,   {Strom} A.~L.,  2018,
  \mn@doi [\apj] {10.3847/1538-4357/aaed28}, \href
  {http://adsabs.harvard.edu/abs/2018ApJ...869..123S} {869, 123}

\bibitem[\protect\citeauthoryear{{Tamura} et~al.,}{{Tamura}
  et~al.}{2018}]{tamura18}
{Tamura} Y.,  et~al., 2018, preprint, \href
  {http://adsabs.harvard.edu/abs/2018arXiv180604132T} {} (\mn@eprint {arXiv}
  {1806.04132})

\bibitem[\protect\citeauthoryear{{Teyssier}}{{Teyssier}}{2002}]{teyssier02}
{Teyssier} R.,  2002, \mn@doi [\aap] {10.1051/0004-6361:20011817}, \href
  {http://adsabs.harvard.edu/abs/2002A%26A...385..337T} {385, 337}

\bibitem[\protect\citeauthoryear{{Thornton}, {Gaudlitz}, {Janka}  \&
  {Steinmetz}}{{Thornton} et~al.}{1998}]{thornton98}
{Thornton} K.,  {Gaudlitz} M.,  {Janka} H.-T.,   {Steinmetz} M.,  1998, \mn@doi
  [\apj] {10.1086/305704}, \href
  {http://adsabs.harvard.edu/abs/1998ApJ...500...95T} {500, 95}

\bibitem[\protect\citeauthoryear{{Topping} \& {Shull}}{{Topping} \&
  {Shull}}{2015}]{topping15}
{Topping} M.~W.,  {Shull} J.~M.,  2015, \mn@doi [\apj]
  {10.1088/0004-637X/800/2/97}, \href
  {http://adsabs.harvard.edu/abs/2015ApJ...800...97T} {800, 97}

\bibitem[\protect\citeauthoryear{{Toro}, {Spruce}  \& {Speares}}{{Toro}
  et~al.}{1994}]{toro94}
{Toro} E.~F.,  {Spruce} M.,   {Speares} W.,  1994, \mn@doi [Shock Waves]
  {10.1007/BF01414629}, \href
  {http://adsabs.harvard.edu/abs/1994ShWav...4...25T} {4, 25}

\bibitem[\protect\citeauthoryear{{Trebitsch}, {Blaizot}, {Rosdahl}, {Devriendt}
   \& {Slyz}}{{Trebitsch} et~al.}{2017}]{trebitsch17}
{Trebitsch} M.,  {Blaizot} J.,  {Rosdahl} J.,  {Devriendt} J.,   {Slyz} A.,
  2017, \mn@doi [\mnras] {10.1093/mnras/stx1060}, \href
  {http://adsabs.harvard.edu/abs/2017MNRAS.470..224T} {470, 224}

\bibitem[\protect\citeauthoryear{{Treu}, {Schmidt}, {Trenti}, {Bradley}  \&
  {Stiavelli}}{{Treu} et~al.}{2013}]{treu13}
{Treu} T.,  {Schmidt} K.~B.,  {Trenti} M.,  {Bradley} L.~D.,   {Stiavelli} M.,
  2013, \mn@doi [\apjl] {10.1088/2041-8205/775/1/L29}, \href
  {http://adsabs.harvard.edu/abs/2013ApJ...775L..29T} {775, L29}

\bibitem[\protect\citeauthoryear{{Tsang} \& {Milosavljevi{\'c}}}{{Tsang} \&
  {Milosavljevi{\'c}}}{2015}]{tsang15}
{Tsang} B.~T.-H.,  {Milosavljevi{\'c}} M.,  2015, \mn@doi [\mnras]
  {10.1093/mnras/stv1707}, \href
  {http://adsabs.harvard.edu/abs/2015MNRAS.453.1108T} {453, 1108}

\bibitem[\protect\citeauthoryear{{Vanzella} et~al.,}{{Vanzella}
  et~al.}{2015}]{vanzella15}
{Vanzella} E.,  et~al., 2015, \mn@doi [\aap] {10.1051/0004-6361/201525651},
  \href {http://adsabs.harvard.edu/abs/2015A%26A...576A.116V} {576, A116}

\bibitem[\protect\citeauthoryear{{Vanzella} et~al.,}{{Vanzella}
  et~al.}{2018}]{vanzella18}
{Vanzella} E.,  et~al., 2018, \mn@doi [\mnras] {10.1093/mnrasl/sly023}, \href
  {http://adsabs.harvard.edu/abs/2018MNRAS.476L..15V} {476, L15}

\bibitem[\protect\citeauthoryear{{Verhamme}, {Schaerer}  \&
  {Maselli}}{{Verhamme} et~al.}{2006}]{verhamme06}
{Verhamme} A.,  {Schaerer} D.,   {Maselli} A.,  2006, \mn@doi [\aap]
  {10.1051/0004-6361:20065554}, \href
  {http://adsabs.harvard.edu/abs/2006A%26A...460..397V} {460, 397}

\bibitem[\protect\citeauthoryear{{Verhamme}, {Schaerer}, {Atek}  \&
  {Tapken}}{{Verhamme} et~al.}{2008}]{verhamme08}
{Verhamme} A.,  {Schaerer} D.,  {Atek} H.,   {Tapken} C.,  2008, \mn@doi [\aap]
  {10.1051/0004-6361:200809648}, \href
  {http://adsabs.harvard.edu/abs/2008A%26A...491...89V} {491, 89}

\bibitem[\protect\citeauthoryear{{Verhamme}, {Dubois}, {Blaizot}, {Garel},
  {Bacon}, {Devriendt}, {Guiderdoni}  \& {Slyz}}{{Verhamme}
  et~al.}{2012}]{verhamme12}
{Verhamme} A.,  {Dubois} Y.,  {Blaizot} J.,  {Garel} T.,  {Bacon} R.,
  {Devriendt} J.,  {Guiderdoni} B.,   {Slyz} A.,  2012, \mn@doi [\aap]
  {10.1051/0004-6361/201218783}, \href
  {http://adsabs.harvard.edu/abs/2012A%26A...546A.111V} {546, A111}

\bibitem[\protect\citeauthoryear{{Verhamme}, {Orlitov{\'a}}, {Schaerer}  \&
  {Hayes}}{{Verhamme} et~al.}{2015}]{verhamme15}
{Verhamme} A.,  {Orlitov{\'a}} I.,  {Schaerer} D.,   {Hayes} M.,  2015, \mn@doi
  [\aap] {10.1051/0004-6361/201423978}, \href
  {http://adsabs.harvard.edu/abs/2015A%26A...578A...7V} {578, A7}

\bibitem[\protect\citeauthoryear{{Verhamme}, {Orlitov{\'a}}, {Schaerer},
  {Izotov}, {Worseck}, {Thuan}  \& {Guseva}}{{Verhamme}
  et~al.}{2017}]{verhamme17}
{Verhamme} A.,  {Orlitov{\'a}} I.,  {Schaerer} D.,  {Izotov} Y.,  {Worseck} G.,
   {Thuan} T.~X.,   {Guseva} N.,  2017, \mn@doi [\aap]
  {10.1051/0004-6361/201629264}, \href
  {http://adsabs.harvard.edu/abs/2017A%26A...597A..13V} {597, A13}

\bibitem[\protect\citeauthoryear{{Vutisalchavakul}, {Evans}  \&
  {Heyer}}{{Vutisalchavakul} et~al.}{2016}]{vutisalchavakul16}
{Vutisalchavakul} N.,  {Evans} II N.~J.,   {Heyer} M.,  2016, \mn@doi [\apj]
  {10.3847/0004-637X/831/1/73}, \href
  {http://adsabs.harvard.edu/abs/2016ApJ...831...73V} {831, 73}

\bibitem[\protect\citeauthoryear{{Walch}, {Whitworth}, {Bisbas}, {W{\"u}nsch}
  \& {Hubber}}{{Walch} et~al.}{2012}]{walch12}
{Walch} S.~K.,  {Whitworth} A.~P.,  {Bisbas} T.,  {W{\"u}nsch} R.,   {Hubber}
  D.,  2012, \mn@doi [\mnras] {10.1111/j.1365-2966.2012.21767.x}, \href
  {http://adsabs.harvard.edu/abs/2012MNRAS.427..625W} {427, 625}

\bibitem[\protect\citeauthoryear{{Weinberger}, {Kulkarni}, {Haehnelt},
  {Choudhury}  \& {Puchwein}}{{Weinberger} et~al.}{2018}]{weinberger18}
{Weinberger} L.~H.,  {Kulkarni} G.,  {Haehnelt} M.~G.,  {Choudhury} T.~R.,
  {Puchwein} E.,  2018, \mn@doi [\mnras] {10.1093/mnras/sty1563}, \href
  {http://adsabs.harvard.edu/abs/2018MNRAS.479.2564W} {479, 2564}

\bibitem[\protect\citeauthoryear{{Weingartner} \& {Draine}}{{Weingartner} \&
  {Draine}}{2001}]{weingartner01}
{Weingartner} J.~C.,  {Draine} B.~T.,  2001, \mn@doi [\apj] {10.1086/318651},
  \href {http://adsabs.harvard.edu/abs/2001ApJ...548..296W} {548, 296}

\bibitem[\protect\citeauthoryear{{Wise} \& {Cen}}{{Wise} \&
  {Cen}}{2009}]{wise09}
{Wise} J.~H.,  {Cen} R.,  2009, \mn@doi [\apj] {10.1088/0004-637X/693/1/984},
  \href {http://adsabs.harvard.edu/abs/2009ApJ...693..984W} {693, 984}

\bibitem[\protect\citeauthoryear{{Wise}, {Demchenko}, {Halicek}, {Norman},
  {Turk}, {Abel}  \& {Smith}}{{Wise} et~al.}{2014}]{wise14}
{Wise} J.~H.,  {Demchenko} V.~G.,  {Halicek} M.~T.,  {Norman} M.~L.,  {Turk}
  M.~J.,  {Abel} T.,   {Smith} B.~D.,  2014, \mn@doi [\mnras]
  {10.1093/mnras/stu979}, \href
  {http://adsabs.harvard.edu/abs/2014MNRAS.442.2560W} {442, 2560}

\bibitem[\protect\citeauthoryear{{Xu}, {Wise}, {Norman}, {Ahn}  \&
  {O'Shea}}{{Xu} et~al.}{2016}]{xu16}
{Xu} H.,  {Wise} J.~H.,  {Norman} M.~L.,  {Ahn} K.,   {O'Shea} B.~W.,  2016,
  \mn@doi [\apj] {10.3847/1538-4357/833/1/84}, \href
  {http://adsabs.harvard.edu/abs/2016ApJ...833...84X} {833, 84}

\bibitem[\protect\citeauthoryear{{Yajima}, {Li}, {Zhu}, {Abel}, {Gronwall}  \&
  {Ciardullo}}{{Yajima} et~al.}{2014}]{yajima14}
{Yajima} H.,  {Li} Y.,  {Zhu} Q.,  {Abel} T.,  {Gronwall} C.,   {Ciardullo} R.,
   2014, \mn@doi [\mnras] {10.1093/mnras/stu299}, \href
  {http://adsabs.harvard.edu/abs/2014MNRAS.440..776Y} {440, 776}

\bibitem[\protect\citeauthoryear{{Yan}, {Sadeghpour}  \& {Dalgarno}}{{Yan}
  et~al.}{1998}]{yan98}
{Yan} M.,  {Sadeghpour} H.~R.,   {Dalgarno} A.,  1998, \mn@doi [\apj]
  {10.1086/305420}, \href {http://adsabs.harvard.edu/abs/1998ApJ...496.1044Y}
  {496, 1044}

\bibitem[\protect\citeauthoryear{{Yan}, {Sadeghpour}  \& {Dalgarno}}{{Yan}
  et~al.}{2001}]{yan01}
{Yan} M.,  {Sadeghpour} H.~R.,   {Dalgarno} A.,  2001, \mn@doi [\apj]
  {10.1086/322775}, \href {http://adsabs.harvard.edu/abs/2001ApJ...559.1194Y}
  {559, 1194}

\bibitem[\protect\citeauthoryear{{Yang}, {Malhotra}, {Gronke}, {Rhoads},
  {Dijkstra}, {Jaskot}, {Zheng}  \& {Wang}}{{Yang} et~al.}{2016}]{yang16}
{Yang} H.,  {Malhotra} S.,  {Gronke} M.,  {Rhoads} J.~E.,  {Dijkstra} M.,
  {Jaskot} A.,  {Zheng} Z.,   {Wang} J.,  2016, \mn@doi [\apj]
  {10.3847/0004-637X/820/2/130}, \href
  {http://adsabs.harvard.edu/abs/2016ApJ...820..130Y} {820, 130}

\bibitem[\protect\citeauthoryear{{de Barros} et~al.,}{{de Barros}
  et~al.}{2016}]{de-barros16}
{de Barros} S.,  et~al., 2016, \mn@doi [\aap] {10.1051/0004-6361/201527046},
  \href {http://adsabs.harvard.edu/abs/2016A%26A...585A..51D} {585, A51}

\makeatother
\end{thebibliography}
\input{ms.bbl}

\end{document}